\documentclass{egpubl}
\usepackage{eg2025}
 
\ConferencePaper

\CGFStandardLicense

\usepackage[T1]{fontenc}
\usepackage{dfadobe}  

\usepackage{cite}
\BibtexOrBiblatex
\electronicVersion
\PrintedOrElectronic
\ifpdf \usepackage[pdftex]{graphicx} \pdfcompresslevel=9
\else \usepackage[dvips]{graphicx} \fi

\usepackage{egweblnk} 
\usepackage{lineno}

\usepackage{booktabs}
\usepackage{subcaption}
\usepackage[export]{adjustbox}
\usepackage[figuresleft]{rotating}
\usepackage{float}
\usepackage{caption}
\usepackage{graphicx}
\usepackage{tabularx}
\usepackage{wrapfig}
\usepackage{color}
\usepackage[utf8]{inputenc}
\usepackage{mathtools}
\usepackage{enumitem}
\usepackage{overpic}
\usepackage{multicol}
\usepackage[ruled]{algorithm2e}
\usepackage[normalem]{ulem}

\SetAlFnt{\small}
\SetAlCapFnt{\small}
\SetAlCapNameFnt{\small}
\SetAlCapHSkip{0pt}

\newenvironment{parWithWrapFigure} %
{\begingroup
\setlength{\columnsep}{1em}%
\setlength{\intextsep}{0em}%
\setlength{\arraycolsep}{0pt}} %
{

\endgroup}

\newcommand{\FLIP}{\protect\reflectbox{F}LIP\xspace}

\begin{document}
\title{FastAtlas: Real-Time Compact Atlases for Texture Space Shading}
\author[Nicholas Vining, Zander Majercik, Floria Gu et al.]
{\parbox{\textwidth}{\centering N. Vining$^{1,2}$, Z. Majercik$^{3,*}$, F. Gu$^{2,*}$, T. Takikawa$^{4}$, T. Trusty$^{4}$, P. Lalonde$^{1}$, M. McGuire$^{5,6}$, A. Sheffer$^{2}$
        }
        \\
{\parbox{\textwidth}{\centering $^1$NVIDIA, Canada\\
$^2$University of British Columbia, Canada\\
$^3$Stanford University, USA\\
$^4$University of Toronto, Canada\\
$^5$Pasteur Labs, Canada\\
$^6$University of Waterloo, Canada\\
$^*$(denotes equal contribution)
       }
}
}

\teaser{
\vspace{-1cm}
\includegraphics[width=\linewidth]{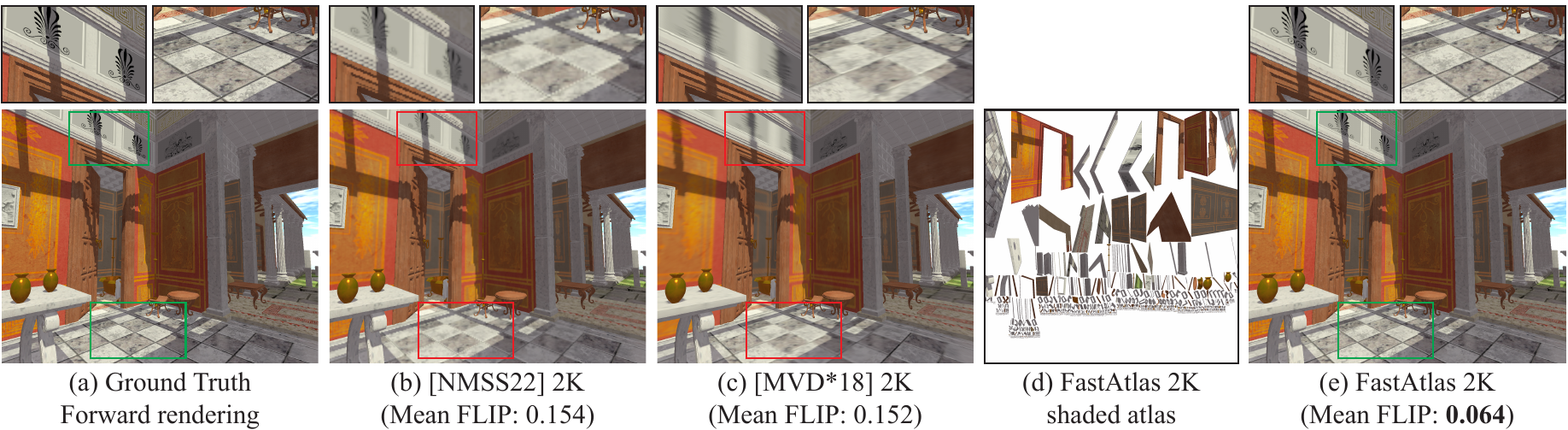}
\caption{FastAtlas tightly packs general charts into atlases in real time (d), facilitating quality texture space shading (TSS) (e) even with highly limited atlas sizes (here $2K\times 2K$). Our TSS renders (e) more closely match ground truth forward rendering (a), qualitatively and quantitatively (measured using the \FLIP perceptual difference estimator\cite{andersson2020FLIP}), than those obtained by state-of-the-art alternatives which rely on less effective atlassing strategies: MSA  \cite{Neff2022MSA} (b), and SAS \cite{mueller2018shading} (c). As the  close-ups demonstrate, outputs rendered using these methods exhibit notable blur (b,c) and smearing (c), while ours remain close to ground truth.  Please zoom in to see fine details.}
\label{fig:teaser}
}

\maketitle

\begin{abstract}
Texture-space shading (TSS) methods decouple shading and rasterization, allowing shading to be performed at a different framerate and spatial resolution than rasterization. TSS has many potential applications, including streaming shading across networks, and reducing rendering cost via shading reuse across consecutive frames and/or shading at reduced resolutions relative to display resolution.  Real-time TSS shading requires texture atlases small enough to be easily stored in GPU memory. Using static atlases leads to significant space wastage, motivating real-time per-frame atlassing strategies that pack only the content visible in each frame.   
We propose {\em FastAtlas}, a novel atlasing method that runs entirely on the GPU and is fast enough to be performed at interactive rates per-frame. Our method combines new per-frame chart computation and parametrization strategies and an efficient general chart packing algorithm. Our chartification strategy removes visible seams in output renders, and our parameterization ensures a constant texel-to-pixel ratio, avoiding undesirable undersampling artifacts.  Our packing method is more general, and produces more tightly packed atlases, than previous work. Jointly, these innovations enable us to produce shading outputs of significantly higher visual quality than those produced using alternative atlasing strategies. 
We validate FastAtlas by shading and rendering challenging scenes using different atlasing settings, reflecting the needs of different TSS applications (temporal reuse, streaming, reduced or elevated shading rates). We extensively compare FastAtlas to prior alternatives and demonstrate that it achieves better shading quality and reduces texture stretch compared to prior approaches using the same settings.

\end{abstract}

\section{Introduction}

\begin{figure}
\includegraphics[width=\linewidth]{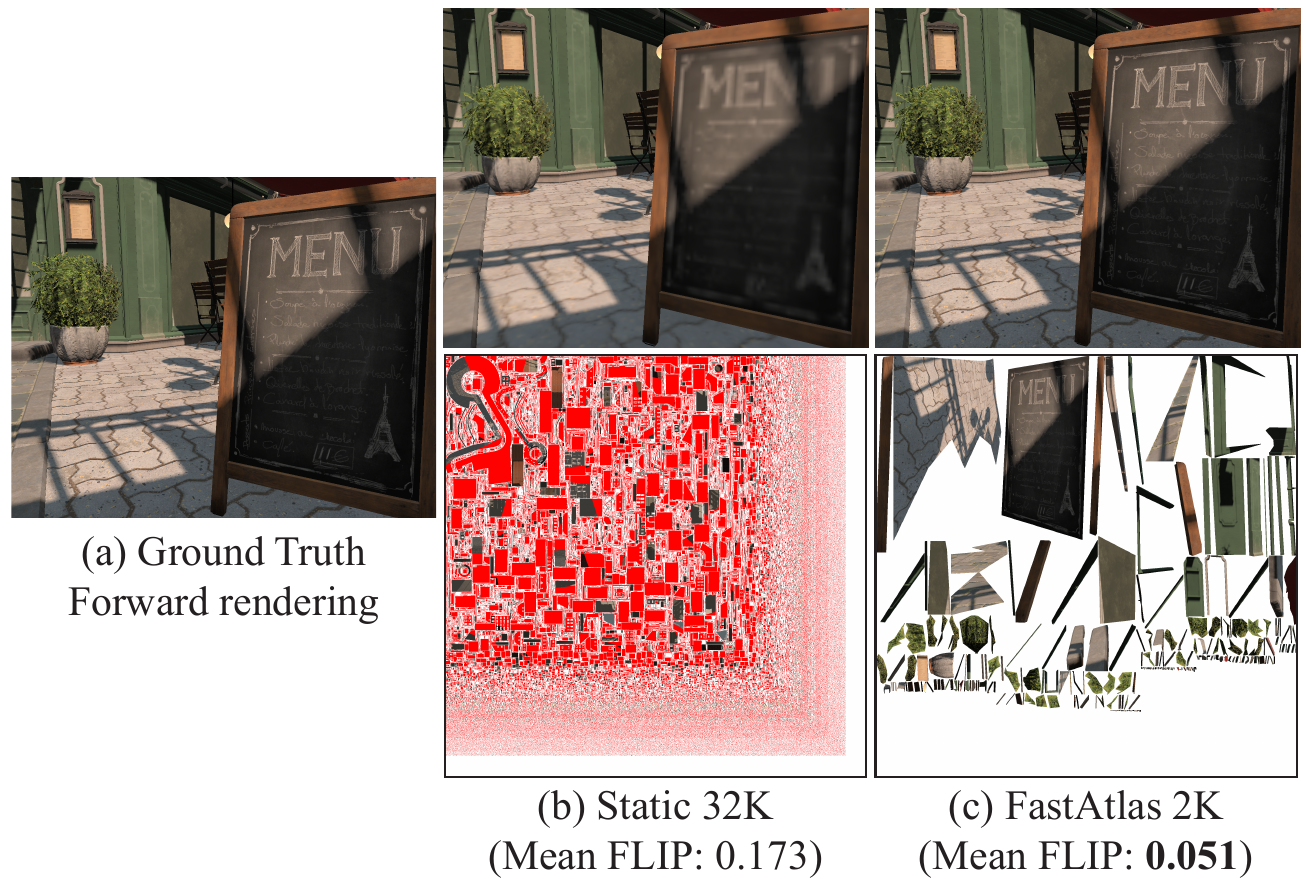}
\caption{(a) Input frame shaded using forward rendering; (b) $32\operatorname{k} \times 32\operatorname{k}$ static atlas of the scene (bottom, content not visible in the frame shaded in red) and TSS render of the input scene using this atlas (top); (c) FastAtlas $2\operatorname{k} \times 2\operatorname{k}$ atlas of the frame (bottom) and corresponding TSS render (top). As reflected by the \FLIP error, using static atlases leads to significant undersampling even at high atlas resolutions. Our results yield higher quality than traditional static atlases at 1/256th of the memory footprint. (Zoom in to check the dessert of the day.)}
\vspace{-6mm}
\label{fig:static}
\end{figure}

In modern rendering, shading is traditionally performed jointly with visibility determination during rasterization or ray tracing, and is typically the most time-consuming part of image synthesis.
With standard forward rendering, the shading rate per-pixel, per-frame, is equal to the number of visible samples. Decoupling shading from visibility computation allows shading to be computed at a different frame rate than rasterization ({\em temporal} decoupling), or at a different spatial resolution ({\em spatial} decoupling). 
Decoupled shading has many applications \cite{ragan2011decoupled,baker2012rock,mueller2021tasa,baker:2016,hillesland2016texel,Neff2022MSA,Baker2022,mueller2018shading,hladky2019tessellated,hladky2021snakebinning,Neff2022MSA}.
Shading at a lower frame rate and/or resolution can yield significant time savings \cite{mueller2021tasa}; alternatively, shading at a higher resolution than display resolution can improve visual quality through more accuracy or reduced flicker \cite{baker2012rock,Baker2022}. 
Decoupled shading can also facilitate applications such as streaming shading across a network, and improve the performance of applications that benefit from shading at a lower resolution than rasterization, such as foveated rendering for near-eye virtual reality displays, and depth-of-field blur. 
Since decoupled shading is not directly supported on commercially available GPUs, the most promising approach for implementing it on existing architectures has historically been Texture-Space Shading (TSS) (e.g. \cite{baker:2016,hillesland2016texel,Neff2022MSA,mueller2018shading,Baker2022}). 
TSS methods first compute shading information and store it in a texture atlas, and then use the shaded atlas during rasterization. We propose {\em FastAtlas}, a new real-time texture atlasing approach that specifically targets TSS applications. FastAtlas computes atlas charts and packing dynamically  and enables better render quality than the state of the art (e.g. Figs. ~\ref{fig:teaser},~\ref{fig:static}).

\begin{figure*}
\includegraphics[width=\linewidth]{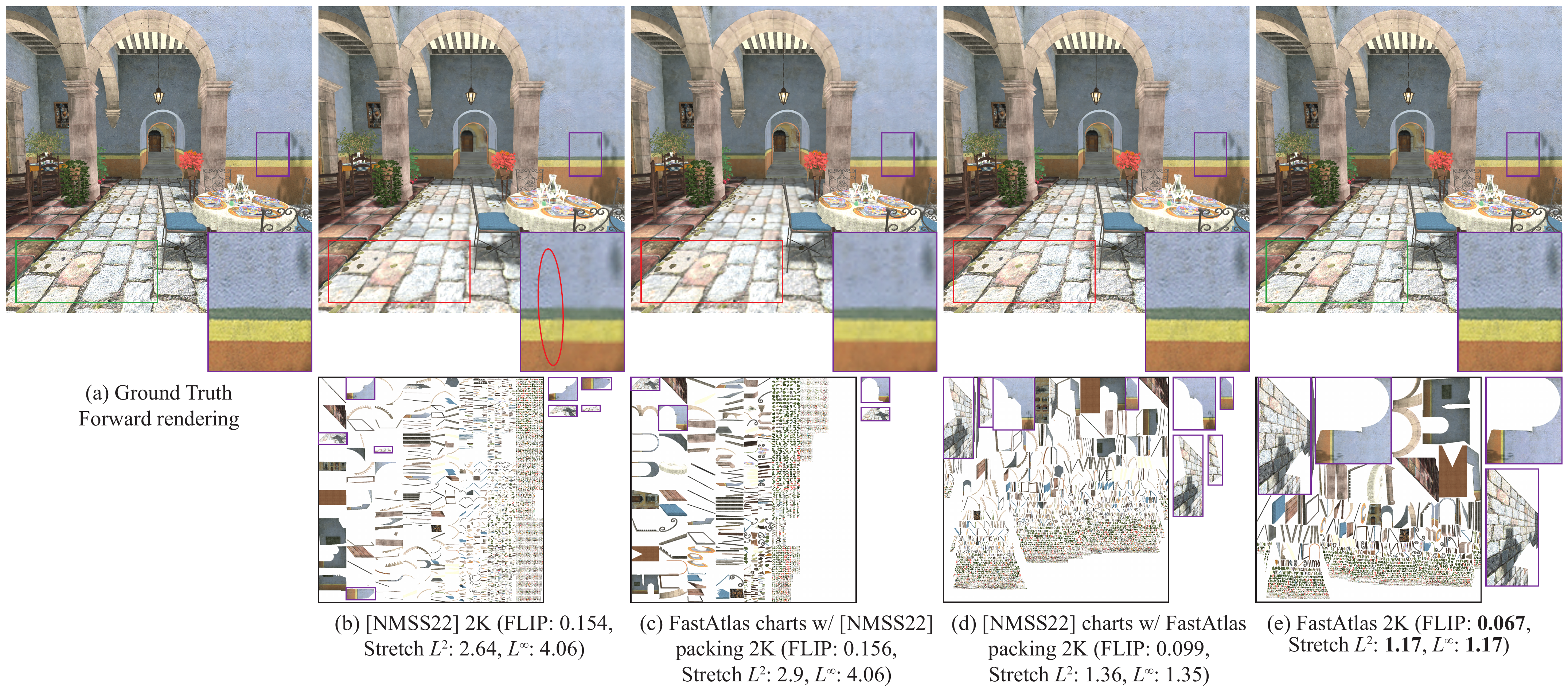}
\vspace{-6pt}
\caption{
FastAtlas vs. \cite{Neff2022MSA}: (a) reference forward renderer; (b) atlas computed using \cite{Neff2022MSA}, and resulting render; (c) atlas computed by packing FastAtlas charts using \cite{Neff2022MSA} packing algorithm and resulting render; (d) atlas computed by applying FastAtlas packing to \cite{Neff2022MSA} charts and resulting render; (e) FastAtlas generated atlas and resulting render. All atlases are $2K \times 2K$. While both methods use projection to parameterize charts, FastAtlas  preserves the input chart scale much better than \cite{Neff2022MSA} ($L^{\infty}$ stretch of 1.17 (e) vs. 4.06 (b)). Lower stretch leads to better render quality (b,d). We further improve visual quality by using charts with no visible seams (a vs b and c). Highlighted regions and their corresponding charts (insets) directly illustrate the impact of stretch (scale) on render quality.}   
\label{fig:atlas}
\label{fig:san_miguel}
\end{figure*}

\begin{figure*}[t]
\includegraphics[width=\linewidth]{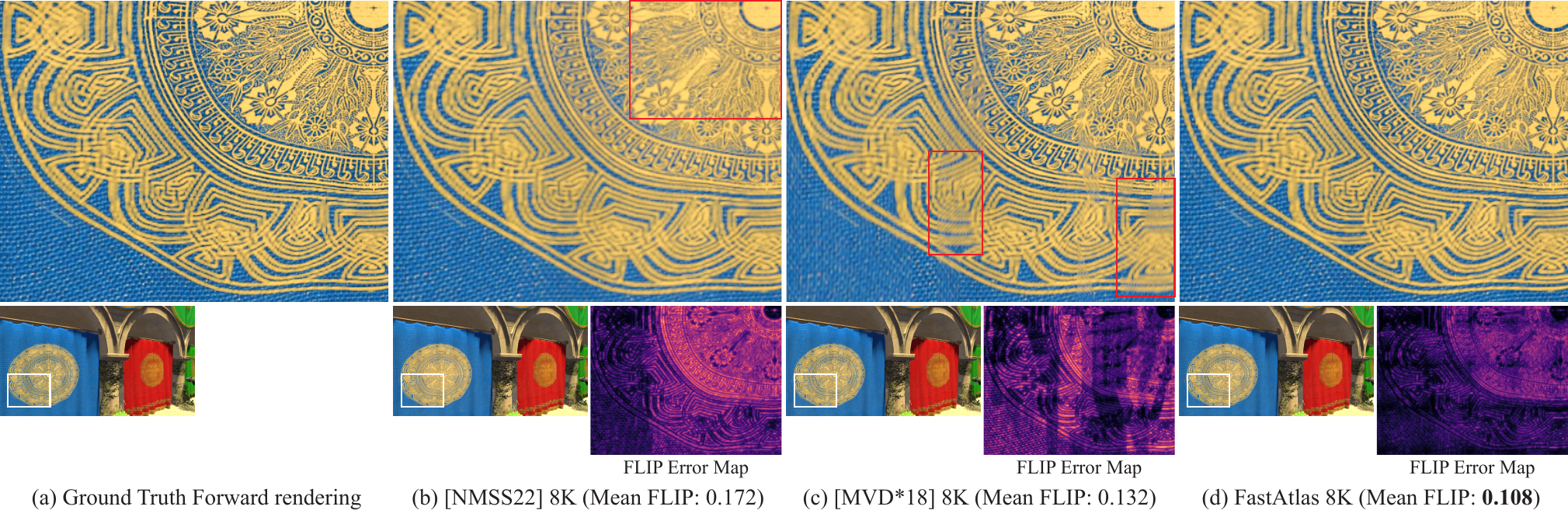}
\caption{Comparing FastAtlas (d) to representative TSS atlassing alternatives using same atlas size (here $8K\times 8K$). (a) Ground truth forward rendering; (b) MSA-P  \cite{Neff2022MSA} notably blurs the intricate banner design; (c) SAS \cite{mueller2018shading} has more localized, but significant, blurring. (d) FastAtlas produces significantly better quality renders than all alternatives. Render quality is reflected in the \FLIP error maps shown next to each output (black low error; purple to pink high).}
\label{fig:prev_work}
\end{figure*}

Early TSS approaches used {\em static}, pre-computed, texture atlases covering {\em entire scenes} \cite{baker:2016, hillesland2016texel,Baker2022}. Consequently, each individual frame only used a small portion of the charts stored in each atlas, leading to significant memory underutilization. Even when the overall atlas size is very large, this workflow often results in undersampling (low texel-to-pixel ratio) of the content visible in individual frames. This undersampling is particularly pronounced on content close to the camera \cite{Neff2022MSA,Karis:NaniteTalk} (Sec. ~\ref{sec:results}, Fig.~\ref{fig:static}b).
More recent {\em dynamic} atlasing methods \cite{mueller2018shading,hladky2019tessellated,hladky2021snakebinning,Neff2022MSA} pre-compute atlas charts once, and then each frame generate a new atlas containing only fully or partially visible charts. Dynamic atlassing significantly reduces per-frame atlas memory waste, and thus helps reduce undersampling. 
While static methods can use time-consuming CPU-based atlassing strategies (Sec.~\ref{sec:related}), dynamic ones must form atlases in real-time. Existing dynamic methods generate atlases with relatively low packing efficiency, negatively impacting sampling rate (Figs. ~\ref{fig:teaser}bc, ~\ref{fig:atlas}b, ~\ref{fig:prev_work}bc). Moreover, most dynamic methods have uneven sampling rate, and often exhibit extreme and highly visible undersampling across portions of rendered scenes (see floor and over-door frieze in Fig. ~\ref{fig:teaser}bc). This makes them ill-suited for applications which seek to control, or at least bound, sampling rate (Sec~\ref{sec:related}).
While the method of \cite{Neff2022MSA} allows better control of sampling rate than the alternatives, it produces renders with notable shading artifacts along visible chart seams (Figs.~\ref{fig:seams}b, ~\ref{fig:atlas}b (wall)); these artifacts become increasingly noticeable as the resolution of the texture atlas and/or sampling rate decreases. 

Avoiding these pitfalls while supporting a broad range of TSS applications requires a dynamic atlassing method that is real-time; can operate across a range of atlas sizes, including ones small enough to be efficiently transmitted over a network; allows control of texel-to-pixel ratio; does not introduce seam artifacts; and lastly, yet critically, is bijective across the visible portion of the rendered scenes. FastAtlas addresses all of these goals through a combination of new GPU-based chartification, parameterization, and atlas packing methods (Sec.~\ref{sec:overview}). Unlike prior approaches, which use offline pre-computed charts, we use the connected components of the visible surfaces in each frame as the texture atlas charts, eliminating visible seams and consequently seam artifacts. We then bijectively parameterize the visible portions of our charts in real-time via simple perspective projection. Projection evenly distributes undersampling, both within each chart and across different charts, making extreme blurring of the type that prior methods are prone to (Fig. ~\ref{fig:teaser}, Fig. ~\ref{fig:prev_work}) highly unlikely. 

Our remaining, and core, challenge is to quickly and efficiently pack the parameterized charts into an atlas (higher packing efficiency allows for a higher sampling rate; Fig.~\ref{fig:atlas}). Traditional packing methods \cite{levy2002least,igarashi2001adaptive,Noll2011,sander2002signal} are too slow for our needs. Prior dynamic atlasing methods rely on strong assumptions about the shape and size of the parameterized charts, which either make them inapplicable in our setting (\cite{mueller2018shading,hladky2019tessellated,hladky2021snakebinning}) or result in highly inefficient packing on our more general charts (\cite{Neff2022MSA}; Fig \ref{fig:atlas}c). We pack the parameterized charts, or more specifically their bounding boxes, using a new and efficient GPU-based parallel algorithm which wastes significantly less space than prior real-time approaches (Sec.~\ref{sec:results}). We compute conservative bounding boxes for the parameterized charts directly in homogenous clip space, avoiding compute-heavy explicit clipping, by reviving a 1978 method by Blinn \cite{Blinn:CalculatingScreenCoverage}. Our method is deterministic, generating identical atlases for identical frames, thus avoiding undesirable flicker. It can be used in other settings where real time packing of general 2D charts or boxes is useful.

We validate our method by generating and shading texture atlases, in real-time, with settings that reflect the requirements of different decoupled shading applications, including spatial and temporal decoupling, and streaming (Sec~\ref{sec:results}). 
Our method is able to generate atlases for most scenes in under 1.5 msec per frame, and is fast enough to be performed every frame on complex scenes (e.g. San Miguel 9M triangles; Fig. ~\ref{fig:san_miguel}).
We compare our method against existing alternatives in the context of reduced rate shading and memory limited applications such as atlas streaming.  When shading at reduced rate, FastAtlas outperforms the closest alternative by 25\% on average in terms of visual quality (perceptual difference vis-a-vis forward rendering). When using fixed size atlases with dimensions ranging from 2K to 8K, we outperform the closest alternative by 35\% on average in terms of visual quality. A major factor in these improvements is the quality of the atlases we produce, measured via the stretch metric \cite{sander2001texture}. Our average $L^2$ stretch, across all atlas sizes and screen resolutions, is 1.66 (a value of 1 is ideal); this value is 17.8 for the closest alternative.
We further demonstrate the versatility of our approach by using FastAtlas atlases for overshading, temporal reuse, and other rendering applications (depth-of-field and foveated rendering); see supplementary.

\section{Related Work}
\label{sec:related}

Our work builds on existing research on texture space shading, texture and shading atlas generation, and atlas packing. 

\paragraph*{Texture-Space Shading.} 
TSS methods decouple shading from rendering by storing the output of a shading computation pass in a texture atlas, then treating it as a texture at render time. TSS has been hailed as a solution for several different problems, including improving render quality via oversampling \cite{Baker2022,baker:2016}, minimizing client compute effort by shading atlases on servers and streaming them to the client \cite{mueller2018shading,hladky2019tessellated,hladky2021snakebinning,Neff2022MSA}, and improving rendering speed by shading atlases at resolutions lower than screen resolution or at temporal frequency lower than rendering frequency \cite{mueller2021tasa}. In all these settings, the goal is to maximize the visual quality of the output renders. For oversampling, the aim of decoupling is to improve rendering quality compared to forward rendering; for other applications, the goal is to generate renders which are visually close to ground truth forward/deferred rendering outputs.

The main sources of visual artifacts in TSS rendering are undersampling and shading discontinuities along visible texture seams. Undersampling blurs fine details in output renders and occurs when a triangle, or any portion of it, is allocated a much smaller space in the atlas than on screen. Some undersampling is inevitable (e.g. if atlas size is smaller than screen size) or desirable (reducing shading time); however, maximizing visual quality requires controlling undersampling as much as possible. Sampling mismatches across visible seams lead to noticeable shading discontinuities \cite{liu2017seamless} which become increasingly noticeable when combined with undersampling (Fig~\ref{fig:seams}b). 

For real-world TSS applications, visual quality is typically balanced against constraints on atlas size (which, at the very least, needs to fit into GPU memory) and runtime. 
As such, uncontrollable oversampling is also undesirable as it decreases rendering speed and wastes memory \cite{hillesland2016texel}.
As demonstrated in Sec. ~\ref{sec:results}, our atlasing strategy provides a high degree of sampling control and can be effectively used for all applications above.

\paragraph*{Traditional Atlas Generation.}
Atlas generation for storing surface signals (such as albedo or normal maps) is a well-researched problem~\cite{sheffer2007mesh}, involving three steps that can sometimes be performed in tandem \cite{BoundedDistortParam:2002,Li:2018:OptCuts}: cutting input surfaces into charts \cite{julius2005d,zhou2004iso}; parameterizing these charts in 2D \cite{levy2002least,sheffer2005abf++,jiang2017simplicial}; and finally packing the 2D charts into an atlas \cite{levy2002least,Limper18BoxCutter,igarashi2001adaptive,liu2019atlas}. The key considerations driving the methods involved are parameterization distortion; output chart seam length; bijectivity guarantees; atlas memory footprint; and runtime.
 
Traditional atlasing methods are CPU-based, and take multiple minutes to generate results for typical scenes; we aim for a GPU-based real-time method with related, but not identical, desiderata. Like traditional methods, we seek to control the atlas's memory footprint by optimizing packing efficiency; aim for a bijective mapping between the {\em visible} 3D content in each frame and the atlas; and aim to minimize the length of visible seams.  While seams can be partly hidden using seam-hiding techniques \cite{liu2017seamless,Ray2010invisible}, these techniques are far from real time and thus unsuitable for our needs. The distortion metrics minimized by the different atlassing methods are application-dependent~\cite{sheffer2007mesh,sander2002signal,sander2001texture}. For TSS, the dominant consideration is sampling rate. Our projection-based parameterization and subsequent packing step ensures a constant sampling rate across all charts. By placing seams only along the boundaries of visible regions, we avoid {\em all} rendering artifacts associated with visible seams. Our atlases can be computed near-instantaneously for charts of any shape or size, and our parameterization is bijective across the visible portions of our charts.

\paragraph*{Traditional Atlas Packing.} Textures and other surface signals are stored as 2D rectangular images defined over a bounding box of the packed atlas charts (where the box and the image have the same aspect ratio).  Any space within this box that is not occupied by the atlas charts is wasted as it contains no actual signal; packing the parameterized charts into a tight bounding box is a key element of any atlasing scheme \cite{levy2002least,Limper18BoxCutter}. Packing efficiency (the ratio between the areas of the packed atlas and its bounding box) is therefore a critical consideration for practical atlasing applications. In the context of TSS, packing efficiency directly impacts sampling rate. Unfortunately, maximizing the packing efficiency for a given set of charts is NP-hard \cite{Milenkovic1999}. Traditional \cite{levy2002least,igarashi2001adaptive} and recent \cite{Limper18BoxCutter,zhangatlasgeneration,knodt2023uvopt,liu2019atlas} methods for packing general charts have multi-minute runtimes. 

Packing methods that attempt real-time performance rely on strong assumptions about the shape and size of the charts packed. At the extreme end, some assume that all charts are same size right angle isosceles triangles \cite{Maruya:1995,Carr:2002} or squares \cite{purnomo2004seamless}. While these methods can achieve high packing efficiency, generating parameterized charts that satisfy these assumptions requires introducing extremely long seams, and high parametric distortion as arbitrarily shaped triangles or charts are forced to map to highly constrained parameter domains. Like the latter methods FastAtlas packing runs in real-time; however in contrast to them it makes no assumptions about parameterized chart shape or connectivity. 

\paragraph*{Atlas Generation and Packing for TSS.} 
Many TSS approaches use {\em static} pre-computed texture atlases covering {\em entire scenes} \cite{baker:2016, Baker2022, hillesland2016texel},
and compute these atlases using traditional time-consuming atlasing strategies discussed above.
Even if the overall atlas size is large, the atlas resolution of the content visible in individual frames is often low; this leads to visible undersampling, particularly on content close to the camera \cite{Neff2022MSA,Karis:NaniteTalk} (Fig.~\ref{fig:static}b). Using smaller atlas sizes, even relatively far objects can be highly undersampled (Sec.~\ref{sec:results}). This approach is therefore largely used in settings where strong assumptions can be made about the rendered content \cite{baker:2016}, or where atlas size exceeds display resolution by orders of magnitude (e.g. \cite{Baker2022}, which uses a $256K \times 256K$ atlas stored in virtualized texture memory.) FastAtlas makes no such assumptions and produces high quality renders using atlases small enough to fit in GPU memory (Fig.~\ref{fig:static}c).

\begin{figure}
\includegraphics[width=\linewidth]{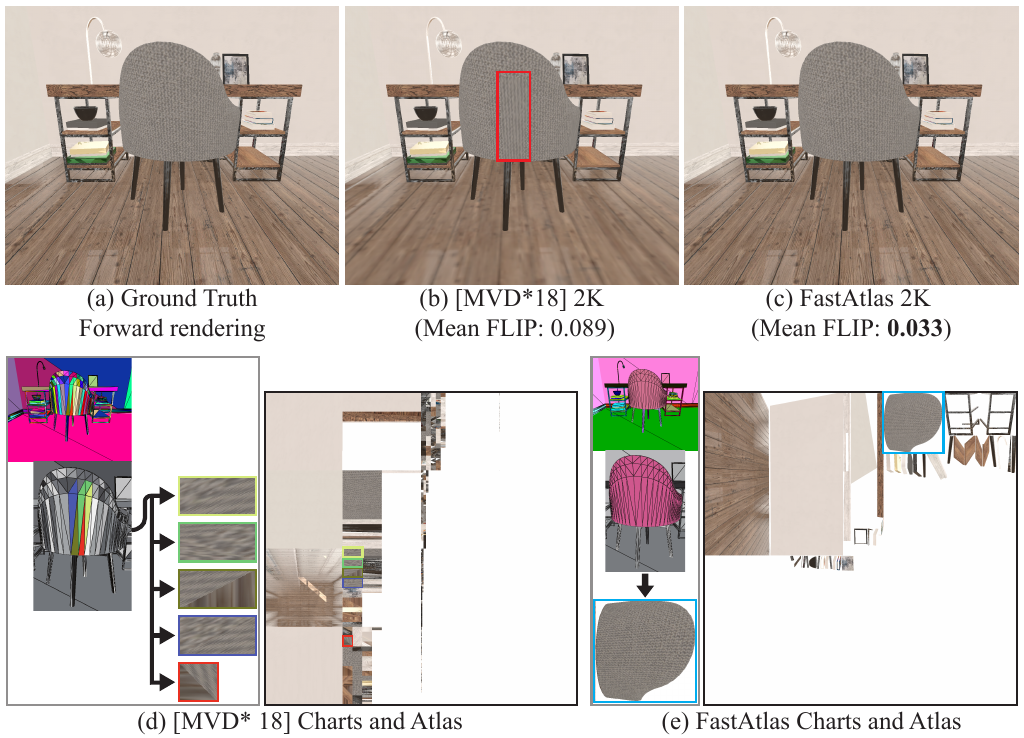}
\caption{SAS ~\cite{mueller2018shading} vs FastAtlas:  (a) Reference forward rendering. (b,d) SAS can form charts which are far from rectangular (d, left); mapping such charts into rectangular parameter domains introduces excessive stretch (d,center/right) leading to blurring/smearing artifacts (b, highlighted). (c,e) FastAtlas projection based parameterization preserves the screen-space triangle shape (e,left), and distributes undersampling uniformly leading to better visual outputs (c).  SAS (d) and FastAtlas (e,right) atlases ($2K \times 2K$). Please zoom in to see details.}
\label{fig:sas_issues}
\vspace{-5mm}
\end{figure}

Several recent TSS methods that target streaming applications \cite{mueller2018shading,hladky2019tessellated,hladky2021snakebinning,Neff2022MSA} compute per-scene static charts, then form per-frame atlases which include only fully or partially visible charts. Shading Atlas Streaming (SAS) \cite{mueller2018shading} pre-segments scenes into charts with 2 to 3 triangles, and parameterizes each chart into a rectangle. They constrain these rectangles' dimensions to be powers of 2, and propose a packing strategy designed for such rectangular charts. These constraints allow for space-efficient, real-time packing, but can introduce significant parametric distortion (in both world and screen space) which results in highly notable, unevenly distributed  undersampling (Fig~\ref{fig:sas_issues}). SAS followups use triangle strips as charts \cite{hladky2019tessellated,hladky2021snakebinning} combined with a similar parameterization and packing strategy. As noted by \cite{hladky2022quadstream}, based on experiments with their code, these methods lack robustness and fail to generate atlases for complex scenes such as San Miguel (Fig.~\ref{fig:san_miguel}). 

Neff et al. \shortcite{Neff2022MSA} use an atlassing framework based on mesh shaders (Fig.~\ref{fig:prev_work}b).
They segment input scenes offline into roughly evenly-sized charts with at most 84 triangles (the largest meshlet size supported by mesh shaders), then parameterize these charts either offline, or dynamically by projecting each chart to view space; and finally form per-frame atlases by packing visible charts using a grid-based method that works best when presented with evenly sized charts. 
As Neff et al. note, both theirs and prior dynamic atlassing methods \cite{mueller2018shading,hladky2019tessellated} perform best when presented with meshes with roughly uniformly sized triangles. In the experiments in their paper, they refined their input meshes to evaluate both their method and SAS, making them more uniform. Our experiments (Sec~\ref{sec:results}) show that the visual quality of the renders produced using these methods significantly degrades on unevenly tessellated inputs, typical of real-life game and other environments. 
To ensure that the parameterized charts can be packed into an atlas of a given size, all above methods  \cite{mueller2018shading,hladky2019tessellated,hladky2021snakebinning,Neff2022MSA} cap the dimensions of the charts they pack, and scale charts down if they exceed these ``superblock'' dimensions (e.g. Fig~\ref{fig:atlas},~\ref{fig:sas_issues}). Sec~\ref{sec:results} provides extensive comparisons of FastAtlas, showing we outperform these alternatives. 

\begin{figure}
\includegraphics[width=\linewidth]{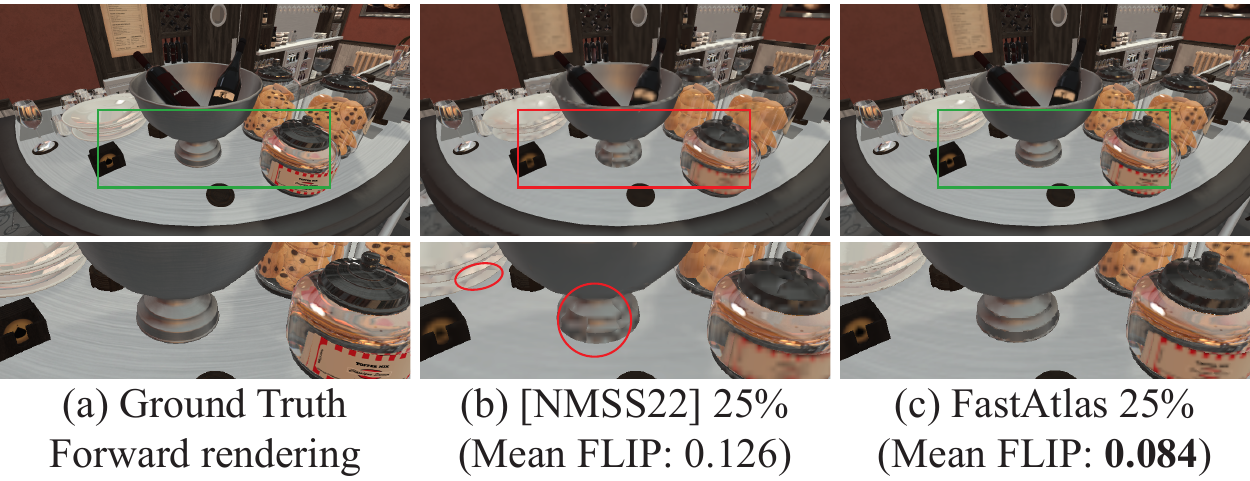}
\caption{MSA-P ~\cite{Neff2022MSA} use charts whose seams when visible can lead to noticeable undesirable shading discontinuities (highlighted in the zoomed image)  (b). FastAtlas charts have no visible seams, and are better packed leading to superior render quality. In this example, we use 25\% shading rate and an atlas size of $8K \times 8K$.}
\label{fig:seams}
\vspace{-5mm}
\end{figure}

\paragraph*{TSS Alternatives and Temporal Reuse.}
While hardware-level decoupled shading \cite{ragan2011decoupled, burns2010lazy,clarberg2013sort} is a promising concept, it is not supported by current commercial hardware. 
In addition to TSS, several other approaches explore decoupling on standard GPU hardware.
Liktor and Dachsbacher~\shortcite{liktor2012decoupled} decouple shading and rasterization by shading directly into a linear buffer which stores an array of shading samples per triangle. This precludes efficient hardware texture sampling and filtering, slowing computation and preventing real-time performance. QuadStream \cite{hladky2022quadstream} decouples shading between client and server; it addresses a different problem than a fully decoupled shading method (reconstructing occluded geometry and shading, seen from novel viewpoints on a client) and operates directly on already-shaded framebuffers.
Researchers have explored screen-space reverse reprojection between frames, or between views, as a pseudo-shading cache \cite{nehab2007accelerating} or for temporal anti-aliasing \cite{karis2014hqaa, Yang2020TAA}.  Reprojection often introduces rendering artifacts such as ghosting and flicker, and flicker, and can fail for content moving in and out of occlusion and for first frames after a camera cut \cite{Scherzer11asurvey}. FastAtlas can be used to facilitate selective shading reuse across frames and does not suffer from these artifacts (see supplemental).

\begin{figure*}[t]
\begin{center}
\includegraphics[width=.85\linewidth]{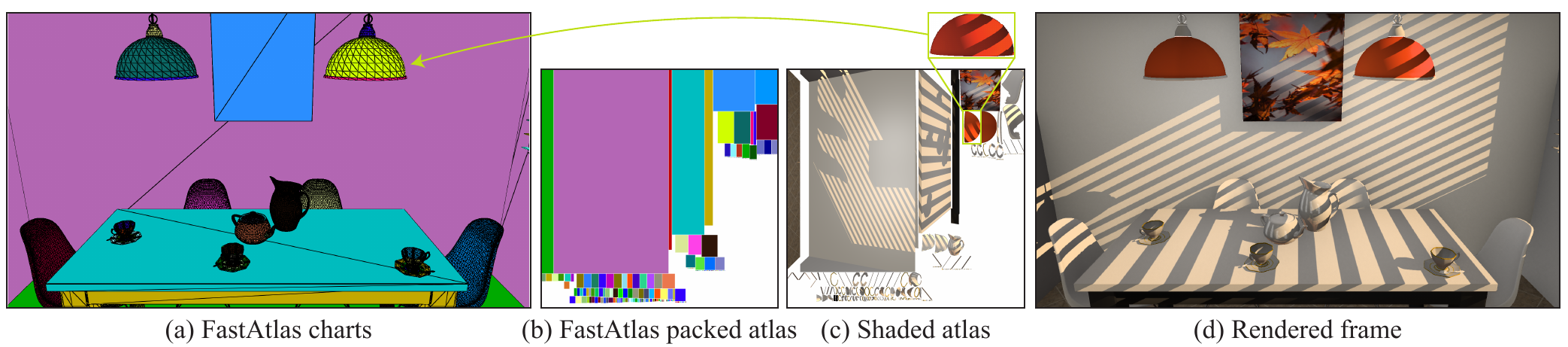}
\end{center}
\caption{
FastAtlas Overview: In each frame, we compute charts spanning fully or partially visible triangles (a), determine texture space bounding boxes for the visible portions of the view-space projections of each chart, and tightly pack these boxes into atlases (b, here $2K \times 2K$). We simultaneously bijectively parameterize and shade the charts into their atlas boxes, obtaining high quality texture space shading (c), and use this shading to render the shaded frames (d).}
\label{fig:overview}
\label{fig:alg_overview}
\end{figure*}

\section{Overview}
\label{sec:overview}
Our work has two core contributions: a real-time, GPU-based algorithm for tight packing of general parameterized charts into compact atlases; and a real-time TSS method that
utilizes this packing.  

\paragraph*{FastAtlas Packing.}
FastAtlas runs entirely on the GPU as a series of compute shaders. It takes the bounding boxes of parameterized charts as input, and packs them into an atlas (Fig~\ref{fig:overview}b, Sec.~\ref{sec:pack}). As such, the only input it requires are the dimensions of the bounding boxes.
Its outputs are deterministic; identical input charts are packed into identical atlases. This is critical for TSS and similar applications, as it ensures that consecutive frames taken from the same camera view have the same shading. Even minute shading differences across such frames can cause sampling jitter, leading to undesirable flicker \cite{baker2012rock}. 
While prior methods such as \cite{mueller2018shading,hladky2019tessellated,hladky2021snakebinning,Neff2022MSA} cap the dimensions of the charts that can be packed as-is for a given atlas size, and scale down all charts that exceed these dimensions, we scale all charts by the same factor, and do so only when strictly necessary to achieve packing success (Figs~\ref{fig:atlas},~\ref{fig:sas_issues}). 

\paragraph*{TSS using FastAtlas.}
Our end-to-end TSS atlas generation method combines the packing method above with a novel approach for computing seamless per-frame charts. 
We define our charts as the connected components of the visible surfaces in each frame (Fig.~\ref{fig:overview}a), and efficiently compute them using a parallel union-find algorithm (Sec.~\ref{sec:visible}). Since the boundaries of these charts coincide with the contours of the rendered surface, they are {\em invisible} to the viewer. This approach 
eliminates the artifacts caused by shading discontinuities along visible seams (Fig.~\ref{fig:seams}). 

\begin{parWithWrapFigure}
\begin{wrapfigure}{l}{.27\columnwidth}%
\includegraphics[width=\linewidth]{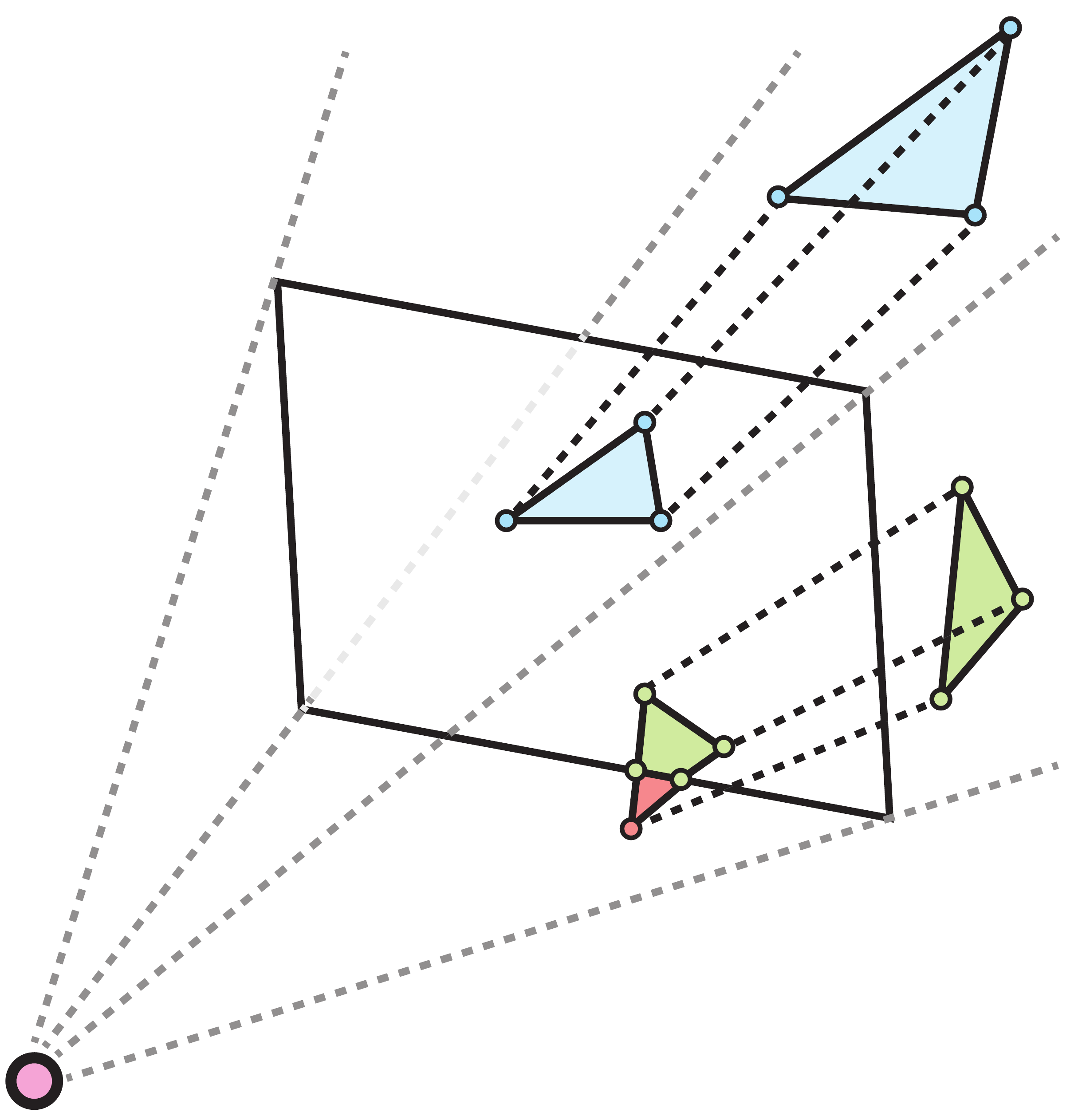}%
\end{wrapfigure}
We bijectively parametrize the {\em visible portions} of our charts by projecting them to view space (inset). This maps a constant number of texels to each pixel in the final rendered output, evenly distributing residual undersampling error across all image pixels. While conceptually straightforward, efficiently parameterizing charts containing partially visible triangles using viewspace projection is non-trivial, as the visible portions may no longer be triangular (e.g. green triangle in the inset); applying naive projection to triangles with vertices behind the camera may produce ill-posed results. Clipping triangles before projection is both computationally expensive and significantly complicates downstream operations. We avoid explicit clipping by observing that all that is required for atlas packing is the dimensions of, potentially conservative, bounding boxes of these projected visible portions. We compute such bounding boxes without explicit chart clipping by adapting a conservative screen coverage estimator \shortcite{Blinn:CalculatingScreenCoverage} (Sec.~\ref{sec:box}). We then pack the computed boxes using FastAtlas. 
\end{parWithWrapFigure}

Finally, we shade the visible portion of each chart into its corresponding atlas bounding box (Fig~\ref{fig:overview}c). 
The resulting texture is then used during rasterization as a standard texture map (Fig. ~\ref{fig:overview}d). 
Our framework is compatible with all existing approaches for texture space shading, including forward shading, raytraced illumination, or deferred shading in texture space \cite{baker:2016}. In the examples shown, we use the standard forward shading based rendering pipeline included in the G3D Innovation Engine \cite{G3D17}, a commercial grade renderer.

\begin{figure*}[t]
\vspace{-10pt}
\includegraphics{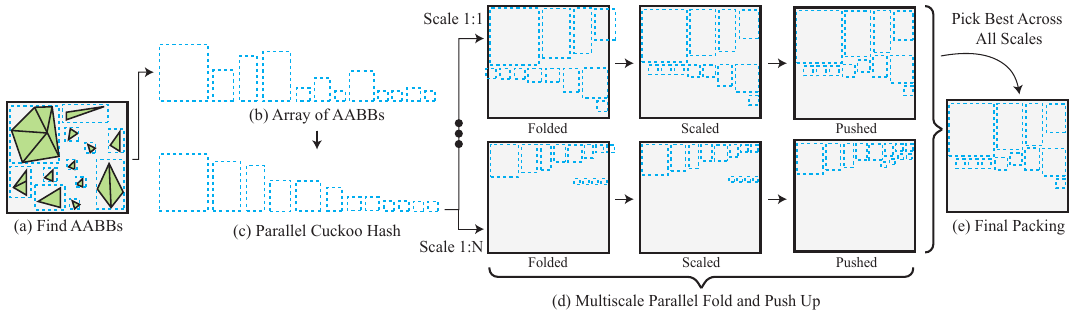}
\caption{FastAtlas box packing: Given a set of boxes (a), we rotate them so that they are at least as tall as they are wide (b), and then hash sort them in descending size while maintaining a deterministic order (c). We then find a scale factor $S$ such that the boxes, scaled and integerized, will fit in the atlas (d). We do this by testing candidate scales in parallel. For each scale we {\em fold} boxes into rows that fit the atlas width. Boxes are placed alternately in one row from left to right, and two rows from right to left, to avoid congestion on the sides of the atlas. We scale all boxes to avoid horizontal overflow. We {\em push} boxes upwards in the atlas to minimize wasted space. The largest scale that does not overflow the atlas is selected (e). All steps are executed on the GPU, and evaluated in parallel.}
\vspace{-10pt}
\label{fig:igarashi_packing}
\end{figure*}

\section{FastAtlas Packing}
\label{sec:pack}
\label{sec:atlasing}
\label{sec:packing}

Given a set of charts with associated bounding boxes, FastAtlas packs these bounding boxes into texture atlases of fixed user-specified dimensions $\omega \times \omega$ texels. We assume that chart bounding boxes have target dimensions in texel space that are specified relative to $\omega$. For TSS applications, these are typically the chart's pixel-space dimensions in the frame buffer, which captures that each chart should ideally have a 1:1 texel-to-pixel ratio.

Our packing must address two inter-related sub-problems: a scale factor such that, after scaling the boxes by this factor and rounding their size to integer coordinates, they can indeed be packed into our atlas; and computing the actual packing. We seek atlases that are packed as efficiently as possible, minimizing memory waste. A key additional requirement for TSS applications is that atlases for identical charts must be strictly identical; this avoids subtexel jitter across frames if the scene and camera remain stationary. This requirement means that we must use purely deterministic algorithms when computing our atlases, including consistent handling of identically sized boxes.
 
We address this challenging problem using a two-step process inspired by traditional packing methods \cite{levy2002least,igarashi2001adaptive}. These methods start by ordering charts by size using traditional sorting methods, then place charts one at a time, choosing the optimal location for each chart based on the locations of the previously placed charts. This ordering is motivated by the observation that larger charts get harder and harder to place as the atlas fills up, and thus should be packed first. The chart placement strategies used by these methods are too slow for real-time computation and cannot be directly parallelized across multiple threads, as placing one chart requires awareness of the placement of all prior charts. We adopt the general two-step framework these methods use (sorting charts first, then placing them leveraging the sorted order), but execute both steps using parallizable variants of the respective algorithms.

We first order our boxes in a deterministic, monotonically decreasing size order (Sec.~\ref{sec:order}). We then leverage GPU parallelism to generate atlases with different scaling factors, and pick the atlas and factor that provides the best packing efficiency as the solution (Sec~\ref{sec:scale}). To generate the actual atlases we introduce a parallel framework inspired by the packing method of \cite{igarashi2001adaptive} (Sec~\ref{sec:actual_packing}).

\subsection{Box Ordering via Cuckoo Hashing.}
\label{sec:order}
We orient all per-chart boxes so that they are at least as tall as they are wide (rotating them by 90 degrees if necessary), and then order them by height. To ensure deterministic results across multiple identical frames, we use height as the primary ordering criterion, and the lowest triangle index in each box's corresponding chart as the secondary ordering criterion (Fig.~\ref{fig:igarashi_packing}bc).

Sorting on the GPU can be efficiently implemented by a prefix sum-based radix sort (e.g. \cite{merrill2011high}). However, ordering our boxes lexicographically requires 36-bit keys: 13 bits for the box height (assuming box height is integer and the height of a box is at most equal to the maximal atlas height, here $\max_H=8192$), and 23 bits for the per-box root triangle index (assuming a maximum triangle index of $~8M = 2^{23}$). Using a GPU radix sort, which can sort 4 bits per pass, ordering such keys would require 9 radix sort passes, which is prohibitively expensive even with hand-optimized compute shaders. We avoid explicit lexicographic sorting by leveraging two observations. First, as we have a fixed number of box heights (1 to $\max_H$), we can simply bucket charts by height into $\max_H$ buckets, rather than sort them. Second, when bucketing charts, it is not necessary for boxes of the same height to be strictly ordered by triangle ID, so long as the ordering of boxes inside a bucket is identical across identical frames.

We place boxes into one of $\max_H$ possible buckets by employing a {\em cuckoo hashing} scheme \cite{pagh2004cuckoo}. At the start of program execution, we allocate a scratch buffer on the GPU which contains enough memory for four times the maximum number of charts on screen. Each frame we compute, for each bucket, the total number of boxes it will contain; this can be done efficiently by using GPU atomic operations. Boxes are then assigned to a bucket, based on their screen-space integer height. Each bucket is allocated enough scratch buffer memory for four times its number of boxes; this prevents hash contention and unnecessary evictions during the cuckoo hashing. Using the lowest triangle ID of each chart, we attempt to place each box in its bucket using GPU atomic compare-and-swap operations, initially at position $(4 \cdot \operatorname{chart\ ID})\ \mathrm{mod}\operatorname{bucket\ size}$ in the bucket. If a box is already placed in a specific bucket location, it is evicted; the evicting thread in the compute shader is then responsible for placing the evicted box in an empty location in the same bucket. A final compute shader removes empty buckets and empty space in buckets. This approach reduces overall runtime by one order of magnitude versus a radix-based sort, while still being deterministic. 

\subsection{Parallel Atlas and Scale Factor Computation.}
\label{sec:scale}
Packing boxes into a fixed size atlas requires us to find a suitable scale factor $S$ such that the boxes can indeed be packed once their dimensions are scaled by this factor and rounded upward to the closest integer values. While one can easily pick a conservative scale factor for which packing is guaranteed, minimizing undersampling (or oversampling) requires searching for a scale factor $S$ that is as close as possible to one.

We compute the suitable factor by leveraging GPU parallelism. We observe that on the GPU we can compute packings in parallel at multiple possible scales with the same efficiency as computing one box packing at a fixed scale. 
We consequently find the desired scale-factor through brute-force search over a range of scales, choosing the scale that maximizes packing efficiency while ensuring that all boxes fit in the atlas. We test box packing over a set of 64 scale factors (ranging from $\frac{1}{64}, \frac{2}{64}, ..., \frac{64}{64}=1$) in parallel as this provided a suitable speed versus quality trade-off.
Candidate scale factors are rejected if not all boxes fit in the fixed-size atlas, and the largest scale factor not rejected is chosen.

\subsection{Box Packing.}
\label{sec:actual_packing}
Given a candidate scale factor $S$, we scale all bounding boxes by this factor, then round box sizes upwards to integer coordinates for packing. We then attempt to pack the boxes into the $\omega \times \omega$ texture atlas. We achieve this goal by observing that while it is notoriously challenging to parallelize standard box packing problems, it {\em is} possible to solve this problem by relaxation. Specifically, we parallelize the 2D box packing problem by allowing the atlas to slightly overflow. We then decrease the scale factor $S$ to scale the packed boxes down so as to fit back in the atlas. While this is not an acceptable trade-off for most bin-packing problems (as bins are not normally allowed to shrink), it is acceptable in our problem context. Our relaxed packing method extends the approach of Igarashi and Cosgrove ~\shortcite{igarashi2001adaptive}; one advantage of this method is that much of it can be parallelized and only one data structure - the array of charts, or boxes - is operated on, reducing cache and register pressure. While it might be possible to parallelize the more popular Tetris Packing method \cite{levy2002least} on the GPU, it is not clear how to do so as Tetris Packing requires maintaining multiple sets of active horizons during packing. 

Given appropriately scaled, rotated and pre-ordered charts, the CPU-based algorithm of Igarashi and Cosgrove ~\shortcite{igarashi2001adaptive} 
proceeds using two operations: {\em folding} and {\em pushing}. During folding, charts are placed sequentially, from the tallest to shortest, starting from the top left corner of the atlas ($X=0,Y=0$) and advancing along the X axis until there is no room left for the next chart. The algorithm then forms the next row by placing charts in a reverse direction from right to left: the first chart in the second row is placed at the $Y$ value directly below the tallest chart placed and at $X= \omega - \operatorname{chartWidth}$ (flush with the right side of the atlas); the rest of the charts are placed one after the other to the left of it. This alternating row layout is repeated until all charts are placed. After the initial placement is complete, a final {\em pushing} operation moves charts upwards in the atlas as far as possible to minimize wasted space. The key idea behind this heuristic approach is that large charts, which are most likely to be difficult to pack efficiently, are placed first; the alternating row layout serves to ensure that the charts with the tallest heights in each row do not bunch on the left hand side of the atlas, increasing packing efficiency during the pushing operation. 

Our key challenge is to efficiently implement the folding step on the GPU: placing boxes in a linear fashion, one at a time, is too slow for real-time performance. Unfortunately, computing the exact location where the algorithm needs to switch rows is an inherently sequential operation: detecting when a box being placed triggers an overflow, necessitating starting a new row, requires knowing where all previous boxes were placed. 

We overcome this challenge and parallelize folding by employing a relaxation strategy which is based on prefix-sums and which provides a good approximation of folding for our purposes. We place our boxes in order, left to right, along a straight line, with the top left corner of each box snug against the right top corner of the previous one.  
Recall that the {\em prefix sum} of a set of numbers $w_1,...,w_n$ is $w_1,(w_1+w_2),...,\sum_{i=1}^{n}  {w_i}$. The prefix sums of the horizontal locations of the top-left corners of our laid-out boxes (represented as 32-bit integers) can be efficiently computed in parallel on the GPU \cite{harris2007parallel}. We observe that, if we restrict our texture atlas width to be a power of two (as most textures already are!), and writing $\omega=2^k$, the prefix sum yields an approximate folding for all boxes at once: the top $32-k$ bits of its corner's prefix-sum determine the row each box belongs to, and the bottom $k$ bits determine each box's horizontal position within the row (mirrored for odd rows). The only caveat is that the right-most box on each row may overflow by some amount. We therefore use prefix sums to place boxes into the atlas (Fig~\ref{fig:igarashi_packing}) expecting an overflow, which we account for by reducing the tested scale factor $S$: we compute the maximum horizontal overflow $m$ (the largest amount, in texels, that any row is over $\omega$), and multiply $S$ by the ratio $\frac{\omega}{\omega + m}$), ensuring that all boxes fit horizontally into the given size atlas. In practice we found our atlases were dominated by a few extremely large charts, and therefore alternate one left-starting row with {\em two} right-starting rows to accomodate this domain-specific knowledge.
 
After folding, we "push up" each box in a row-by-row fashion, using an advancing front approach. We start by setting the vertical offset of each box to zero. Each box in the row then writes its height to a buffer of size $\omega$, for each of the texels that it covers horizontally; this represents the current frontline. Subsequent rows then set their vertical offset to the maximum value that they cover with the current frontline, then write that value, plus their height, to any frontline texels that they cover. 

\section{Charting and Atlas Bounding Box Computation for TSS}
FastAtlas packing operates on parametrized charts generated using any existing chartification and parametrization method, and only requires as input the bounding boxes of the parametrized charts. Our end-to-end TSS framework combines it with our new chartification and bounding box computation methods, described below.

\subsection{Chart Computation}
\label{sec:charts}
\label{sec:visible}

To compute our charts, we first perform a depth pre-pass; a second pass then marks visible triangles in a {\em visible triangle (VT) buffer} containing an entry for each triangle in the scene \cite{burns2013visibility,kubisch2014opengl}. Triangles are visible if they cover at least one sample in the output framebuffer.
We then form charts by grouping visible triangles sharing common edges, forming connected components. While numerous algorithms exist for finding connected components on the CPU, we require an algorithm with a parallel, GPU-only implementation. We therefore use a union-find approach \cite{skiena1998algorithm} using the {\em ECL-CC} method of Jaiganesh and Burtscher \shortcite{Jaiganesh:2018}: for each visible triangle $i$, we check if any of its neighbors are visible by querying the VT buffer. If so, we set the $i$'th entry in the buffer to the minimum of the triangle's and its visible neighbors' buffer entries; we repeat until all pairs of visible triangles sharing common edges have the same ID. Once the process converges, we use the IDs to form connected charts. For more details and pseudocode, see the supplemental material.

\subsection{Computing Bounding Boxes of Projected Visible Sub-Charts}
\label{sec:box}

The last step before calling our packing method is generating the target 2D axis-aligned bounding boxes for the visible portions of each projected (parameterized) chart. 
We would like these boxes to tightly bound the visible projected chart portions, or {\em visible sub-charts}, ensuring shading is captured for any visible geometry while also being as small as possible for packing efficiency. Computing bounding boxes by naively projecting chart vertices to the view plane is not only wasteful, but more importantly can produce ill-posed results for charts containing points in front of the near clip plane. We recall that the standard rendering pipeline uses homogeneous coordinates, where each three-dimensional point $p=(x,y,z)$ is represented as a four dimensional vector $p^h=(x,y,z,1)$, and the perspective transformation is represented as a $4\times4$ matrix $P$. To apply a perspective transformation to $p$, we must first compute $p^p=(x^{p},y^{p},z^{p},w^{p}) = P p^h$ and then apply a perspective divide $p' = (x^{p}/w^{p},y^{p}/w^{p},z^{p}/w^{p})$. For points in front of the near plane ($w^{p}\leq0$) the projective divide can produce ill-defined values, projecting them way outside of the screen, or even folding triangles over of the rest of the chart \cite{blinn1978clipping}. Accordingly, our bounding box computation must account for geometry outside the view frustum.

Explicitly clipping charts against the view frustum in a compute shader prior to packing fails to effectively utilize the GPU's fixed-function clip hardware, and also requires keeping track of the clipped geometry for every visible triangle on screen through the rest of the pipeline. As a triangle clipped by a frustum can have up to 7 sides, this is highly time consuming and adds significant complexity to our shading and rasterization steps. We therefore reduce waste and ensure valid bounding boxes by taking inspiration from a classic article by Blinn \shortcite{Blinn:CalculatingScreenCoverage}, which shows that conservative screen-space extents for an object can be found by clamping each post-projection vertex $p^p = (x^p,y^p,z^p,w^p)$ to the range $[-w^p,w^p$], performing the perspective divide by $w^p$, and then computing the bounding box for all resulting points $p'$. This produces results that are correct but conservative, especially when a triangle intersects the near plane. We further observe that while clipping against all planes is complicated and compute-heavy, clipping against {\em one plane} produces either triangles or quads, which allows for straightforward processing by cases (see supplementary for details). Therefore, we clip each triangle of a chart against the near clip plane if it intersects it, and against the frustum plane that reduces the size of the bounding box the most otherwise, prior to using Blinn's coverage estimator. 

The output of this step is a bounding box for each sub-chart in NDC space on the screen, whose coordinates range from $[-1...1]$. We find an integer size for each sub-chart by applying the standard viewport transform which maps $[-1...1,-1...1]$ to $[0...\operatorname{screenWidth},0...\operatorname{screenHeight}]$, and this becomes the target bounding box size for FastAtlas packing.

\section{Results and Validation}
\label{sec:results}
\label{sec:eval}

We evaluate the suitability of FastAtlas for TSS applications by assessing both baseline texture-space shading settings (one-to-one texel-to-pixel shading rate) and settings reflective of potential decoupled shading application categories: using atlases to control per-frame sampling rates and streaming shading across networks. Figs. \ref{fig:alg_overview} and \ref{fig:igarashi_packing} show the stages of the method, and Figs. \ref{fig:teaser},~\ref{fig:static},~\ref{fig:atlas}, and~\ref{fig:sas_issues} show typical FastAtlas atlases. In each category we compare the results obtained with FastAtlas against those generated using the most relevant alternatives. Detailed statistics, renders and comparisons are provided in the supplementary document, supplemental galleries, and video. Please zoom in on images throughout the paper to see fine details. For additional ablations, including experiments with temporal re-use and other applications such as depth-of-field and foveated rendering, please see our supplemental material.

\paragraph*{Setup.}

In the experiments below, we evaluate FastAtlas and alternatives by atlassing, shading and rasterizing thirteen representative flight paths of complex scenes. Nine of the paths are generated by rotating a camera from a fixed position in the scene, and four are generated by moving the camera along a trajectory specified by a spline flight path. All flight paths use frequently rendered scenes from the Computer Graphics Archive \cite{McGuire2017Data} (see supplemental for more detail). We rasterize scenes at two window resolutions: $1920 \times 1080$, the size of a standard high-definition display, and $3840 \times 2160$, used by high-resolution UHD displays. 

Regardless of application, TSS aims to produce rendered outputs that are as visually similar as possible to those obtained via standard forward rendering. We hence measure visual quality as the difference between TSS (ours or alternative) and forward rendering outputs, generated using the standard forward renderer in the G3D engine \cite{G3D17} which uses the same ubershader as TSS. We quantify this difference using the \FLIP metric which measures the perceptual difference between a frame generated by the renderer being tested and a ``ground truth'' reference frame \cite{andersson2020FLIP}.  \FLIP values range from $0$ to $1$, with 0 being best. We expect \FLIP values to always be non-zero, as even at 100\% shading rate there will always be small differences between pixel and texel space shading due to aliased sampling with arbitrary phase offsets. We report the average mean \FLIP error across all frames in the path. 

We also report the average $L^{2}$ and $L^{\infty}$ texture stretch metrics of Sander et al. ~\shortcite{sander2001texture} between the triangles on screen and in the atlas for all of our frames; $L^{2}$ measures the root-mean-squares stretch of the texture atlas over a frame, and $L^{\infty}$ measures the worst per triangle stretch. These metrics assess sampling uniformity at a triangle level. $L^{2}$ and $L^{\infty}$ of 1 mean that the sampling rate is one-to-one, with values above 1 indicating undersampling, and below 1 oversampling. A divergence between $L^{2}$ and $L^{\infty}$ indicates different sampling rates in different parts of the scene. 

\vspace{-3mm}
\paragraph*{Comparison Setup.}

We compare the visual quality of FastAtlas TSS to other popular TSS methods. First, we compare against fixed, or static atlases, generated using the {\em xatlas library}. Until recently, texture space shading has always been performed with static atlases; this establishes a baseline against our method. Second, we compare against Shading Atlas Streaming (SAS) \cite{mueller2018shading}, which packs small 1-3 triangle charts into regular grids using a ``superblock'' scheme inspired by memory management. Finally, we compare our visual quality to the meshlet shading atlases of Neff et al. \shortcite{Neff2022MSA}, the current state-of-the-art for dynamic texture space shading. Neither SAS nor Neff et al. \shortcite{Neff2022MSA} provide code; while the authors of Neff et al. generously shared their meshlet generation code with us, they were unable to share the rest of the codebase. We therefore reimplemented these two methods as discussed in the supplemental material (App. B) and confirmed the correctness of our implementations as much as possible with the authors. In the experiments below we use the MSA-P method of Neff et al. (using screen projection to parameterize charts). In our experiments the MSA-UV method (parameterizing charts in advance using \cite{levy2002least}) led to higher \FLIP error and introduced folded/non-bijectively mapped charts for many scenes, leading to catastrophic failures for frames containing these charts.  

\subsection{Texture-Space Shading Baseline}
\begin{table}
\scriptsize
\setlength{\tabcolsep}{2pt}
\centering
\begin{tabular}{l|cccc|ccc}
& \multicolumn{4}{c|}{\FLIP $\downarrow$} & \multicolumn{3}{c}{Improvement vs.} \\
& \textbf{Static} & \textbf{SAS} & \textbf{MSA-P} & \textbf{Ours} & \textbf{Static} & \textbf{SAS} & \textbf{MSA-P}\\
\hline
$ 1920\times1080 $ & 0.090 & 0.062 & 0.061 & \textbf{0.036} & $53\%$ & $43\%$ & $42\%$ \\
$ 3840\times2160 $ & 0.123 & 0.070 & 0.061 & \textbf{0.033} & $70\%$ & $53\%$ & $46\%$ \\
\end{tabular}
\caption{Average visual quality of FastAtlas and alternatives at baseline 8K atlas resolution; left to right: \FLIP for static atlases; SAS \cite{mueller2018shading};  MSA-P \cite{Neff2022MSA}, and FastAtlas, followed by relative improvement vis-as-vis these methods.  We report mean values across all scenes, at both 1920 $\times$ 1080 and 3840 $\times$ 2160 screen resolutions. See supplementary for details.}
\label{tab:flip_tss_overview}
\vspace{-3mm}
\end{table}

As a baseline experiment we use FastAtlas to shade our input corpus, aiming for a 100\% shading rate. We use a fixed atlas size of $8K \times 8K$ which in our experiments was sufficient for FastAtlas to achieve this goal at both standard and high display resolutions (Tab.~\ref{tab:flip_tss_overview}). We ran all alternative methods using the same settings. Using this baseline setting, our \FLIP error is $42\%$ lower than that of the best performing alternative (MSA-P) at standard display resolution, and $46\%$ lower than this best alternative at 4K resolution.
  
We compare our baseline results against static atlases with an extra-large atlas size of $32K \times 32K$; this serves as a proxy for the virtualized texture-space shading methods of Baker et al. \cite{baker:2016,Baker2022}. We show representative outputs of our method and this alternative in Fig. \ref{fig:static}. 
Using static 32K atlases on a regular size display, our average \FLIP with an 8K atlas is 0.038, versus an average \FLIP using a 32K static atlas of 0.055. Our method outperforms static 32K atlases on all inputs except {\em Breakfast Room}, where both \FLIP errors are negligible - ours is 0.002, and the alternative is 0.001. As the results show, even without increasing shading rate, FastAtlas achieves higher rendering quality than static atlases at both 8K and 32K, and can further improve rendering quality when shading rate is increased. We note that, unlike our method, static atlases do not allow control over shading rate. 

\begin{table}
\scriptsize
\setlength{\tabcolsep}{2pt}
\centering
\begin{tabular}{l|cccc|cccc}
  & \multicolumn{4}{c|}{Mean $L^2$} & \multicolumn{4}{c}{Mean $L^\infty$} \\
  & \textbf{Static} & \textbf{SAS} & \textbf{MSA-P} & \textbf{Ours} & \textbf{Static} & \textbf{SAS} & \textbf{MSA-P} & \textbf{Ours}\\
\hline
\textbf{1920 $\times$ 1080} & & & & & & & & \\
8K avg. 	& 3.05 & 3.58 & 1.54 & \textbf{1.01} & 11.19 & 23.63 & 3.66 & \textbf{1.01} \\
4K avg.	& 7.9 & 3.58 & 1.96 & \textbf{1.04} & 26.75 & 23.67 & 3.82 & \textbf{1.04} \\
2K avg.	& 11.85 & 6.91 & 16.77 & \textbf{1.59} & 37.26 & 33.39 & 37.61 & \textbf{1.59} \\
\hline
\textbf{3840 $\times$ 2160} & & & & & & & & \\
8K avg. 	& 6.13 & 6.41 & 1.97 & \textbf{1.04} & 22.43 & 30.37 & 3.81 & \textbf{1.04} \\
4K avg.	& 15.9 & 6.86 & 11.36 & \textbf{1.55} & 53.59 & 34.67 & 25.88 & \textbf{1.55} \\
2K avg.	& 23.71 & 17.14 & 73.33 & \textbf{3.21} & 74.47 & 77.71 & 146.00 & \textbf{3.21}
\end{tabular}
\caption{Texture stretch  \cite{sander2001texture} measurements for FastAtlas renders versus static atlases (Static), \cite{mueller2018shading} (SAS), and \cite{Neff2022MSA} (MSA-P), when shading using fixed-size atlases. For each atlas size, we report the mean $L^2$ and $L^\infty$ stretch across all scenes, at both 1920 $\times$ 1080 and 3840 $\times$ 2160 screen resolutions. $L^2$ measures root-mean-squared stretch across all directions in the frame; $L^{\infty}$ the greatest stretch in any direction. 
A value of 1 is ideal; when stretch is uniform, $L^2=L^{\infty}$.}
\label{tab:stretch_fixed_atlas}
\vspace{-3mm}
\end{table}

Table ~\ref{tab:stretch_fixed_atlas} reports stretch metrics versus prior art at $8K \times 8K$ for both $1920 \times 1080$ and $3840 \times 2160$ screen resolutions. At both screen resolutions, FastAtlas's $L^2$ and $L^\infty$ metrics are very close to 1, reflecting an ideal one-to-one sampling rate. MSA-P has reasonable $L^2$ stretch with an 8k atlas size, but has higher $L^\infty$ stretch than FastAtlas (most likely due to undersampling of meshlets in specific superblocks). Both static atlases and SAS have higher $L^2$ stretch, and much higher $L^\infty$ stretch, than FastAtlas; the discrepancy between the $L^2$ and $L^\infty$ stretches is due to both methods having high parametric distortion. SAS in particular has the highest $L^\infty$ stretch of all methods, consistent with our observation, illustrated in Figs. \ref{fig:teaser} and \ref{fig:sas_issues}, that constraining parameterized charts to rectangular patches can introduce excessive stretch. Fig~\ref{fig:prev_work} shows a representative example of our and alternative renders generated using this baseline setting. 

\subsection{Controlled Shading Rate}
\label{sec:shading_rate}
\begin{table}
\scriptsize
\setlength{\tabcolsep}{2pt}
\centering
\begin{tabular}{l|ccc|cc}
  & \multicolumn{3}{c|}{\FLIP $\downarrow$} & \multicolumn{2}{c}{Improvement vs.} \\
   & \textbf{Upsampling} & \textbf{MSA-P} & \textbf{Ours} & \textbf{Upsampling} & \textbf{MSA-P} \\
\hline
\textbf{1920 $\times$ 1080} & & & & \\
 50\%   &  0.103 & 0.110 & \textbf{0.076} & 29\% & 31\% \\
 25\%    &  0.158 & 0.161 & \textbf{0.123} & 25\% & 25\% \\
 12.5\%  &  0.217 & 0.209 & \textbf{0.169} & 25\% & 20\% \\
\hline
\textbf{3840 $\times$ 2160} & & & & \\
 50\%    &  0.088 & 0.098 & \textbf{0.069} & 26\% & 31\% \\
 25\%    &  0.140 & 0.149 & \textbf{0.116} & 21\% & 24\% \\
 12.5\%  &  0.194 & 0.194 & \textbf{0.162} & 20\% & 18\% \\
\end{tabular}
\caption{Average visual quality of FastAtlas renders versus \cite{Neff2022MSA}  and a forward upsampling baseline  targeting reduced shading rates. For each shading rate, we report mean \FLIP error vs. forward rendering for all methods and our mean relative improvement across all scenes, at standard and high screen resolutions.}
\label{tab:flip_fixed_sr}
\vspace{-3mm}
\end{table}

\begin{table}
\scriptsize
\setlength{\tabcolsep}{2pt}
\centering
\begin{tabular}{l|cc|cc||cc|cc}
& \multicolumn{4}{c||}{\textbf{1920 $\times$ 1080}} & \multicolumn{4}{c}{\textbf{3840 $\times$ 2160}} \\
& \multicolumn{2}{c|}{Mean $L^2$} & \multicolumn{2}{c||}{Mean $L^\infty$} & \multicolumn{2}{c|}{Mean $L^2$} & \multicolumn{2}{c}{Mean $L^\infty$} \\
& \textbf{MSA-P} & \textbf{Ours} & \textbf{MSA-P} & \textbf{Ours} & \textbf{MSA-P} & \textbf{Ours} & \textbf{MSA-P} & \textbf{Ours} \\
\hline
100\% SF & 1.0 & \textbf{1.0} & 1.0 & \textbf{1.0} & * & \textbf{1.0} & * & \textbf{1.0} \\
50\% SF & 3.2 & \textbf{2.0} & 7.2 & \textbf{2.0} & 3.1 & \textbf{2.0} & 7.3 & \textbf{2.0} \\
25\% SF & 6.6 & \textbf{4.0} & 14.2 & \textbf{4.0} & 6.4 & \textbf{4.0} & 14.5 & \textbf{4.0} \\
12.5\% SF & 14.0 & \textbf{8.0} & 27.7 & \textbf{8.0} & 13.3       & \textbf{8.0} & 28.4 & \textbf{8.0} \\ 
\end{tabular}
\caption{Texture stretch measurements \cite{sander2001texture} for FastAtlas renders versus \cite{Neff2022MSA} (MSA-P) when shading at fixed shading reduction rates. For each atlas size, we report the mean $L^2$ and $L^\infty$ errors across all scenes, at both standard and 4K (high) resolution. A value of * means that atlas packing failed.}\
\label{tab:stretch_fixed_sr}
\vspace{-3mm}
\end{table}

\begin{figure*}
   \includegraphics[width=\linewidth]{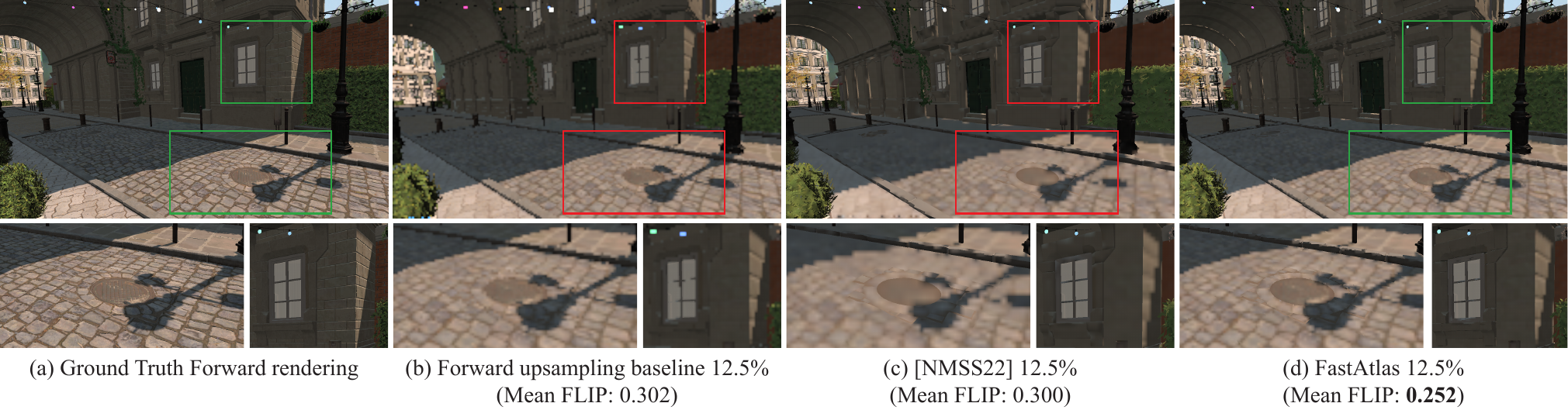}
    \caption{Comparing FastAtlas to \cite{Neff2022MSA} and a forward upsampling baseline when targeting reduced shading rates. Here, all methods target a shading rate of 12.5\% ($0.125^2=0.016$ or 1.6\% of samples). The forward upsampling baseline (b) introduces jagged artifacts while \cite{Neff2022MSA} (c) exhibits undersampling (highlighted on zoomed images). FastAtlas (d) achieves the most similar results to ground truth forward rendering (a) and preserves important details even at the extremely low 12.5\% shading rate.}
    \label{fig:reduced_sr}
\end{figure*}

We evaluate the use of FastAtlas to generate atlases with a specified target texel to pixel ratio. This setting enables  undersampling, where the shading rate is lower than rasterization rate, reducing overall render time. 
We evaluate FastAtlas suitability for this application both qualitatively, through visual inspection, and quantitatively, by measuring perceptual differences vis-a-vis forward rendering outputs.
For all examples, we use an atlas size of $8K^2$ and set the shading rate by pre-scaling the target bounding box sizes passed to FastAtlas. We evaluate FastAtlas performance in these settings by uniformly scaling the boxes to a fixed percentage of their original dimensions (50\%, 25\%, 12.5\%); target effective shading rate is then the {\em square} of this scale factor (a 50\% scale factor yields a target shading rate of 25\%). We test this setup on both standard and high-resolution display settings. Examples of output renders generated with these settings are shown in Figs.~\ref{fig:seams} and~\ref{fig:reduced_sr}. We showcase two key applications of controlled shading rate with FastAtlas (depth-of-field and foveated rendering) in our supplemental material.
An example of oversampling (multiple texels correspond to a single pixel) is provided in the supplementary.

\paragraph*{Comparisons} We compare our render quality using FastAtlas for reduced shading rates against potential alternatives (Tab. ~\ref{tab:flip_fixed_sr}). To establish a baseline, we shade a reduced-size framebuffer using the forward renderer, then upsample it to our screen resolution using simple bilinear upsampling. We outperform this method by 26\% and 22\% on average at 1080p and 4K resolutions respectively (Tab.\ref{tab:flip_fixed_sr}).
We compare our results to those of the MSA-P method of \cite{Neff2022MSA}  using the scale factors above; after computing the screen space bounding box of each meshlet, we multiply both box dimensions by the corresponding scale factor. We then pass the bounding boxes to their packing method, which may further scale them down if needed to fit in the atlas. The error introduced by their method is significantly higher than ours (Tab. ~\ref{tab:flip_fixed_sr}, ``MSA-P''), but the difference decreases as scale factor goes down. This is predictable, as at lower scales the error is dominated by undersampling; however, even at 12.5\% scale factor our error is 20\% smaller (18\% on a 4K display.) These numbers are consistent with visual inspection (Figs.~\ref{fig:seams},~\ref{fig:reduced_sr}), and our texture stretch measurements (Tab.~\ref{tab:stretch_fixed_sr}).

Shading Atlas Streaming (SAS) \cite{mueller2018shading} and its followups \cite{hladky2019tessellated,hladky2021snakebinning} are designed for streaming as the target application, and do not provide a way to control the scale of individual charts relative to their screen resolution. As such, they are inherently ill suited for applications that wish to control shading rates, and we cannot compare against them for this test.

\subsection{Limited Atlas Size/Streaming}
\label{sec:compare_streaming}
\begin{table}
\scriptsize
\setlength{\tabcolsep}{2pt}
\centering
\begin{tabular}{l|cccc|ccc}
  & \multicolumn{4}{c|}{\FLIP $\downarrow$} & \multicolumn{3}{c}{Improvement vs.} \\
  & \textbf{Static} & \textbf{SAS} & \textbf{MSA-P} & \textbf{Ours} & \textbf{Static} & \textbf{SAS} & \textbf{MSA-P} \\
\hline
\textbf{1920 $\times$ 1080} & & & & & & & \\
 4K  & 0.139 & 0.062 & 0.070 & \textbf{0.039} & 68\% & 39\% & 45\% \\
 2K  & 0.158 & 0.083 & 0.159 & \textbf{0.064} & 67\% & 26\% & 58\% \\
\hline
\textbf{3840 $\times$ 2160} & & & & & & & \\
 4K  & 0.171 & 0.077 & 0.126 & \textbf{0.055} & 68\% & 32\% & 57\% \\
 2K  & 0.187 & 0.118 & 0.238 & \textbf{0.102} & 54\% & 18\% & 56\% \\
\end{tabular}
\caption{Visual quality of FastAtlas vs. \cite{Neff2022MSA} (MSA-P), \cite{mueller2018shading} (SAS), and static atlases (Static) when shading using fixed-size atlases. For each atlas size, we report mean \FLIP error vs. forward rendering and our mean improvement across all scenes, at both 1920 $\times$ 1080 and 3840 $\times$ 2160 screen resolutions. See Tab.~\ref{tab:flip_tss_overview} for statistics on 8K atlases.}
\label{tab:flip_fixed_atlas}
\vspace{-6mm}
\end{table}

Our next experiment represents a typical use case for games and streaming applications, where users seek both runtime and memory savings and rely on a fixed budget of texture memory to determine shading rate.  We test this setup using both standard and 4K displays. In addition to the baseline $8K \times 8K$ atlas size reported earlier, we generate and shade texture atlases at $4K \times4K$, and $2K\times2K$ resolutions. We do not evaluate higher resolutions since, as reported above, FastAtlas can pack almost all frames at 100\% shading rate, using 8K atlases, even for high-resolution displays. Tab. ~\ref{tab:flip_fixed_atlas} and Figs. ~\ref{fig:teaser}, \ref{fig:static}, \ref{fig:atlas},~\ref{fig:sas_issues},~\ref{fig:more_lowres} demonstrate FastAtlas's performance in this setting. As expected, \FLIP error increases as atlas size decreases, leading to undersampling. However, this increase is relatively minor, with the visual quality remaining high even for 2K atlases (e.g. Fig. ~\ref{fig:teaser}).

\subsubsection{Comparisons} 
\begin{figure*}
\includegraphics[width=\linewidth]{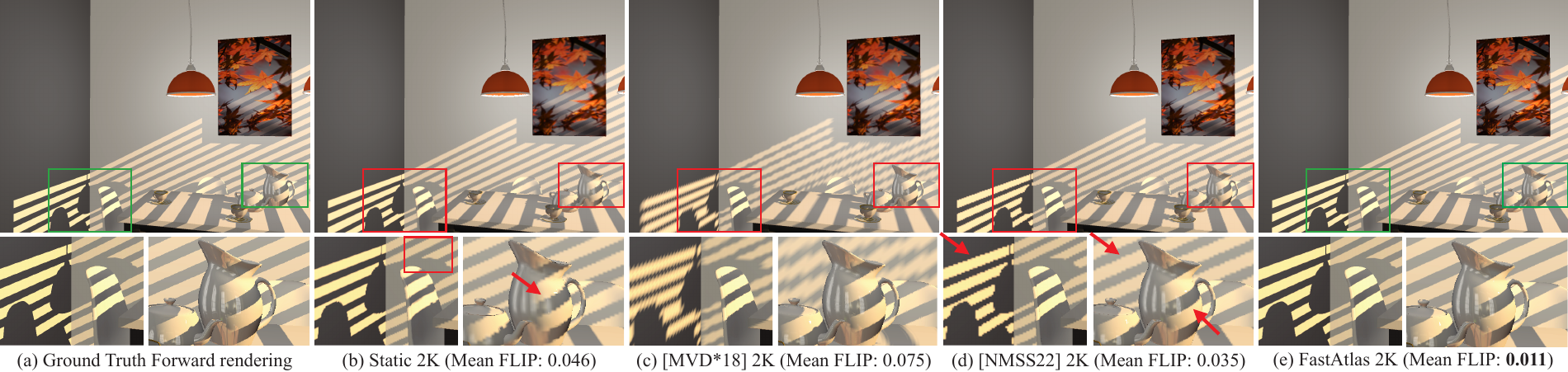}
\caption{Additional comparisons with prior art ($2K \times 2K$ atlases). Left to right:  forward render, static atlasing, SAS \cite{mueller2018shading}, MSA  \cite{Neff2022MSA}, FastAtlas. While all prior method outputs exhibit notable undersampling, our results remain visually close to forward rendering.}
\label{fig:more_lowres}
\end{figure*}

We compare the visual quality of our results generated using fixed atlas sizes to static atlases; shading atlas streaming (SAS) \cite{mueller2018shading}; and MSA \cite{Neff2022MSA}. We show representative outputs of our method, and these alternatives, in Figs. ~\ref{fig:teaser}, \ref{fig:atlas},~\ref{fig:sas_issues}, and ~\ref{fig:more_lowres}.  As documented in Tabs. ~\ref{tab:flip_tss_overview} and ~\ref{tab:flip_fixed_atlas}, the visual error (\FLIP) of outputs produced using fixed size static atlases (8K, 4K, and 2K) is a factor of 2 to 3.6 times higher than ours, and significantly higher than that of dynamic atlasing alternatives. With smaller fixed atlas sizes, our texture stretch is also much lower (Tab.~\ref{tab:stretch_fixed_atlas}).

Our comparisons (Fig. \ref{fig:atlas},~\ref{fig:sas_issues},~\ref{fig:more_lowres}, Tab.~\ref{tab:flip_fixed_atlas}) show that FastAtlas consistently outperforms SAS across the input scenes tested. As detailed in the appendix, we outperform SAS on all scenes using 8K and 4K atlases for both standard and 4K displays. At 1080p, we reduce \FLIP error by 43\% with 8K atlases and 39\% with 4K atlases on average compared to SAS. At 4K, we reduce \FLIP error by 53\% and 32\% respectively. Using 2K atlases, we outperform SAS on 11 out of 13 scenes at 1080p (average improvement 26\%), and 10 out of 13 scenes on a 4K display (average improvement 18\%). SAS has worse texture stretch (both $L^2$ and $L^\infty$) than FastAtlas on all scenes at all fixed atlas resolutions, with $L^\infty$ texture stretch being particularly severe: at $2K$, our mean $L^2$ texture stretch is $1.59$ versus SAS's $L^2$ of $16.77$, and our mean $L^\infty$ texture stretch across all frames is $1.59$ versus their $L^\infty$ of $33.39$. This confirms our visual assessment that their method has significant patches of severe parametric distortion; we note that while these may be immediately visible to a human viewer (Figs. ~\ref{fig:teaser} (frieze); ~\ref{fig:sas_issues}), this sort of visual error may not be captured by \FLIP on regions of low-contrast texture.

Using fixed atlas sizes on both regular and high-res displays, the ratio of MSA-P's \FLIP to ours ranges between 1.7 and 1.8; a very significant difference. These numbers are consistent with visual inspection. As Figs. ~\ref{fig:teaser}, ~\ref{fig:seams}, ~\ref{fig:more_lowres} show, at resolutions of 4K and less, the seam artifacts in the outputs of Neff et al become increasingly pronounced. On the {\em Robot Lab} scene which Neff et al. use in their paper (albeit with refined tessellation), our method's improvement is notable across all resolutions. With a $8\operatorname{k}^{2}/4\operatorname{k}^{2}$ texel atlas and a 1080p framebuffer, MSA-P has the best texture stretch of previous work; however, we still outperform it by a significant margin (4K 1080p: MSA-P mean $L^2$ $1.96$ vs FastAtlas $1.04$; $L^\infty$ $3.82$ versus FastAtlas $1.04$). However, with a fixed size $2\operatorname{k}^2$-texel atlas, or on a 4K HDMI display, FastAtlas outperforms it by an order of magnitude (4K HDMI: MSA-P mean $L^2$ 11.36 versus FastAtlas $1.55$; MSA-P mean $L^\infty$ 25.88 versus FastAtlas $1.55$.)

\begin{figure*}
\includegraphics[width=\linewidth]{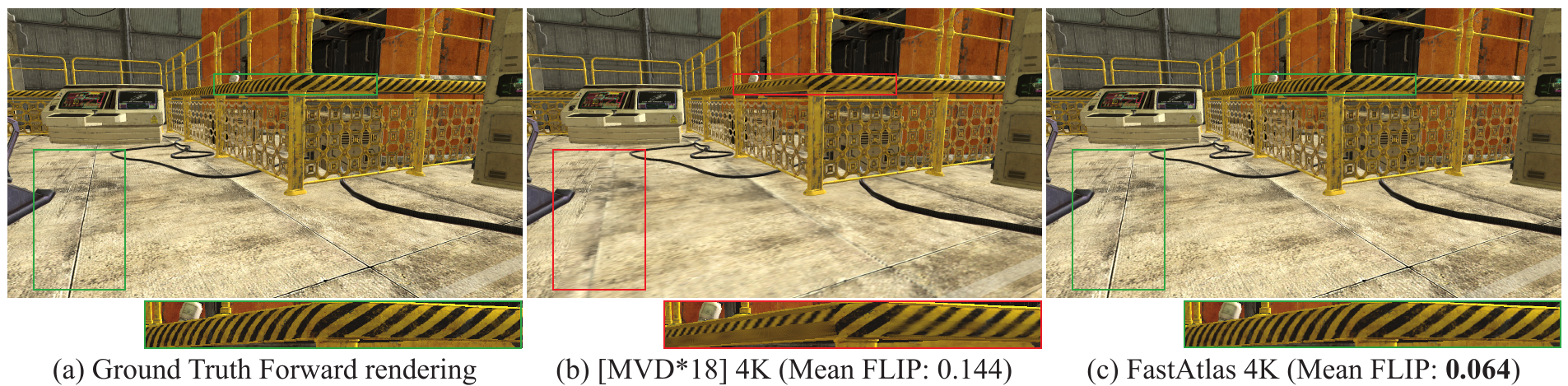}
\caption{Additional comparisons with SAS \cite{mueller2018shading}  using $4K \times 4K$ atlases. Left to right: forward render, SAS \cite{mueller2018shading}, FastAtlas. While SAS outputs exhibit notable undersampling and smudging artifacts, our results remain visually close to forward rendering. }
\label{fig:more_lowres_sas}
\end{figure*}

\begin{figure*}
\includegraphics[width=\linewidth]{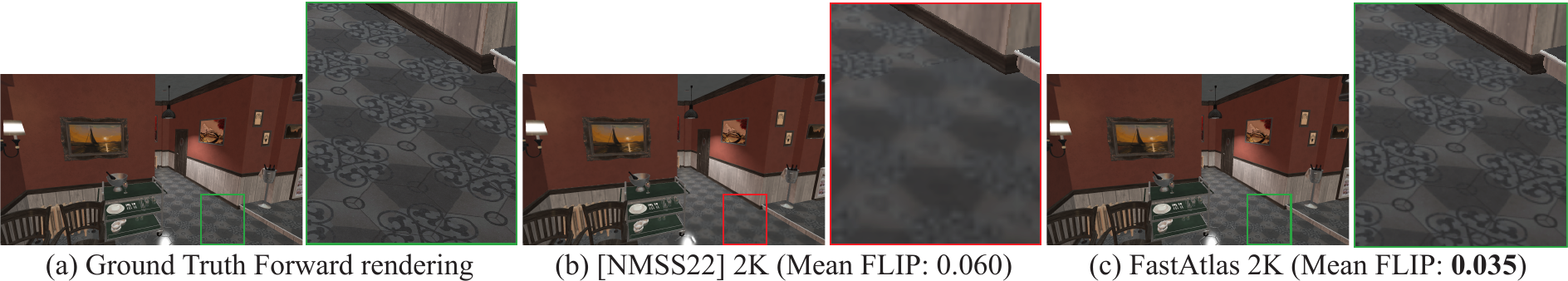}
\caption{Additional comparisons with MSA \cite{Neff2022MSA}  using $2K \times 2K$ atlases. Left to right: forward render, MSA-P \cite{Neff2022MSA}, FastAtlas. While MSA outputs exhibit notable undersampling and seam artifacts, our results remain visually close to forward rendering. }
\label{fig:more_lowres_meshlet}
\end{figure*}

\begin{table}
\vspace{1mm}
\scriptsize
\setlength{\tabcolsep}{1pt}
\tiny
\centering
\begin{tabular}{l|cccc|c|ccc|c||c}
& Chart    & AABB  & AABB      & Atlas   & Total    & Shading & Raster & Other & Total   & Total   \\
& Extract. & Comp. & Pre-Order & Packing & Atlasing &         &        & Rend. & Rend.   &         \\
\hline
\multicolumn{11}{l}{\textbf{1920 $\times$ 1080}} \\
8K        & 0.249 & 0.040 & 0.608 & 0.691 & 1.589 & 63.703  & 0.469 & 0.623 & 64.795  & 66.384  \\
4K        & 0.248 & 0.039 & 0.425 & 0.684 & 1.396 & 62.370  & 0.465 & 0.625 & 63.460  & 64.856  \\
2K        & 0.249 & 0.040 & 0.362 & 0.662 & 1.313 & 47.105  & 0.446 & 0.625 & 48.176  & 49.488  \\
\hline
8K 50\%   & 0.248 & 0.040 & 0.599 & 0.687 & 1.574 & 43.630  & 0.441 & 0.625 & 44.696  & 46.270  \\
8K 25\%   & 0.249 & 0.040 & 0.605 & 0.686 & 1.581 & 37.387  & 0.424 & 0.625 & 38.435  & 40.016  \\
8K 12.5\% & 0.249 & 0.040 & 0.602 & 0.689 & 1.580 & 35.003  & 0.414 & 0.624 & 36.041  & 37.621  \\
\hline
\multicolumn{11}{l}{\textbf{3840 $\times$ 2160}} \\
8K        & 0.255 & 0.040 & 0.605 & 0.766 & 1.666 & 133.230 & 0.768 & 1.371 & 135.369 & 137.034 \\
4K        & 0.254 & 0.041 & 0.421 & 0.714 & 1.430 & 82.692  & 0.667 & 1.373 & 84.731  & 86.162  \\
2K        & 0.254 & 0.041 & 0.356 & 0.657 & 1.308 & 49.205  & 0.627 & 1.371 & 51.203  & 52.512  \\
\hline
8K 50\%   & 0.256 & 0.041 & 0.611 & 0.765 & 1.671 & 66.680  & 0.647 & 1.371 & 68.698  & 70.369  \\
8K 25\%   & 0.254 & 0.040 & 0.609 & 0.765 & 1.669 & 45.820  & 0.622 & 1.369 & 47.811  & 49.480  \\
8K 12.5\% & 0.254 & 0.041 & 0.612 & 0.764 & 1.672 & 39.569  & 0.604 & 1.376 & 41.549  & 43.221 
\end{tabular}
\vspace{1mm}
\caption{Average runtimes across experimental setups. All times in milliseconds. Left to right: chart extraction, AABB computation, AABB pre-ordering, atlas packing, total atlasing (sum of previous four columns), shading, rasterization, other rendering/presentation, total rendering (sum of previous three columns), total time (sum of total rendering \& atlasing). }
\label{tab:runtimes}
\vspace{-3mm}
\end{table}

\subsection{Runtime.} 
Tab.~\ref{tab:runtimes} reports average runtimes for each part of our atlas generation and shading methods. Times were measured on an Intel Core i9-13900k CPU with 32 GB of RAM and an NVIDIA GeForce RTX 4090; see supplementary for exact experiment protocol. Total FastAtlas atlas generation time remains steady across all experiments, averaging $1.4ms$ per frame. Atlassing time is highest (3ms on average) on {\em San Miguel} (over 9M triangles). This number is negligible compared to shading time, which averages between 45 and 134 ms across the different experiments, and accounts for roughly 95\% of the total time per frame. As desired, shading time decreases with a decrease in shading rate. This is most notable in our experiments where shading rate is explicitly controlled, especially when going from 100\% to 50\% (decrease is not linear, as smaller charts do not shrink below a $2 \times 2$ pixel minimal size). 

Atlas computation time is dominated by the cost of ordering and packing boxes; ordering cost is roughly constant for each atlas resolution, ranging from 0.6ms for 8K to 0.3ms for 2K. Packing cost is largely resolution independent, but increases with increased scene complexity, and thus chart count, ranging from 0.5ms for simpler scenes to 1.1ms for {\em San Miguel}; as this number is dominated by the cost of evaluating the multiple packings, it can be ``tuned'' for target applications by trading packing tightness for runtime speed. 

We have no access to code for any of the prior methods we compare against, nor to the exact inputs they used, making runtime comparisons infeasible. As such, our re-implementation may not have the same performance as their original code. Moreover, as noted by Neff et al. ~\shortcite{Neff2022MSA}, when evaluating on standard scenes such as {\em Robot Lab} both they and SAS refine their meshes to demonstrate cases where their method and SAS perform best; these refined meshes are not readily available. 
As such, we cannot compare our runtimes to their reported ones for the same inputs. Overall, however, we believe our method is both competitive in terms of run-time and suitable for real-time applications on commercially available hardware. 

Finally, we compare the runtimes of our unoptimized research code to those of the G3D forward renderer, which has been highly optimized and hand-tuned over several years. G3D forward rendering takes $25.3$ msec at 1080p on average, and $71.4$ msec at 4K on average across our test scenes. While this is faster than our method at 1080p across all experiments, at 50\% shading rate on 4K monitors we outperform the forward renderer (70.4 msec) and are significantly faster at 25\% shading rate (49.5 msec.) Improvement is most pronounced on scenes with lower geometric and material complexity (at 50\% SR we outperform the forward renderer on all scenes except {\em bistro exterior}, {\em San Miguel}, and {\em modern house}); this and profiling suggests that our primary performance disadvantage during shading is due to unoptimized handling of scenes with multiple materials, which can be improved for commercial applciations. Overall, this confirms that our method is practical for reducing shading cost on high-resolution displays, or where reduced shading rate is desireable.

\section{Conclusions}

We present FastAtlas, a novel GPU-based method for the generation of dynamic texture atlases for texture-space shading. As our experiments demonstrate, FastAtlas can be used across the entire spectrum of TSS applications. Our method addresses the main sources of visual artifacts present in prior methods: uneven sampling rate, undersampling, and shading discontinuities along visible texture seams. It generates charts with invisible seams, and bijectively parameterizes and packs their visible fragments into texture atlases with high packing efficiency in real time. We validate our method by performing texture-space shading on complex scenes, and demonstrating its superiority over the state of the art in terms of shading quality given the same application driven setup. 

\paragraph*{Limitations and Future Work} Like all projection-based TSS methods, when charts are downscaled triangles on screen with near-subpixel geometry may not be rasterized into the shading atlas due to the hardware rasterizer's crack-free rendering rules. In rare cases, this can lead to tiny but distinct details being removed. A potential solution could be to support per chart scaling that prevents visible triangles from shrinking to size below one pixel.

Like prior work, FastAtlas only considers the geometry of the scenes being rendered and is agnostic to the texture or material properties of the rendered content. However, as pointed out by \cite{sander2002signal}, the impact of atlas stretch on actual visual output quality depends on the properties of the data stored in the atlas. A data (shading) dependent atlasing strategy is therefore a promising area of future work. It is also interesting to further explore GPU based texture atlasing algorithms that further increase packing efficiency.

\paragraph*{Acknowledgements} The authors wish to thank the authors of Neff et al. \shortcite{Neff2022MSA} and Mueller et al. \shortcite{mueller2018shading}, especially Markus Steinberger, for their assistance and meshlet generation source code. We also thank Jinfan Yang and Suzuran Takikawa for their help with figures. We acknowledge the support of the Natural Sciences and Engineering Research Council of Canada (NSERC) grant RGPIN-2024-03981. F. Gu was supported by an NSERC CGS-M scholarship. Finally, this work is supported in part by the Institute for Computing, Information and Cognitive Systems (ICICS) at UBC.

\bibliographystyle{eg-alpha-doi}

\newcommand{\etalchar}[1]{$^{#1}$}

\appendix
\section{Implementation Details}
\label{app:implementation_details}

We discuss GPU-specific implementation details of our algorithm below.

\paragraph*{Detecting Visible Triangles.} To identify all triangles which are visible, we first perform a depth pre-pass from the camera's view, filling the depth buffer. We then perform a second pseudo-rendering pass with conservative rasterization enabled, testing each triangle against the depth buffer using the {\em less than or equal} depth test condition. For any fragment that passes the depth test, we mark the triangle from which that fragment originated as visible in a {\em visible triangle buffer} (VT buffer) which contains an entry for each triangle in the scene. This buffer is initialized with all null values; once a triangle is determined to be visible we record its ID in the corresponding entry. This method of recording visibility is inspired by visibility buffers \cite{burns2013visibility} and the observation that efficient GPU occlusion testing can be performed using buffer objects \cite{kubisch2014opengl}.  

\paragraph*{GPU Union Find.} We implement GPU union find using the {\em ECL-CC} method of Jaiganesh and Burtscher \shortcite{Jaiganesh:2018}, which implements the union-find algorithm on the GPU in three stages: initialization, a union-find algorithm based on pointer hooking, and finalization. As Jaiganesh and Burtscher's algorithm was originally designed for arbitrary graphs of high degree, we specialize it for our problem case as follows. First, their original paper initializes node component IDs (in our case, triangle IDs) by setting them to the value of their first connected neighbour with a lower value. For arbitrary graphs, this produces a good initialization while not visiting all edges, which is important for nodes of high degree. Since our highest degree is 3, we initialize the triangle component IDs directly using the visibility shader, and simply visit every edge subsequently during initialization. Second, they split their hooking step into two distinct kernels to account for vertices of high degree; as our highest degree is 3, we do not do this. We implement the entire algorithm as three separate GLSL compute shaders, synchronized with memory barriers to achieve synchronization between all threads and not just those in a given work group.

\begin{figure*}
\vspace{-10pt}
\includegraphics[width=\linewidth]{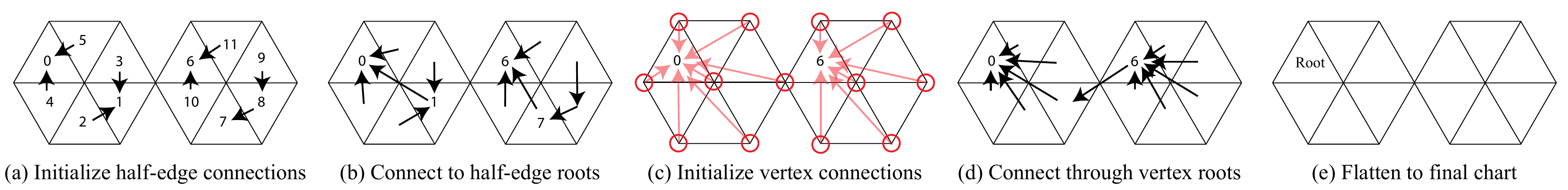}
\caption{GPU Union Find. Given each triangle labelled with its starting index, we first initialize by linking each triangle to its neighbor of lowest index (a) through the half-edge data structure. Triangles are then hooked together such that each has a path to the root triangle (b). We then initialize a mapping from vertices to triangle roots by writing to each vertex the minimum root of its connected triangles (c). Finally, we perform another connection step through the initialized vertex roots (d) and then flatten the entire data structure to point to the global root (e), which is then recorded as the root triangle for the newly formed chart.}
\vspace{-10pt}
\label{fig:union_find}
\end{figure*}

In order for chart IDs to be read by the vertex shader during shading and rasterization without changing the mesh topology in the triangle index buffer, we require each {\em vertex} to be mapped to a unique visible chart; this is made possible by our charts having seams at visible borders. We therefore run an additional compute pass which merges any two formed charts that share a common vertex at their boundary. The resulting set of charts then satisfies this requirement, and we generate a {\em vertex-to-chart map} to be used during shading and rasterization. Pseudocode is provided in Algorithm 1 in this document. An overview diagram is provided in Figure \ref{fig:union_find}.

\paragraph{Discontinuity Processing at Chart Boundaries.} Our chartification method does not form visible seams on continuous surfaces, allowing us to effortlessly avoid artifacts such as those in Fig. 2c. However, like all standard shading approaches that use bilinear sampling, we must take care to avoid sampling artifacts along geometric discontinuities; in other words, even though our seams are invisible, any sample at the edge of a chart still must still have a plausible color for bilinear filtering to work correctly. Following Andersson et al. \cite{andersson2014adaptive}, we use a two-pass approach with conservative rasterization. We first render each triangle with conservative rasterization disabled; we then render each triangle again with depth compare set to {\em strictly less} and conservative rasterization enabled. This ensures that a bilinear tap occuring at the boundaries of a chart has shaded texels to read from, without the problems of image dilation methods.

\paragraph{Alpha Tested Materials.} We handle alpha tested materials conservatively by not writing any triangle during the depth pre-pass that is not fully opaque (ie. has alpha=$1$ everywhere), and by testing triangle visibility for alpha-tested materials as though alpha testing was disabled. In effect, we never consider materials with alpha to be occluders, and we never test alpha value during rasterization. In practice, we found that this approach avoids numerous edge cases with little to no impact on rendering speed. 

\paragraph*{Shading and Rasterization.}

Our framework is compatible with all existing approaches for texture space shading, including forward shading, raytraced illumination, or deferred shading in texture space \cite{baker:2016}. For demonstration purposes, in the examples shown we implement a standard forward shading based rendering pipeline: we transform the vertices of our chart triangles within the vertex shader, clip these triangles using the standard GPU triangle setup engine, and finally shade the fragments within a dedicated fragment shader.  We implemented our method in the G3D Innovation Engine \cite{G3D17}, a commercial grade renderer which provides a suitable forward shading pipeline for texture-space shading. 

To position our charts in the texture atlas for shading, we run a standard vertex shader on the vertices of each chart. Our shader first applies the standard camera and perspective transformation to all vertices for the current view; it then scales all charts by the scale factor $S$ and finally, for the vertices within each chart, it applies a per-chart rigid transformation that maps the chart's bounding box in post-perspective divide NDC space to its corresponding previously computed bounding box in the shading atlas. When shading into the atlas, we clip triangles using the GPU triangle setup engine: we compute the distance in homogenous clip space between each vertex and the homogenous clip planes of the original, screen-space projection in the vertex shader, and pass them to the GPU as user clip planes  to discard any fragment that needs to be clipped. 

\paragraph*{Filtering Across Hard Edges.} To render the frame of interest as seen from the camera using a FastAtlas generated atlas, we use the same logic as used for shading, except that the chart UV coordinates are passed to the fragment shader verbatim post-division with no need to invoke the user clip planes. (Any fragment being rasterized where $w<0$ during rasterization is off-screen by definition, and will be culled as part of the normal culling pipeline.) When the input geometry contains sharp edges or discontinuities, our method will by default perform bilinear filtering across the boundary; however, some applications may prefer shading to be discontinuous across such edges. We propose a simple modification to the final rasterization pass to achieve this. For each fragment, we consider the 4 texels that would be addressed by a standard bilinear fetch. We determine whether each texel should contribute to the final fragment color with two simple tests: if it is in the same triangle as the current fragment; or if it is in a triangle adjacent to the current fragment, but the angle between the two faces is less than a user-defined threshold. We find empirically that a threshold of 75 degrees performs well, effectively preventing sampling across sharp edges and depth discontinuities within the same chart. To support these queries, we also render a 32-bit triangle ID texture when shading to the atlas.

Our conditions may yield anywhere between 0 and 4 valid texels for each fragment. To reconstruct shading from a variable number of samples, we proceed case-by-case. If 0 texels are valid, we use the nearest texel, and if only 1 texel is valid, we use it. If 2 texels are valid, we linearly interpolate their values. If 3 texels are valid, we perform barycentric interpolation using the barycentric coordinate formula derived by Sen et al. \cite{Sen2004SilmapTex}. However, it is possible for the result to fall outside the range of the three input colors if the sample location is outside the triangle formed by the three valid texels, yielding shading that appears incorrect \cite{Sen2004SilmapTex}. Slightly modifying their proposed solution, we clamp the resulting color to the per-component minimum and maximum of the input colours. Finally, if all 4 texels are valid, we use standard bilinear interpolation. All results in the paper are generated with our modified sampling enabled.

\begin{algorithm}
\caption{GPU Union-Find (modifed from ECL-CC)}
\KwIn{Frontfacing triangles initialized with triangle indices}
\KwOut{Mapping T from triangle index to chart index (initialized to triangle index)}
\BlankLine

// Initialize connected components \\
\ForPar{each frontfacing triangle i}{
\For{each connected triangle j}{
    \If{j is frontfacing}{
        T[i] = min(i,j)
    }
}
}

// Hook all connected components to roots \\
\ForPar{each frontfacing triangle i}{
currentRoot = findRootNode(i)\;
\For{each connected triangle j}{
	connectedRoot = findRootNode(j)\;
	\While{currentRoot != connectedRoot}{
		\eIf{currentRoot < connectedRoot}{
			connectedRoot = atomicCompSwap(T[connectedRoot], connectedRoot, currentRoot)
		}{
			currentRoot = atomicCompSwap(T[currentRoot], currentRoot, connectedRoot)
		}
		// Loop will repeat if the atomicCompSwap failed
	}
}
}

// Initialize vertices from connected components \\
\ForPar{each frontfacing triangle i}{
root = findRootNode(i)\;
\For{each vertex v of i}{
	atomicMin(v, root)
}
}

/* Repeat hooking step for vertices */

// Flatten vertex to chart mappings \\
\ForPar{each frontfacing triangle i}{
currentParent = T[i]\;
nextParent = T[currentParent]\;
\While{currentParent > nextParent}{
	currentParent = nextParent\;
	nextParent = T[currentParent]\;
}
T[i] = currentParent 
}
\end{algorithm}

\paragraph*{Data Structures.} Table \ref{tab:data_structures} provides a summary of all data structures and intermediate buffers used by our method, as well as their memory requirements.

\begin{table}[]
\scriptsize
\setlength{\tabcolsep}{2pt}
\centering
\begin{tabular}{p{0.3\linewidth} | p{0.65\linewidth}}
\textbf{Data Structure}          & \textbf{Description} \\
\hline
\textbf{Geometry Representation} & \\
halfEdges                        & 3 half-edges per triangle (start/end, face, next, twin). \\
\hline
\textbf{General}                 & \\
charts                           & List of charts; can hold a user-defined maximal number of charts. \\
visibleTriangles                 & Visible triangle (VT) buffer; one boolean value per triangle indicating whether it is visible. \\
metadata                         & Small metadata buffer for number of charts in this frame and chosen scale factor out of 64. \\
\hline
\textbf{Union-Find}              & \\
triangleToRoot                   & Mapping from each triangle to its "root" in union-find. \\
rootToChart                      & Mapping from each root triangle to its chart index in the charts buffer. \\
vertToChart                      & Mapping from each vertex to its chart index in the charts buffer. \\
\hline
\textbf{Cuckoo Hash Sorting}     & \\
chartCountAndOffsetPerSize       & Two integers per chart height (starting point in even and odd positions). \\
chartIndicesSparse               & Scratch buffer with enough space for $4\times$ the maximal number of charts. \\
chartIndicesCompacted            & Compacted scratch buffer with enough space for $4\times$ the maximal number of charts. \\
sortedCharts                     & Mapping from sorted index to chart index in the charts buffer. \\
\hline
\textbf{Folding and Pushing}     & \\
prefixSum                        & Prefix sum of the sorted chart widths. \\
pivotIndices                     & Index of the last chart in each row, per parallel scale factor. \\
overfillScaleFactor              & One scale per parallel scale factor (the scale required to fit all rows in the atlas width). \\
offsets                          & Intermediate x, y offsets in the atlas for each chart, per parallel scale factor. \\
numPivotsAndOverfill             & Two integers per parallel scale factor (number of folding points and maximum vertical height reached by charts). \\
frontLine                        & Total chart height at each x-position in the atlas during pushing, per parallel scale factor.
\end{tabular}

\vspace{1mm}
\caption{Data structures used in each stage of FastAtlas atlas computation. Our implementation maintains each buffer separately for readability, but we note that in particularly memory-constrained environments, further optimizations are viable by reusing intermediate buffer storage between stages.}
\label{tab:data_structures}
\end{table}

\section{Performance Measurement Details.}

\paragraph*{Test Corpus.} Our test corpus includes detailed exterior (\textit{bistro exterior}, \textit{Greek Villa}, \textit{San Miguel}) and interior spaces (\textit{bistro interior, sibenik, robot lab, breakfast room, modern house}), with highly detailed geometry and high triangle counts (2.9 million triangles for \textit{bistro exterior}, over nine million triangles for {\em San Miguel}). The circular paths have 300 frames, and target 5 seconds of rendered footage at 60 frames per second. The spline paths have 1,320 frames.

\paragraph*{Timing Methodology.} To ensure accurate and representative timings, the GPU and memory clocks were locked using the {\em nvidia-smi} tool prior to measurement. We first render one full flight path without taking measurements to ensure that caches are heated and all shaders required for the scene are compiled. We then render all frames again and report the measurements across all frames. To produce a representative measure of algorithm performance in real world scenarios we follow \cite{Neff2022MSA} and add extra lights to the scenes (specifically, we add 100 extra point lights using a loop in the shader). 

\paragraph*{Implementation of Alternative Methods.} To compare our results with \cite{Neff2022MSA} we used the meshlet generation code provided to us by the authors, and re-implemented the rest of their pipeline, including their superblock packing scheme (they were unable to share the rest of the codebase but kindly advised us on implementation details). Our implementation of Shading Atlas Streaming \cite{mueller2018shading} is based on their paper, using the same superblock allocation code we use for \cite{Neff2022MSA}. We note that Hladky et al. \shortcite{hladky2022quadstream}, who had access to the original SAS codebase, report that SAS fails on some models on which our reimplementation of their method succeeds (most notably, {\em San Miguel}). We also note that both \cite{Neff2022MSA} and \cite{mueller2018shading} operate on refined and remeshed versions of the input scenes for their experiments. Our paper operates on the original scenes, taken from the Computer Graphics Archive \cite{McGuire2017Data}, without geometry refinement. Neff et al. originally used a dilation filter to handle bilinear sampling at chart boundaries, rather than conservative rasterization; we tried both approaches and found that conservative raster produced slightly reduced \FLIP error for both methods on our test scenes; therefore we use conservative raster for both their method and ours. For SAS, we handle bilinear filtering by following their methodology and dilating UVs inwards by $0.5$ texels; for 1-pixel charts, which SAS can produce, we simply disable it.

The size of the superblock used for packing is a critical parameter for both of these methods: they can fail to produce atlases if the superblock size is too large and not enough superblocks can be fit into the atlas for a given frame. At the same time, using smaller superblocks requires scaling charts down, reducing sampling rate and thus visual quality. In their work, Neff et al. use $2048 \times 2048$ superblocks and a target atlas size of $16k \times 8k$. The original SAS paper ~\shortcite{mueller2018shading} uses superblock sizes of size $256 \times 256$ for their experiments with 8 and 16 megapixel atlases and $128 \times 128$ for 2 and 4 megapixel atlases. For a maximally fair comparison, we start with each method's largest reported superblock size for all test scenes; if packing fails, we rerun the entire scene using a superblock size that is half the previous size, and repeat until finding a superblock size that fits. We note that this would not be viable in a real-world application, as this comes with a significantly increased computational cost. Unlike our prefix-sum based approach, block allocation does not lend itself to parallelization across all candidate scales; to achieve high performance, block allocation is already parallelized efficiently across all charts \cite{mueller2018shading}, meaning that that the GPU is already fully utilized. (Furthermore, their three-phase allocation scheme would require keeping a full copy of the superblock data structure for each resolution being tested.) The alternative of picking a superblock size small enough to ensure that packing would never fail for any scene in our corpus would result in much more severe undersampling, significantly degrading both methods' outputs. 

\begin{figure}
\includegraphics[width=\linewidth]{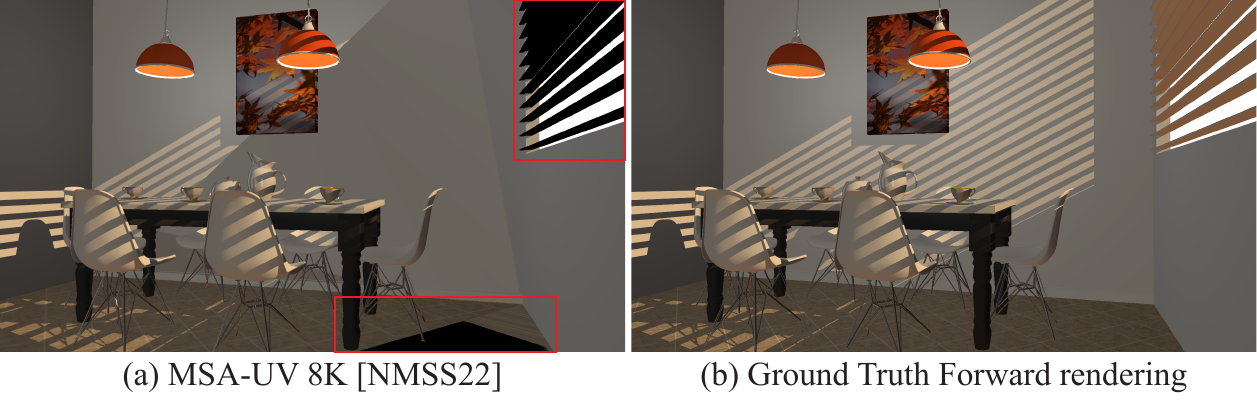}
\caption{Attempting to parameterize meshlets with LSCM parameterization on non-remeshed input geometry results in failure, as LSCM does not guarantee bijectivity for irregular atlases.}
\label{fig:msa_uv_foldover_fail}
\end{figure}

We attempted to compare against UV-based meshlet shading atlases (MSA-UV in \cite{Neff2022MSA}), which use offline LSCM-based UV computation for each meshlet rather than perspective projection, but found that in practice it was impossible to generate LSCM-parameterized meshlets with non-refined geometry without foldovers (Fig. ~\ref{fig:msa_uv_foldover_fail}). 

\section{Additional Gallery.}

\begin{figure*}
   \includegraphics[width=\linewidth]{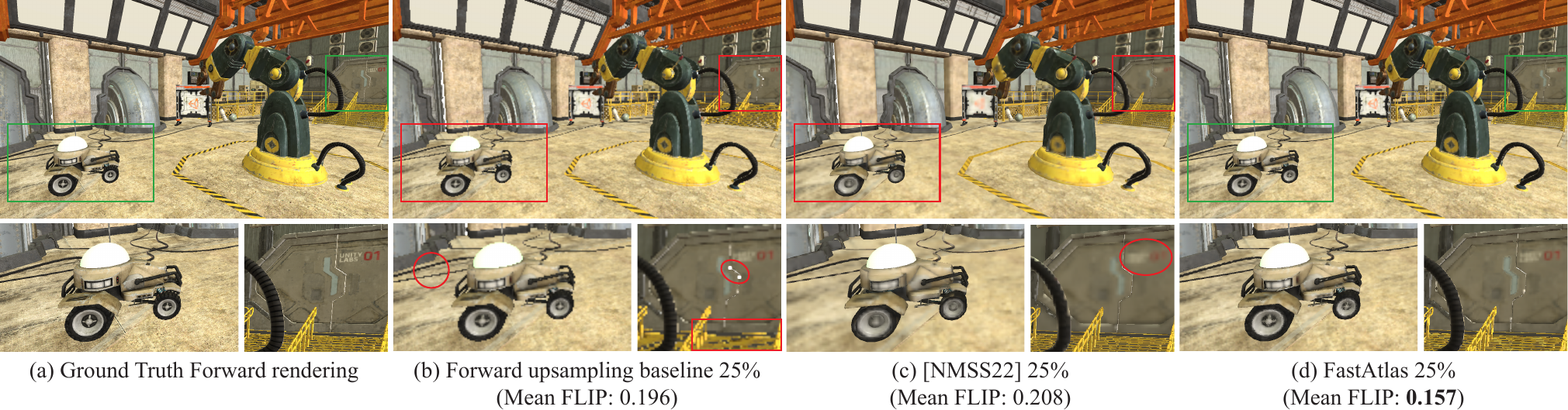}
    \caption{Comparing FastAtlas to \cite{Neff2022MSA} and a forward upsampling baseline when targeting reduced shading rates. Here, all methods target a shading rate of 25\% ($0.25^2=0.0625$, or 6.5\% samples). The forward upsampling baseline (b) introduces jagged artifacts while \cite{Neff2022MSA} (c) exhibits undersampling (highlighted on zoomed images). FastAtlas (d) achieves the most similar results to ground truth forward rendering (a) and preserves important details even at a 25\% shading rate.}
    \label{fig:supp_reduced_sr}
\end{figure*}

\begin{figure*}
\includegraphics[width=\linewidth]{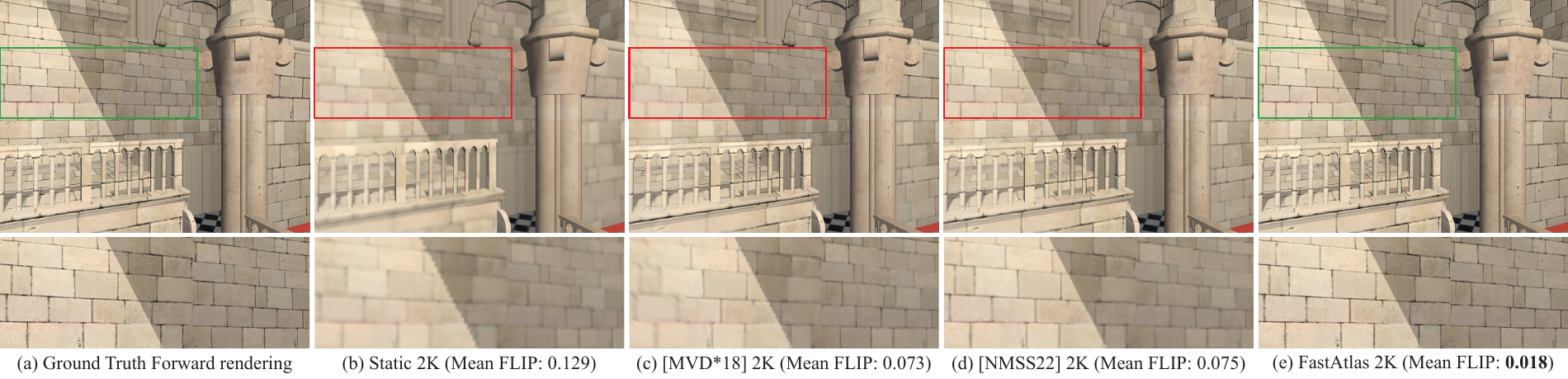}
\caption{Additional comparisons with prior art ($2K \times 2K$ atlases). Left to right: reference forward render, static atlasing, SAS \cite{mueller2018shading}, MSA  \cite{Neff2022MSA}, FastAtlas. While all prior method outputs exhibit notable undersampling, our results remain visually close to forward rendering.}
\label{fig:supp_more_lowres}
\end{figure*}

\begin{figure*}
\includegraphics[width=\linewidth]{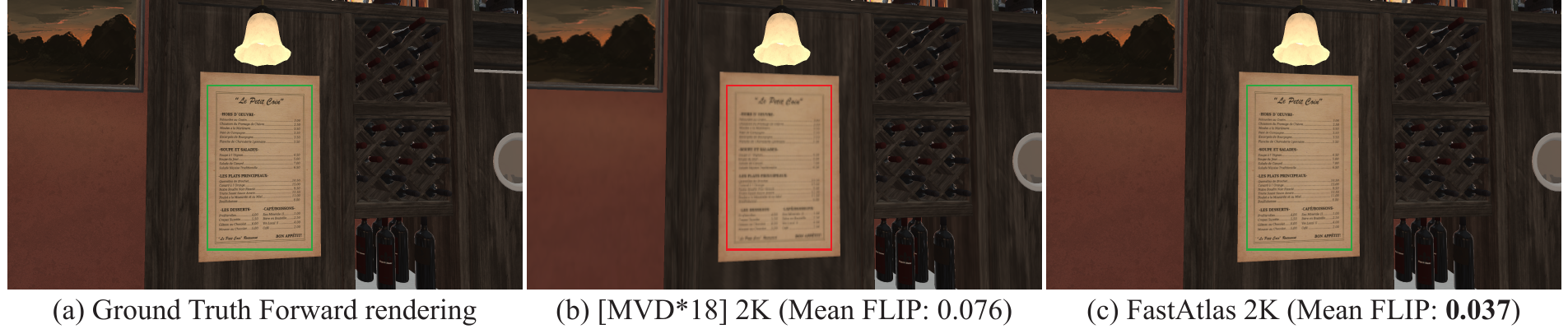}
\caption{Additional comparisons with SAS \cite{mueller2018shading} using $2K \times 2K$ atlases. Left to right: reference forward render, SAS \cite{mueller2018shading}, FastAtlas. While SAS outputs exhibit notable undersampling, our results remain visually close to forward rendering.}
\label{fig:supp_more_lowres_sas}
\end{figure*}

\begin{figure*}
\includegraphics[width=\linewidth]{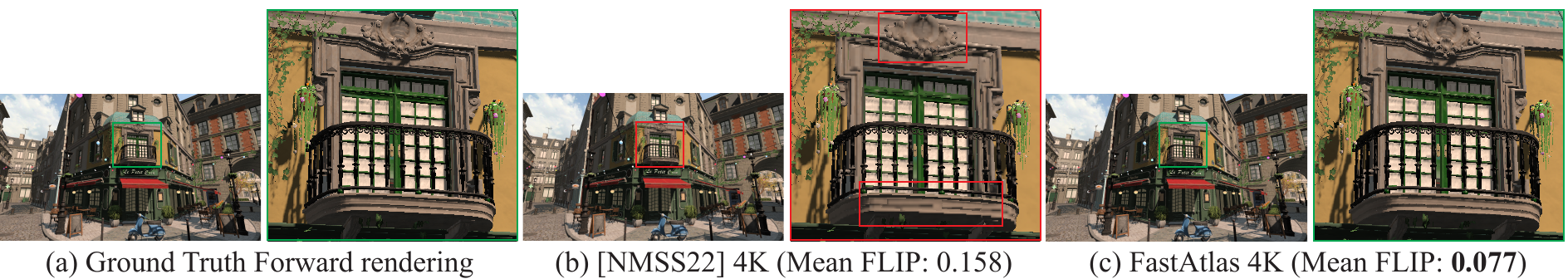}
\caption{Additional comparisons with MSA \cite{Neff2022MSA} using $4K \times 4K$ atlases. Left to right: reference forward render, MSA-P \cite{Neff2022MSA}, FastAtlas. While MSA outputs exhibit notable undersampling and seam artifacts, our results remain visually close to forward rendering.}
\label{fig:supp_more_lowres_meshlet}
\end{figure*}

We provide additional examples of FastAtlas renders generated using reduced shading rates and fixed-size atlases ($2K \times 2K$ and $4K \times 4K$) compared to alternatives in Figs. \ref{fig:supp_reduced_sr}, \ref{fig:supp_more_lowres}, \ref{fig:supp_more_lowres_sas}, and \ref{fig:supp_more_lowres_meshlet}.

\section{Additional Experiments.}

\paragraph*{Temporal Reuse.}
\label{sec:compare_temporal}
\label{sec:temporal}
\label{sec:reuse}

\begin{figure}
\includegraphics[width=\linewidth]{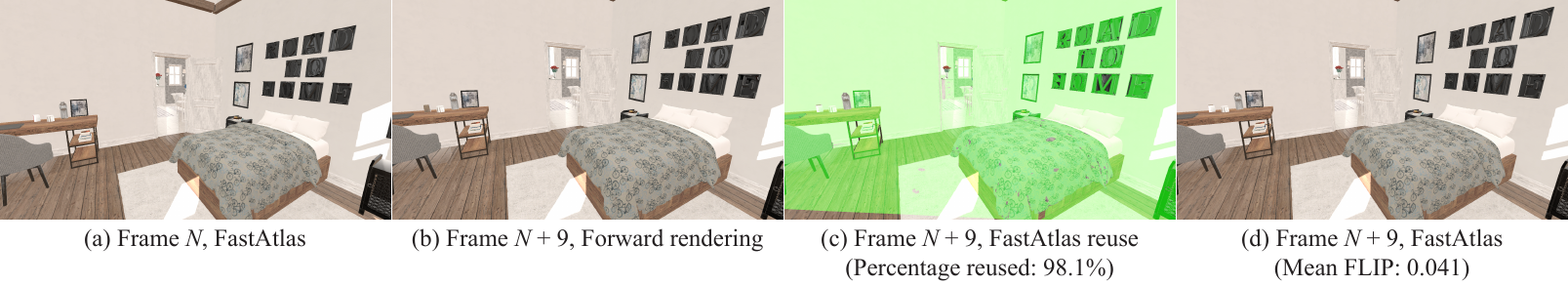}
\caption{Using FastAtlas for temporal re-use: (a) Frame $N$ rendered using FastAtlas; (b) Frame $N+9$ rendered using forward rendering. (c) Our reuse pattern in frame $N+9$ - reused areas greened out. (d) Frame $N+9$ rendered using FastAtlas.}
\label{fig:temporal_compare}
\end{figure}

To evaluate FastAtlas's suitability for temporal decoupling algorithms, we implement a basic temporal re-use algorithm on top of TSS atlases (Fig~\ref{fig:temporal_compare}), and report runtime, reuse, and quality relative to standard forward rendering. 
In the first rendered frame, we chart and shade all visible triangles as discussed above. For subsequent frames, we generate a new atlas for each frame, but only shade triangles in this atlas if they are newly or partially visible (a triangle is \textit{partially visible} if it contains at least one vertex for which $\max(|x^p|, |y^p|) \geq w^p$). During rasterization, we test each fragment in a triangle to see if it was visible in the initial frame, by interpolating the triangle's homogenous clip-space coordinates from the original frame and applying the test above. Depending on if the fragment was visible in the initial frame or not, we read texture information from the appropriate atlas. At an adjustable interval, we re-shade the entire chart atlas from scratch. This approach gracefully handles both sudden camera movement and disocclusions.  On average, our FastAtlas-based temporal re-use method reduces shading time by 33\% across all scenes.

\begin{figure}
\includegraphics[width=\linewidth]{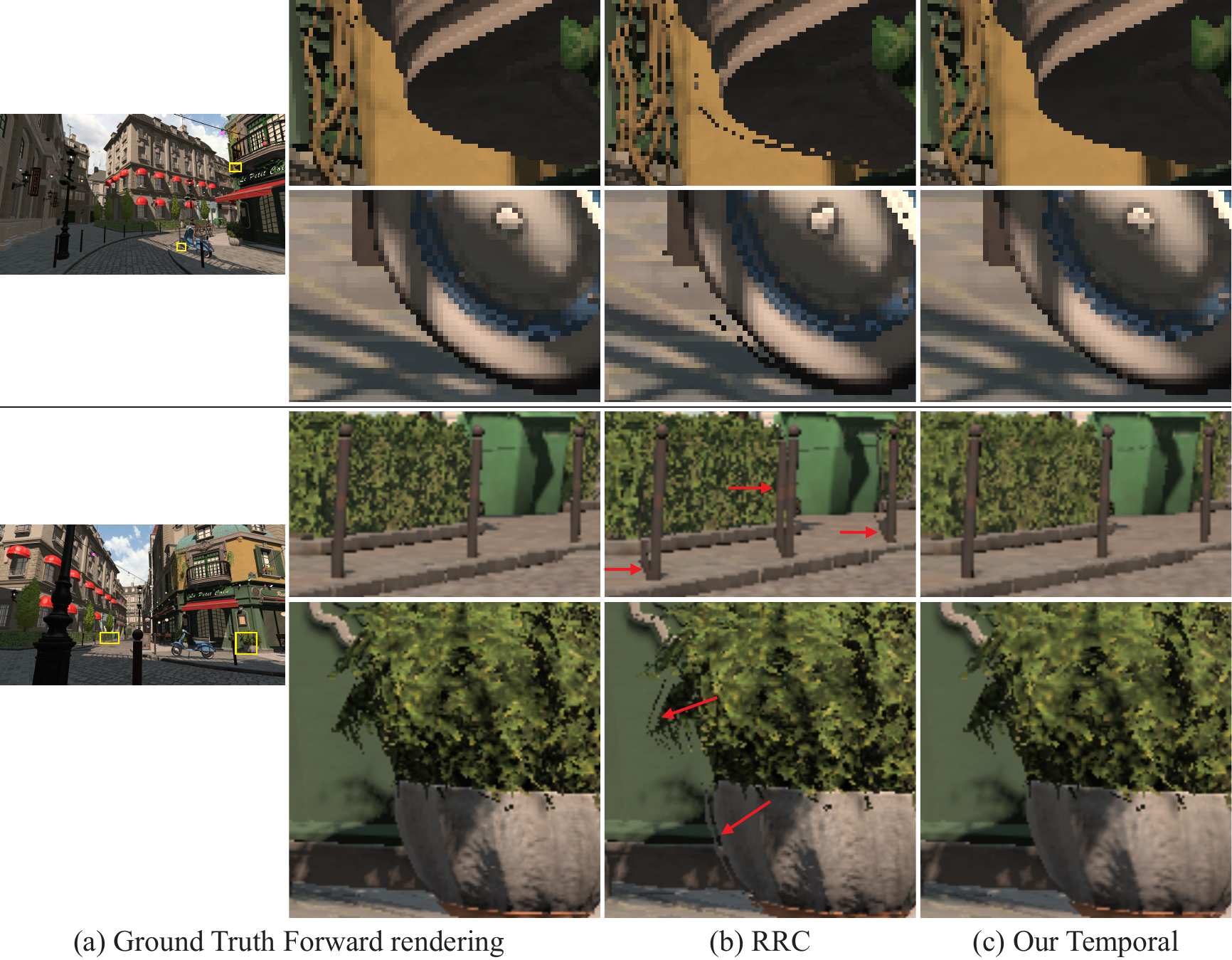}
\caption{Examples of typical ghosting artifacts produced by reverse reprojection cacheing \cite{nehab2007accelerating}(b) and our ghost-free temporal reuse results (c); ground truth forward renders on the left (a).}
\label{fig:ghosts}
\end{figure}

We compare our approach to the standard reverse reprojection caching method used for screen-space shading reuse (RRC)  \cite{nehab2007accelerating} (Fig. ~\ref{fig:temporal_compare}). We refresh both methods every ten frames, and exclude frames from reporting where both methods regenerate the entire shading cache or atlas from scratch.  
For both methods we measure the minimum and mean percentages of fragments reused, and mean \FLIP error compared to forward rendering. Both methods achieve comparable fragment reuse on average (90\% for both RRC and FastAtlas) and in the worst case (RRC: 78\%; FastAtlas: 79\%).  We achieve both better average mean \FLIP (RRC: 0.05; FastAtlas: 0.04) and maximum (RRC: 0.07; FastAtlas: 0.05) error across all test scenes; our error with temporal re-use is in fact very close to our baseline TSS error without it (comparing our temporal re-use method to a forward renderer naturally means that \FLIP will measure some error on this test that is not due to temporal re-use, but due to signal aliasing differences between the forward and TSS renderers). While these metrics show moderate improvement, the key advantage of the simple reuse approach we propose is that, by construction, it removes ``ghosting'' artifacts which are prevalent 
in RRC outputs, despite our efforts to tune its parameters for our scenes (Fig. ~\ref{fig:ghosts}). Visual inspection suggests that ghosting shows up in a large percentage of the RRC outputs (see supplementary gallery).

\begin{table}
\scriptsize
\setlength{\tabcolsep}{2pt}
\centering
\begin{tabular}{l|ccc|ccc|ccc}
 & \multicolumn{3}{c|}{\textbf{SAS}} & \multicolumn{3}{c|}{\textbf{MSA-P}} & \multicolumn{3}{c}{\textbf{Ours}} \\
\textbf{Atlas Size - Path} & Orig.        & Simp.     & Diff.     & Orig.          & Simp.     & Diff.     & Orig.         & Simp.     & Diff.                \\
\hline
\textbf{1920 $\times$ 1080} & & & & & & & & & \\
8K - Rotation Path         & 0.063        & 0.065     & 0.002     & 0.087          & 0.085     & -0.002    & 0.051         & 0.050     & -0.001               \\
4K - Rotation Path         & 0.063        & 0.065     & 0.002     & 0.095          & 0.098     & 0.003     & 0.052         & 0.052     & -0.001               \\
2K - Rotation Path         & 0.090        & 0.098     & 0.009     & 0.301          & 0.195     & -0.106    & 0.095         & 0.094     & -0.001               \\
8K - Flight Path           & 0.066        & 0.071     & 0.005     & 0.085          & 0.084     & -0.001    & 0.050         & 0.049     & -0.001               \\
4K - Flight Path           & 0.066        & 0.072     & 0.006     & 0.098          & 0.102     & 0.004     & 0.052         & 0.051     & -0.001               \\
2K - Flight Path           & 0.094        & 0.105     & 0.011     & 0.266          & 0.268     & 0.002     & 0.094         & 0.093     & -0.001               \\
\hline
\textbf{3840 $\times$ 2160} & & & & & & & & & \\
8K - Rotation Path         & 0.057        & 0.068     & 0.011     & 0.076          & 0.081     & 0.005     & 0.042         & 0.043     & 0.001                \\
4K - Rotation Path         & 0.078        & 0.089     & 0.011     & 0.134          & 0.142     & 0.009     & 0.078         & 0.079     & 0.002                \\
2K - Rotation Path         & 0.126        & 0.141     & 0.015     & 0.413          & 0.408     & -0.005    & 0.137         & 0.136     & -0.001               \\
8K - Flight Path           & 0.063        & 0.077     & 0.014     & 0.079          & 0.084     & 0.005     & 0.041         & 0.042     & 0.001                \\
4K - Flight Path           & 0.080        & 0.092     & 0.012     & 0.178          & 0.185     & 0.007     & 0.076         & 0.077     & 0.001                \\
2K - Flight Path           & 0.132        & 0.148     & 0.016     & 0.452          & 0.452     & 0.000     & 0.135         & 0.135     & -0.001              
\end{tabular}
\vspace{1mm}
\caption{Visual quality for FastAtlas versus \cite{Neff2022MSA} (MSA-P) and \cite{mueller2018shading} (SAS) on the original and simplified (50\% of original face count) Bistro Exterior models on both rotation and spline flight paths. While FastAtlas \FLIP remains virtually unchanged, simplification increases the error for the alternatives.}
\label{tab:bistro_ext_simplified}
\end{table}

\paragraph*{Level of Detail Ablation.} To examine how well FastAtlas and other methods perform on coarser geometry that is optimized for reduced mesh size and vertex/triangle count, we simplified the {\em bistro exterior} scene (2.9M triangles), which has very regular mesh complexity, using the commercially available {\em RapidPipeline} level-of-detail software package. Tab. ~\ref{tab:bistro_ext_simplified} shows the results of our method, SAS, and MSA-P \cite{Neff2022MSA}, on this simplified model with a triangle count reduced by 50\%. Both SAS and MSA-P exhibit higher stretch on this coarser model, which in turn leads to notably lower \FLIP; this observation is supported by Neff's experimental methodology, in which they refine their input meshes before packing them, increasing triangle count in the process. Conversely, FastAtlas produces almost the same \FLIP errors on the unsimplified and simplified model, showcasing robustness.

\begin{table}
\scriptsize
\setlength{\tabcolsep}{2pt}
\centering
\begin{tabular}{l|ccc|c}
  & \textbf{MSA-P} & \textbf{Our Charts w.} & \textbf{MSA-P Charts} & \textbf{Ours} \\
  &  & \textbf{MSA Packing} & \textbf{w. Our Packing} & \\
\hline
\textbf{1920 $\times$ 1080} & & & & \\
 8K avg. & 0.061 & 0.061 & 0.038 & \textbf{0.036} \\
 4K avg. & 0.070 & 0.065 & 0.044 & \textbf{0.039} \\
 2K avg. & 0.159 & 0.126 & 0.082 & \textbf{0.064} \\
\hline
\textbf{3840 $\times$ 2160} & & & & \\
 8K avg. & 0.061 & 0.057 & 0.037 & \textbf{0.033} \\
 4K avg. & 0.126 & 0.106 & 0.072 & \textbf{0.055} \\
 2K avg. & 0.238 & 0.212 & 0.123 & \textbf{0.102} \\
\end{tabular}

\vspace{1mm}
\caption{Visual quality of FastAtlas versus MSA \cite{Neff2022MSA}, MSA charts with our packing, and our charts with MSA packing.
Using our packing decreases the error compared to MSA packing, and using our full method further reduces it.}
\label{tab:ablation}
\end{table}

\begin{table}
\scriptsize
\setlength{\tabcolsep}{2pt}
\centering
\begin{tabular}{l|ccc|c|ccc|c}
  & \multicolumn{4}{c|}{Mean $L^2$} & \multicolumn{4}{c}{Mean $L^\infty$} \\
  & \textbf{MSA-P} & \textbf{Our} & \textbf{MSA-P} & \textbf{Ours} & \textbf{MSA-P} & \textbf{Our} & \textbf{MSA-P} & \textbf{Ours} \\
  &  & \textbf{Charts} & \textbf{Charts} & & & \textbf{Charts} & \textbf{Charts} & \\
  &  & \textbf{w. MSA} & \textbf{w. Our} & & & \textbf{w. MSA} & \textbf{w. Our} & \\
  &  & \textbf{Packing} & \textbf{Packing} & & & \textbf{Packing} & \textbf{Packing} & \\
\hline
\textbf{1920 $\times$ 1080} & & & & & & & & \\
8K avg. & 1.54  & 1.55  & \textbf{1.01} & \textbf{1.01} & 3.66   & 3.54  & \textbf{1.01} & \textbf{1.01} \\
4K avg. & 1.96  & 1.69  & 1.11          & \textbf{1.04} & 3.82   & 3.55  & 1.11          & \textbf{1.04} \\
2K avg. & 16.77 & 8.41  & 2.05          & \textbf{1.59} & 37.61  & 17.89 & 2.05          & \textbf{1.59} \\
\hline
\textbf{3840 $\times$ 2160} & & & & & & & & \\
8K avg. & 1.54  & 1.55  & \textbf{1.01} & \textbf{1.01} & 3.66   & 3.54  & \textbf{1.01} & \textbf{1.01} \\
4K avg. & 1.96  & 1.69  & 1.11          & \textbf{1.04} & 3.82   & 3.55  & 1.11          & \textbf{1.04} \\
2K avg. & 16.77 & 8.41  & 2.05          & \textbf{1.59} & 37.61  & 17.89 & 2.05          & \textbf{1.59} \\
\end{tabular}

\vspace{1mm}
\caption{Texture stretch measurements \cite{sander2001texture} for FastAtlas renders versus two ablations. For each atlas size, we report the mean $L^2$ and $L^\infty$ stretch across all scenes, at both standard and high screen resolutions.}
\label{tab:ablation_stretch}
\vspace{-5mm}
\end{table}

\paragraph*{Packing vs Chartifiction Ablation.} To quantify what percentage of our improvement over prior art is due to seam avoidance versus better atlas packing, we perform two experiments. In the first experiment, we pack the meshlet charts of Neff et al. ~\cite{Neff2022MSA} using our packing method; in the second, we pack FastAtlas charts using Neff et al.'s method. We compare the results of these experiments to Neff et al. and standard FastAtlas in Tab.~\ref{tab:ablation}. Using our charts and their packing, the visual quality is notably worse than ours at all resolutions. At higher resolutions the error in this setting is similar to Neff et al., but at low resolutions this ablation notably outperforms their method. When packing meshlets using FastAtlas with 8K atlases the visual quality is nearly-identical to ours; however, as our fixed atlas size reduces to $2K$, we reduce perceptual error by 20.5\% compared to FastAtlas-packed Meshlet charts. These experiments confirm our hypotheses that packing is a key component of our method, that correct handling of seams is critical for good texture-space shading performance, and that the importance of seams increases as chart resolution decreases.

\begin{figure}
\includegraphics[width=\linewidth]{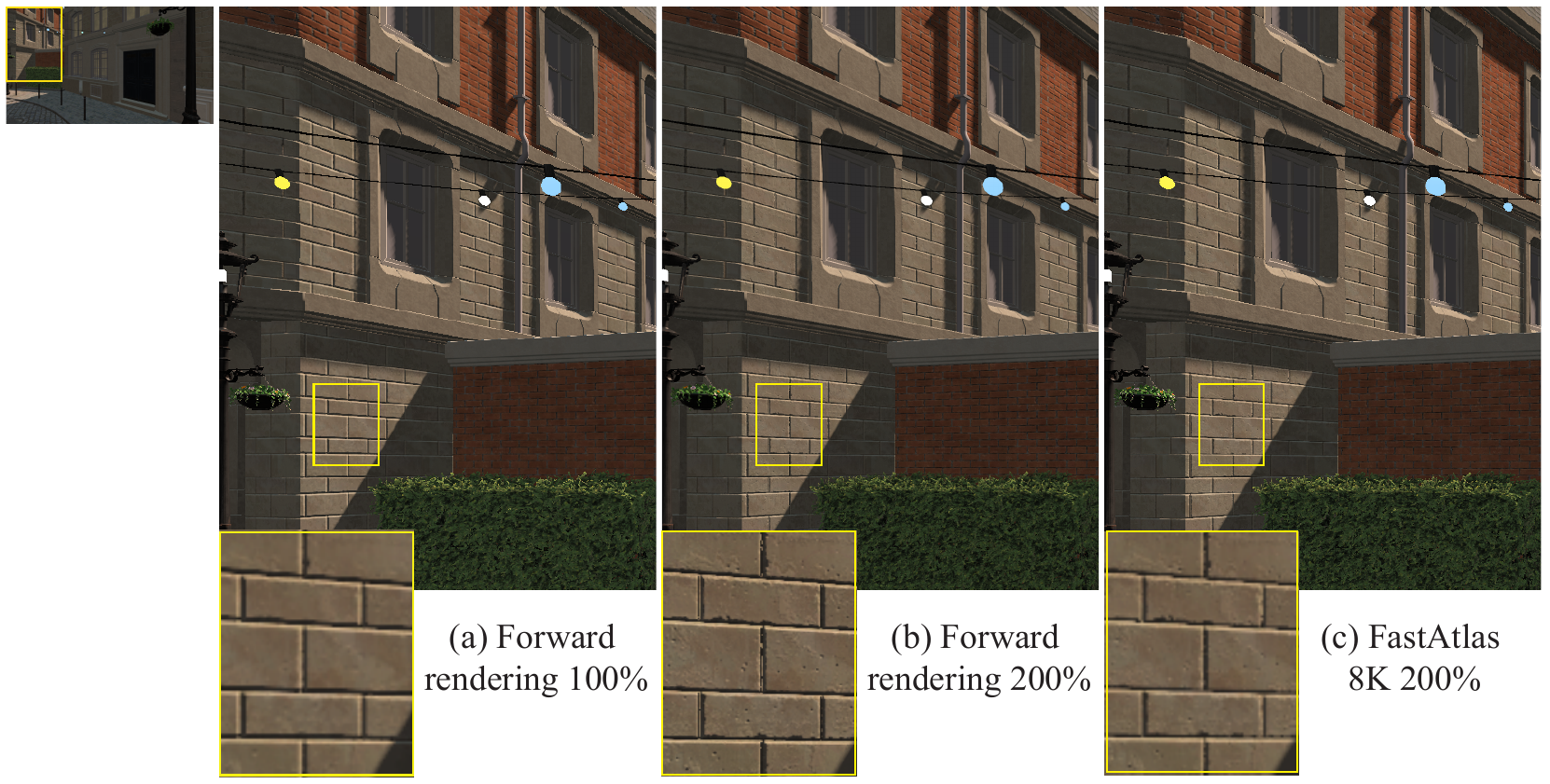}
\caption{Overshading using FastAtlas: (a) Forward render, (b) Forward rendering to a 3840 $\times$ 2160 display and then downsampled, (c) Same frame rendered using an $8K\times 8K$ FastAtlas with 200\% shading rate.}
\label{fig:overshading}
\end{figure}

\paragraph*{Oversampling/Overshading. }
FastAtlas supports oversampling, where multiple texels correspond to a single pixel. It had been suggested \cite{Baker2022} that oversampling can improve visual quality beyond that of standard forward rendering.  Fig~\ref{fig:overshading} shows an example where we shaded the atlas at 200\% shading rate (4 texels per pixel) by scaling all charts up by factor 2 in each direction. Notably in our experiments we can still successfully pack such charts in an $8K \times 8K$ atlas with only minimal downscaling for all frames of the Bistro Exterior scene; our average chosen scale factor out of 64 is 99.5\% of the target.

\begin{figure*}
\includegraphics[width=\linewidth]{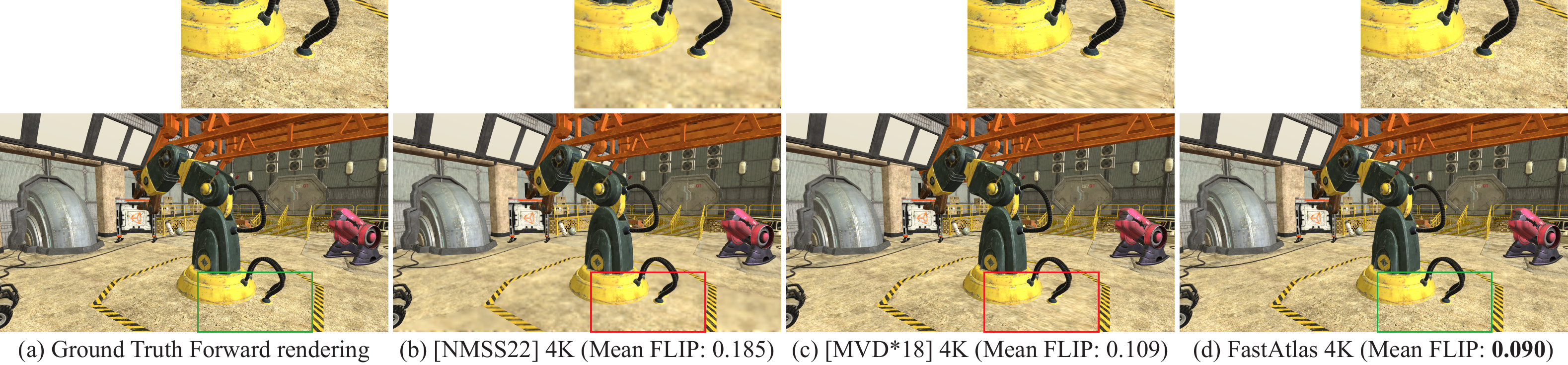}
\caption{On 4K displays (3840 $\times$ 2160), MSA \cite{Neff2022MSA} cannot faithfully preserve even large-scale details such as the floor (b); SAS \cite{mueller2018shading} exhibits severe texture stretch artifacts (c). FastAtlas successfully handles this challenging case (d). Please zoom in for 4K pictures.}
\label{fig:high_res}
\end{figure*}

\paragraph*{High-resolution Displays.}  Fig~\ref{fig:high_res} compares FastAtlas, SAS \cite{mueller2018shading}, and MSA \cite{Neff2022MSA} outputs for high resolution displays.

\begin{table}
\scriptsize
\setlength{\tabcolsep}{2pt}
\centering
\begin{tabular}{lcccc|cccc}
\textbf{Atlas Size} & \multicolumn{4}{c|}{\textbf{4K}} & \multicolumn{4}{c}{\textbf{2K}} \\
\textbf{Scene}     & \textbf{CRF 5}       & \textbf{CRF 10}      & \textbf{CRF 25}      & \textbf{CRF 35}      & \textbf{CRF 5}       & \textbf{CRF 10}      & \textbf{CRF 25}      & \textbf{CRF 35}      \\
\hline
Breakfast Room     & 0.10\%               & 0.10\%               & 0.06\%               & 0.03\%               & 1.00\%               & 0.92\%               & 0.20\%               & 0.08\%               \\
Sibenik            & 0.47\%               & 0.47\%               & 0.34\%               & 0.12\%               & 2.03\%               & 2.03\%               & 0.77\%               & 0.26\%               \\
Bistro Interior    & 0.51\%               & 0.51\%               & 0.47\%               & 0.19\%               & 2.04\%               & 2.04\%               & 0.90\%               & 0.30\%               \\
Bistro Exterior    & 0.54\%               & 0.54\%               & 0.52\%               & 0.34\%               & 2.21\%               & 2.21\%               & 1.68\%               & 0.59\%               \\
San Miguel         & 0.55\%               & 0.55\%               & 0.55\%               & 0.47\%               & 2.20\%               & 2.20\%               & 2.14\%               & 0.97\%               \\
Robot Lab          & 0.52\%               & 0.52\%               & 0.52\%               & 0.27\%               & 2.13\%               & 2.13\%               & 1.23\%               & 0.42\%               \\
Sponza             & 0.51\%               & 0.51\%               & 0.51\%               & 0.18\%               & 2.08\%               & 2.08\%               & 0.90\%               & 0.27\%               \\
Greek Villa        & 0.49\%               & 0.49\%               & 0.40\%               & 0.14\%               & 2.07\%               & 2.07\%               & 0.80\%               & 0.27\%               \\
Modern House       & 0.52\%               & 0.52\%               & 0.52\%               & 0.23\%               & 2.12\%               & 2.12\%               & 1.05\%               & 0.39\%               \\
Bistro Interior FP & 0.49\%               & 0.50\%               & 0.41\%               & 0.15\%               & 2.02\%               & 2.02\%               & 0.77\%               & 0.26\%               \\
Bistro Exterior FP & 0.51\%               & 0.51\%               & 0.51\%               & 0.33\%               & 2.04\%               & 2.04\%               & 1.52\%               & 0.54\%               \\
Robot Lab FP       & 0.51\%               & 0.51\%               & 0.51\%               & 0.31\%               & 2.04\%               & 2.04\%               & 1.33\%               & 0.47\%               \\
Modern House FP    & 0.50\%               & 0.50\%               & 0.50\%               & 0.24\%               & 2.03\%               & 2.03\%               & 1.08\%               & 0.41\%               \\
\hline
\textit{avg.}      & \textit{0.48\%}      & \textit{0.48\%}      & \textit{0.45\%}      & \textit{0.23\%}      & \textit{2.00\%}      & \textit{1.99\%}      & \textit{1.11\%}      & \textit{0.40\%}      \\
\end{tabular}
\vspace{1mm}
\caption{Per-scene breakdown of compression ratio achieved by our atlases at four different constant rate factor (CRF) values for 4K and 2K atlas sizes. We encode our atlases as a video stream using the nvenc H.264 encoder and measure compression ratio on the outputs. Note that atlas sizes of 8K by 8K were not supported by the encoder. Results at 1920 $\times$ 1080 screen resolution.}
\label{tab:supp_compression}
\vspace{-3mm}
\end{table}

\paragraph*{Compression Ratios.} A key measure of spatiotemporal stability for streaming is measured using compression ratio under a fixed constant rate factor (CRF); this measures temporal coherency between atlas frames, which in turn determines how efficiently an atlas can be streamed over a network. Across our scenes, our average compression ratios are 0.48\% (CRF=5), 0.48\% (CRF=10), 0.45\% (CRF=25), and 0.23\% (CRF=35) with 4K atlases, and 2.00\% (CRF=5), 1.99\% (CRF=10), 1.11\% (CRF=25), and 0.40\% (CRF=35) with 2K atlases (Table \ref{tab:supp_compression}). We do not compare to our implementation of SAS or MSA, as we do not implement their temporal reuse scheme, which would result in an unfair-to-them comparison. However, we note that Hladky et al. ~\shortcite{hladky2021snakebinning}, who had access to the source code for \cite{mueller2018shading}, report CRF numbers for SAS. Our compression ratios are consistently better than their reported numbers, and at low CRF is an order of magnitude better than their reported results (SAS: CRF=5: 14.66\%; CRF=10: 12.12\%; CRF=25: 4.80\%; CRF=35: 2.71\%).

\paragraph*{Effective Shading Rate.} We further evaluate the utility of FastAtlas for spatial decoupling by measuring the effective shading rate of our method. At 1920x1080 with fixed-size atlases, our mean effective shading rates are 104\%, 98\%, and 50\% for 8K, 4K, and 2K atlases respectively. For a 4K framebuffer, our mean effective shading rates are 0.96 (8K), 0.49 (4K), and 0.13 (2K). Using fixed reductions in chart scale, rates at 1920x1080 are 29.4\% (50\% scale factor; 25\% target shading rate); 9.1\% (25\% SF; 6.25\% target)/3.5\% (12.5\% SF; 1.5\% target); at 4K, 26.9\% (50\% SF; 25\% target), 7.6\% (25\% SF; 6.25\% target); 2.5\% (12.5\% SF; 1.5\% target). We attribute our slight overshading versus ideal effective shading rates as due to adding borders for bilinear sampling. Our method uniformly distributes samples (and error) across the screen due to uniform scaling and perspective projection, as confirmed by our texture stretch metric experiments. Table ~\ref{tab:supp_shading_rate} provides a breakdown of effective shading rate across all scenes and test conditions.

\begin{figure}
    \includegraphics[width=\linewidth]{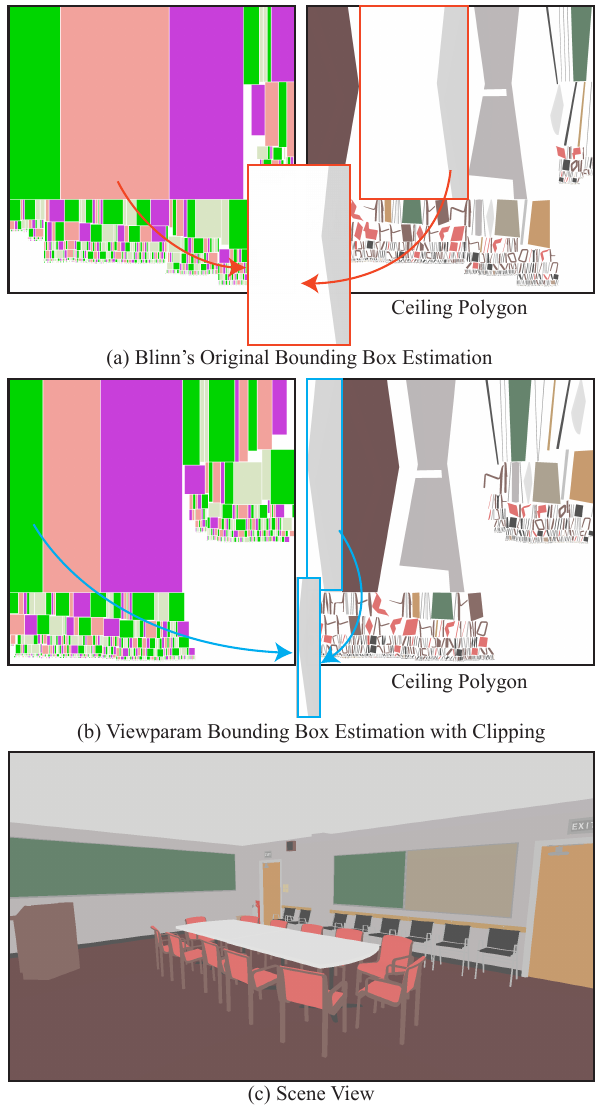}
    \caption{Comparison of our near-plane clipping strategy with Blinn's original bounds estimator. Note how Blinn's estimator can produce badly oriented charts with imprecise bounds (top left).
    The addition of our near plane clipping provides more precise bounds, which allows our chart packing step to pick a larger scale factor, thus decreasing shading error.} 
    \label{fig:bbox_clip_alt}
\end{figure}

\paragraph{Bounding Box Computation Alternatives.}
\label{app:bbox_near_clip}
Fig. ~\ref{fig:bbox_clip_alt} shows the impact of clipping the near clip plane versus Blinn's original formulation for computing chart bounding boxes. As shown in this example, when large triangles intersect the near clip plane, the original method produces large bounding boxes that overestimate the amount of space needed. By selectively clipping line segments that intersect the near clip plane if needed and then applying Blinn's formulation, we produce tighter boxes that occupy less space in the chart. 

\paragraph*{Runtimes on Low-End Hardware.}
While the G3D engine cannot load some of our more complex test scenes on older GPUs due to insufficient video memory, we can measure performance on the remaining scenes using an older graphics card (a 6-year old NVIDIA GeForce RTX2080). Reducing the number of scale factors, we achieve good atlasing runtime performance on this older graphics card. At 16 scales, our average total atlasing times are 1.76ms (8K), 1.49ms (4K), and 1.37ms (2K); at 32 scales, 2.17ms (8K), 1.89ms (4K), and 1.77ms (2K). We note that the number of scale factors is the primary bottleneck of our method, and can be adjusted to increase performance if desired.

\begin{table*}
\scriptsize
\setlength{\tabcolsep}{2pt}
\centering
\tiny
\begin{tabular}{lccccc|ccccc}
& \multicolumn{5}{c|}{\textbf{16 scales}} & \multicolumn{5}{c}{\textbf{32 scales}} \\
& Chart              & AABB           & AABB           & Atlas          & Total          & Chart              & AABB           & AABB           & Atlas          & Total          \\
& Extract.           & Comp.          & Pre-Order      & Packing        & Atlasing       & Extract.           & Comp.          & Pre-Order      & Packing        & Atlasing       \\
\hline
\textbf{8K}  & & & & & & & & & & \\
Breakfast Room                & 0.169              & 0.030          & 0.950          & 0.328          & 1.477          & 0.170              & 0.030          & 1.057          & 0.624          & 1.881          \\
Sibenik                       & 0.079              & 0.026          & 0.953          & 0.338          & 1.396          & 0.079              & 0.027          & 1.073          & 0.635          & 1.814          \\
Bistro Interior               & 0.459              & 0.060          & 0.952          & 0.411          & 1.881          & 0.460              & 0.060          & 1.057          & 0.711          & 2.288          \\
Robot Lab                     & 0.302              & 0.049          & 0.952          & 0.394          & 1.697          & 0.304              & 0.049          & 1.062          & 0.690          & 2.105          \\
Sponza                        & 0.181              & 0.035          & 0.951          & 0.356          & 1.523          & 0.182              & 0.036          & 1.074          & 0.659          & 1.951          \\
Greek Villa                   & 0.383              & 0.044          & 0.949          & 0.363          & 1.738          & 0.380              & 0.044          & 1.068          & 0.660          & 2.152          \\
Modern House                  & 0.695              & 0.081          & 0.954          & 0.412          & 2.143          & 0.698              & 0.081          & 1.071          & 0.711          & 2.561          \\
Bistro Interior - Flight Path & 0.446              & 0.053          & 0.957          & 0.399          & 1.855          & 0.446              & 0.053          & 1.062          & 0.696          & 2.258          \\
Robot Lab - Flight Path       & 0.315              & 0.057          & 0.955          & 0.417          & 1.743          & 0.315              & 0.057          & 1.064          & 0.718          & 2.155          \\
Modern House - Flight Path    & 0.699              & 0.078          & 0.968          & 0.428          & 2.173          & 0.695              & 0.079          & 1.069          & 0.728          & 2.571          \\
\hline
\textit{avg.}                 & \textit{0.373}     & \textit{0.051} & \textit{0.954} & \textit{0.385} & \textit{1.763} & \textit{0.373}     & \textit{0.052} & \textit{1.066} & \textit{0.683} & \textit{2.174} \\
\hline
\textbf{4K}  & & & & & & & & & & \\
Breakfast Room                & 0.168              & 0.029          & 0.679          & 0.325          & 1.201          & 0.169              & 0.030          & 0.791          & 0.623          & 1.612          \\
Sibenik                       & 0.078              & 0.026          & 0.679          & 0.335          & 1.118          & 0.079              & 0.026          & 0.780          & 0.629          & 1.514          \\
Bistro Interior               & 0.459              & 0.060          & 0.677          & 0.408          & 1.604          & 0.460              & 0.060          & 0.783          & 0.707          & 2.010          \\
Robot Lab                     & 0.305              & 0.048          & 0.675          & 0.390          & 1.419          & 0.303              & 0.049          & 0.782          & 0.690          & 1.824          \\
Sponza                        & 0.182              & 0.035          & 0.676          & 0.353          & 1.246          & 0.182              & 0.035          & 0.799          & 0.656          & 1.673          \\
Greek Villa                   & 0.381              & 0.044          & 0.675          & 0.360          & 1.460          & 0.382              & 0.044          & 0.792          & 0.658          & 1.877          \\
Modern House                  & 0.695              & 0.081          & 0.685          & 0.409          & 1.870          & 0.696              & 0.082          & 0.789          & 0.709          & 2.275          \\
Bistro Interior - Flight Path & 0.453              & 0.053          & 0.688          & 0.400          & 1.594          & 0.446              & 0.053          & 0.786          & 0.697          & 1.982          \\
Robot Lab - Flight Path       & 0.315              & 0.057          & 0.679          & 0.414          & 1.465          & 0.317              & 0.057          & 0.787          & 0.715          & 1.877          \\
Modern House - Flight Path    & 0.699              & 0.078          & 0.692          & 0.424          & 1.893          & 0.696              & 0.079          & 0.790          & 0.725          & 2.289          \\
\hline
\textit{avg.}                 & \textit{0.374}     & \textit{0.051} & \textit{0.681} & \textit{0.382} & \textit{1.487} & \textit{0.373}     & \textit{0.052} & \textit{0.788} & \textit{0.681} & \textit{1.893} \\
\hline
\textbf{2K}  & & & & & & & & & & \\
Breakfast Room                & 0.169              & 0.029          & 0.569          & 0.322          & 1.089          & 0.169              & 0.030          & 0.675          & 0.621          & 1.494          \\
Sibenik                       & 0.078              & 0.026          & 0.568          & 0.333          & 1.006          & 0.079              & 0.026          & 0.679          & 0.634          & 1.418          \\
Bistro Interior               & 0.458              & 0.060          & 0.575          & 0.394          & 1.486          & 0.460              & 0.060          & 0.677          & 0.696          & 1.892          \\
Robot Lab                     & 0.302              & 0.048          & 0.571          & 0.374          & 1.295          & 0.302              & 0.049          & 0.674          & 0.677          & 1.702          \\
Sponza                        & 0.180              & 0.035          & 0.569          & 0.345          & 1.129          & 0.181              & 0.035          & 0.671          & 0.649          & 1.536          \\
Greek Villa                   & 0.381              & 0.044          & 0.576          & 0.358          & 1.360          & 0.380              & 0.044          & 0.684          & 0.655          & 1.762          \\
Modern House                  & 0.696              & 0.081          & 0.574          & 0.402          & 1.753          & 0.695              & 0.081          & 0.678          & 0.700          & 2.155          \\
Bistro Interior - Flight Path & 0.446              & 0.052          & 0.574          & 0.385          & 1.457          & 0.446              & 0.053          & 0.676          & 0.685          & 1.860          \\
Robot Lab - Flight Path       & 0.315              & 0.057          & 0.574          & 0.394          & 1.340          & 0.317              & 0.057          & 0.678          & 0.698          & 1.750          \\
Modern House - Flight Path    & 0.697              & 0.078          & 0.576          & 0.412          & 1.763          & 0.696              & 0.078          & 0.678          & 0.713          & 2.164          \\
\hline
\textit{avg.}                 & \textit{0.372}     & \textit{0.051} & \textit{0.573} & \textit{0.372} & \textit{1.368} & \textit{0.373}     & \textit{0.051} & \textit{0.677} & \textit{0.673} & \textit{1.773} \\
\end{tabular}
\vspace{1mm}
\caption{Per-scene FastAtlas atlasing times (milliseconds) on an NVIDIA GeForce RTX2080 when using fixed-size atlases. We report times for both 16 and 32 parallel scale factors. Note that two of our scenes (\textit{Bistro Exterior} and \textit{San Miguel}) could not be loaded, due to insufficient GPU video memory. Column labels from left to right: chart extraction, AABB computation, AABB pre-ordering, atlas packing, total atlasing (sum of previous four columns). Results at 1920 $\times$ 1080 screen resolution.}
\label{tab:supp_runtime_2080}
\vspace{-3mm}
\end{table*}

\section{Applications.}

\begin{figure*}
\includegraphics[width=\linewidth]{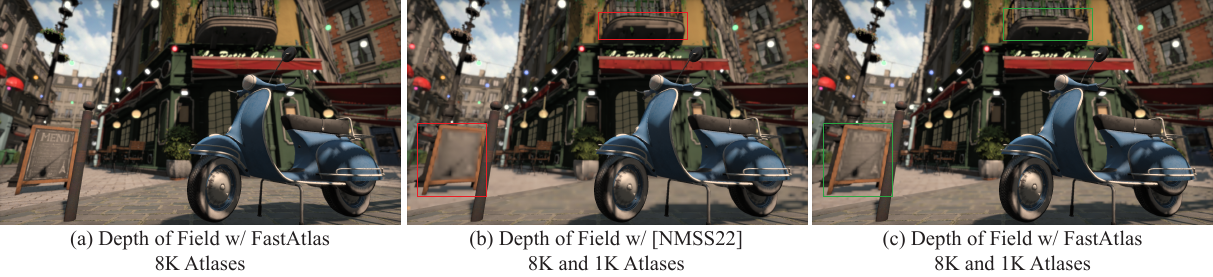}
\caption{Depth of field render using 100\% shading rate (a) and using a combination of 100\% and 12.5\% shading rate (b,c). In (b,c) triangles falling partly inside the focal range are shaded into a 100\% texture atlas; triangles falling outside of the focal range are shaded into a second atlas at 12.5\% shading rate. Using \cite{Neff2022MSA} with the same settings (b) introduces artifacts such as the seams on the awning and the coarse shading on the balcony and menu due to undersampling (highlighted in red). Using FastAtlas (c), the rendered outputs are nearly identical.} 
\label{fig:depthOfField}
\end{figure*}

\begin{figure*}
\includegraphics[width=\linewidth]{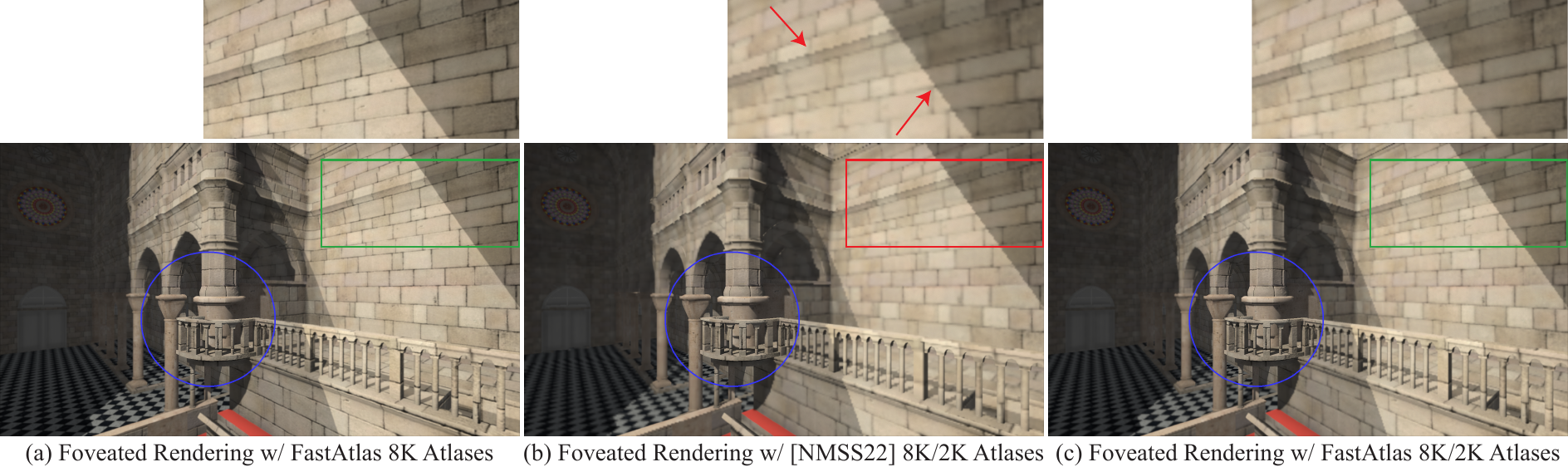}
\caption{Foveated rendering using a 100\% texture atlas (a) and using a combination of 100\% and 25\% shading rate atlases (b,c). In (b,c) triangles falling inside the foveated region are rendered into a high resolution texture atlas; triangles falling outside of the shaded region are rendered into a low-resolution texture atlas and then blurred. (b) Using \cite{Neff2022MSA} with the same setting, the area outside the foveated region has undersampling artifacts on areas such as the wall boundary and shadows (inset). Using FastAtlas (c), we are able to obtain high-quality foveated rendering suitable for VR and AR displays with eye tracking.}
\label{fig:foveated_rendering}
\end{figure*}

\label{sec:applications}
A key advantage of texture-space shading is that shading rate can be decoupled from screen space resolution. This decoupling means that one can generate multiple shading atlases for the same scene and use them to shade different content at different shading resolutions. In particular, one can shade a portion of a scene using a high-resolution atlas, and another portion using a low resolution one. In these scenarios, only a portion of the high-resolution atlas is shaded, significantly reducing computational costs.
Applications that can benefit from this technique are ones where only a portion of a scene requires high quality shading, whereas the rest of the 
scene can be rendered with low resolution or blurry shading. We evaluate the applicability of our method to such scenarios by applying it to two representative examples: depth of field and foveated rendering. 
 
\paragraph*{Depth-of-Field.} 
We modify the depth-of-field method of Bukowski et al. \shortcite{Bukowski2013DepthOfField} to utilize two texture atlases. For the first atlas, we set the target scale to 100\% and use an $8K \times 8K$ atlas that provided enough space to fit all charts at this scale. Our second atlas is allocated to be  $1K \times 1K$ (1/64th the number of original texels). We generate a single packing by computing charts and atlases for the entire visible scene for the $8k$ atlas. 
The packing for the low resolution atlas is generated by simply scaling the coordinates of all charts by factor 0.125 along both axes. When packing the 8K atlas we allocate a gutter size sufficient to ensure that charts continue to not touch if the $8k$ packing is scaled down to fit the $1k$ atlas. We then use this packing to shade both atlases. We shade the entire low-resolution atlas, and only shade a triangle in the $8k$ atlas if it is contained or partly contained within the focal field. We then read from the high-resolution atlas when rasterizing areas in the focal field or its transition region, and use the low resolution one otherwise. The net result is that any shading values outside of the focal field are shaded at a reduced one sixty-fourth shading rate. As  Fig.~\ref{fig:depthOfField}c demonstrates, using our atlases achieves compelling depth-of-field effects while requiring one sixty-fourth the amount of shading for textures outside of the focal field.
We used the exact same setup for \cite{Neff2022MSA} using 8K and 1K atlases. See accompanying video and Fig.~\ref{fig:depthOfField} for a side by side comparison.  

\paragraph*{Foveated Rendering.} We use the same technique for foveated rendering. We generate an $8K$ atlas packing, and then scale it down to create an identical $2K$ atlas packing. We then shade a low-resolution $2K$ atlas for the entire scene, and a matching high-resolution $8K$ atlas with the same packing only for those triangles contained inside the fovea region. The low resolution view is then blurred using a two-pass Gaussian blur and composited with a high-resolution rendering of the foveal region to produce the final image (Fig. \ref{fig:foveated_rendering}c). By providing a fast, robust texture-space shading method, our seamless texture atlases make foveated rendering practical for modern virtual reality systems with eye tracking. We used the exact same setup for \cite{Neff2022MSA} (Fig.~\ref{fig:foveated_rendering}b).

The quality of renders generated using a mix of high and low resolution atlases is clearly contingent on the quality of the shading generated using the low-resolution atlases. Patney et al.~\shortcite{patney2016towards} note that static TSS methods do not provide sufficient quality low resolution results to enable this approach for foveated rendering.
This observation is aligned with our measurements (Tab.~\ref{tab:supp_flip_fixed_atlas}, Tab.~\ref{tab:supp_flip_fixed_sr}) and figures which show that these methods fail to produce adequate quality shading at low resolutions. Shading using the method of \cite{Neff2022MSA} at these resolutions produces undesirable undersampling and highly noticeable visible seams at meshlet boundaries (Figs. \ref{fig:depthOfField}b and \ref{fig:foveated_rendering}b.)  FastAtlas generates atlases of sufficient quality at low resolutions for the needs of these applications.

\section{Additional Statistics.}

The tables below provide detailed scene level statistics for the experiments we ran. We include tables containing a detailed breakdown of \FLIP error metrics across individual scenes for every scene and path combination we tested, both for fixed atlas sizes and for fixed shading rate reductions. We also include similar tables for $L^2$ and $L^{\infty}$ texture stretch metrics. For our packing experiment ablation, where we pack our charts into the superblock packing scheme of \cite{Neff2022MSA}, and pack their charts with our packing scheme, we include a full breakdown across all scenes tested. Finally, we report the number of reused fragments and \FLIP error across all scenes for our temporal reuse comparison to reverse reprojection caching \cite{nehab2007accelerating}, and a breakdown of per-scene effective shading rates for our method across different fixed atlas sizes.

\begin{table}[ht]
\scriptsize
\setlength{\tabcolsep}{2pt}
\setlength{\intextsep}{0pt}
\centering
\begin{tabular}{lccc}
\textbf{Scene}                & \textbf{Static 16K} & \textbf{Static 32K} & \textbf{Ours 8K}        \\
\hline
Breakfast Room                & \textbf{0.001}      & \textbf{0.001}      & 0.002                   \\
Sibenik                       & 0.024               & 0.019               & \textbf{0.018}          \\
Bistro Interior               & 0.029               & 0.035               & \textbf{0.025}          \\
Bistro Exterior               & 0.102               & 0.064               & \textbf{0.051}          \\
San Miguel                    & *                   & *                   & \textbf{0.041}          \\
Robot Lab                     & 0.077               & 0.091               & \textbf{0.057}          \\
Sponza                        & 0.150               & 0.104               & \textbf{0.069}          \\
Greek Villa                   & 0.058               & 0.052               & \textbf{0.038}          \\
Modern House                  & 0.033               & 0.044               & \textbf{0.023}          \\
Bistro Interior - Flight Path & 0.034               & 0.032               & \textbf{0.025}          \\
Bistro Exterior - Flight Path & 0.122               & 0.081               & \textbf{0.050}          \\
Robot Lab - Flight Path       & 0.077               & 0.097               & \textbf{0.056}          \\
Modern House - Flight Path    & 0.037               & 0.036               & \textbf{0.021}          \\
\hline
\textit{avg.}                 & \textit{0.062}      & \textit{0.055}      & \textit{0.036}   \\
\textit{improvement vs.}      & \textit{31\%}       & \textit{25\%}       &                 
\end{tabular} 

\vspace{1mm}
\caption{Per-scene comparison of our visual quality using an 8K atlas against static 16K and 32K atlases (mean \FLIP error vs. forward rendering). Results at 1920 $\times$ 1080 screen resolution. A value of * means that atlas packing failed.}
\label{tab:supp_flip_infinite_atlas}
\vspace{-3mm}
\end{table}

\begin{table*}[t]
\scriptsize
\setlength{\tabcolsep}{3pt}
\centering
\begin{tabular}{l cc | ccc | ccc | ccc}
\textbf{Shading Rate}         & \multicolumn{2}{c|}{\textbf{100\%}} & \multicolumn{3}{c|}{\textbf{50\%}}  & \multicolumn{3}{c|}{\textbf{25\%}}  & \multicolumn{3}{c}{\textbf{12.5\%}}  \\
\textbf{Scene}                & \textbf{MSA-P} & \textbf{Ours}           & \textbf{Baseline} & \textbf{MSA-P} & \textbf{Ours}           & \textbf{Baseline} & \textbf{MSA-P} & \textbf{Ours}           & \textbf{Baseline} & \textbf{MSA-P} & \textbf{Ours}           \\
\hline
\textbf{1920 $\times$ 1080}   & & & & & & & & & & & \\
Breakfast Room                & \textbf{0.002} & \textbf{0.002}          & 0.017             & 0.009          & \textbf{0.006}          & 0.028             & 0.018          & \textbf{0.011}          & 0.046             & 0.031          & \textbf{0.020}          \\
Sibenik                       & 0.020          & \textbf{0.015}          & 0.072             & 0.083          & \textbf{0.054}          & 0.120             & 0.132          & \textbf{0.100}          & 0.169             & 0.173          & \textbf{0.142}          \\
Bistro Interior               & 0.023          & \textbf{0.022}          & 0.076             & 0.078          & \textbf{0.050}          & 0.119             & 0.120          & \textbf{0.084}          & 0.172             & 0.160          & \textbf{0.122}          \\
Bistro Exterior               & 0.051          & \textbf{0.045}          & 0.151             & 0.153          & \textbf{0.099}          & 0.227             & 0.225          & \textbf{0.161}          & 0.311             & 0.289          & \textbf{0.225}          \\
San Miguel                    & 0.038          & \textbf{0.032}          & 0.119             & 0.133          & \textbf{0.089}          & 0.190             & 0.195          & \textbf{0.148}          & 0.263             & 0.250          & \textbf{0.206}          \\
\end{tabular}
\vspace{1mm}
\caption{Per-scene comparison of visual quality (mean \FLIP error vs. forward rendering) for FastAtlas (Ours) against \cite{Neff2022MSA} (MSA-P) and a forward upsampling baseline (Baseline) when targeting fixed shading rates. Note that the forward upsampling baseline at 100\% shading rate is equivalent to forward rendering, and is thus not shown. A value of * means that atlas packing failed.}
\label{tab:supp_flip_fixed_sr}
\vspace{-3mm}
\end{table*}

\begin{table*}[t]
	\scriptsize
	\setlength{\tabcolsep}{3pt}
	\centering
	\begin{tabular}{l cc|cc|cc|cc}
\textbf{Shading   Rate}       & \multicolumn{2}{c|}{\textbf{100\%}}                 & \multicolumn{2}{c|}{\textbf{50\%}}                  & \multicolumn{2}{c|}{\textbf{25\%}}                  & \multicolumn{2}{c}{\textbf{12.50\%}}               \\
& \textbf{MSA-P}       & \textbf{Ours}               & \textbf{MSA-P}       & \textbf{Ours}               & \textbf{MSA-P}       & \textbf{Ours}               & \textbf{MSA-P}       & \textbf{Ours}               \\
\textbf{Scene}                & \textbf{Mean $L^2$/}   & \textbf{Mean $L^2$/}          & \textbf{Mean $L^2$/}   & \textbf{Mean $L^2$/}          & \textbf{Mean $L^2$/}   & \textbf{Mean $L^2$/}          & \textbf{Mean $L^2$/}   & \textbf{Mean $L^2$/}          \\
& \textbf{Mean $L^\infty$} & \textbf{Mean $L^\infty$}        & \textbf{Mean $L^\infty$} & \textbf{Mean $L^\infty$}        & \textbf{Mean $L^\infty$} & \textbf{Mean $L^\infty$}        & \textbf{Mean $L^\infty$} & \textbf{Mean $L^\infty$}        \\
\hline
\textbf{1920x1080}            &                      &                             &                      &                             &                      &                             &                      &                             \\
Breakfast Room                & 1.0 / 1.0            & \textbf{1.0 / 1.0}          & 3.2 / 4.2            & \textbf{2.0 / 2.0}          & 6.4 / 8.0            & \textbf{4.0 / 4.0}          & 13.2 / 16.7          & \textbf{8.0 / 8.0}          \\
            		Sibenik                       & 1.0 / 1.0            & \textbf{1.0 / 1.0}          & 3.1 / 6.9            & \textbf{2.0 / 2.0}          & 6.4 / 14.2           & \textbf{4.0 / 4.0}          & 13.7 / 27.5          & \textbf{8.0 / 8.0}          \\
\end{tabular}
	\vspace{1mm}
	\caption{Per-scene comparison of texture stretch measurements for FastAtlas (Ours) against \cite{Neff2022MSA} (MSA-P), \cite{mueller2018shading} (SAS), and static atlases (Static) when targeting fixed shading rates. A value of * means that atlas packing failed.}
	\label{tab:supp_texture_stretch_fixed_atlas}
	\vspace{-3mm}
\end{table*}

\begin{table*}[t]
\scriptsize
\setlength{\tabcolsep}{3pt}
\centering
\begin{tabular}{l cccc|cccc|cccc}
\textbf{Atlas Size}           & \multicolumn{4}{c|}{\textbf{8K}} & \multicolumn{4}{c|}{\textbf{4K}}  & \multicolumn{4}{c}{\textbf{2K}}  \\
\textbf{Scene}                & \textbf{Static} & \textbf{SAS}   & \textbf{MSA-P} & \textbf{Ours}           & \textbf{Static} & \textbf{SAS}   & \textbf{MSA-P} & \textbf{Ours}           & \textbf{Static} & \textbf{SAS}   & \textbf{MSA-P} & \textbf{Ours}           \\
\hline
\textbf{1920 $\times$ 1080}   & & & & & & & & & & & & \\
Breakfast Room                & \textbf{0.002}  & 0.009          & 0.004          & \textbf{0.002}          & 0.005           & 0.009          & 0.004          & \textbf{0.002}          & 0.011           & 0.017          & 0.011          & \textbf{0.003}          \\
Sibenik                       & 0.044           & 0.042          & 0.039          & \textbf{0.018}          & 0.085           & 0.042          & 0.039          & \textbf{0.022}          & 0.123           & 0.062          & 0.072          & \textbf{0.027}          \\
Bistro Interior               & 0.035           & 0.038          & 0.043          & \textbf{0.025}          & 0.060           & 0.037          & 0.051          & \textbf{0.026}          & 0.108           & 0.048          & 0.089          & \textbf{0.040}          \\
Bistro Exterior               & 0.168           & 0.063          & 0.087          & \textbf{0.051}          & 0.282           & 0.063          & 0.095          & \textbf{0.052}          & *               & \textbf{0.090} & 0.301          & 0.095                   \\
San Miguel                    & *               & 0.093          & 0.072          & \textbf{0.041}          & *               & 0.092          & 0.097          & \textbf{0.046}          & *               & 0.109          & 0.368          & \textbf{0.098}          \\
\end{tabular}

\vspace{1mm}
\caption{Per-scene comparison of visual quality (mean \FLIP error vs. forward rendering) for FastAtlas (Ours) against \cite{Neff2022MSA} (MSA-P), \cite{mueller2018shading} (SAS), and static atlases (Static) when shading using fixed-size atlases. A value of * means that atlas packing failed.}
\label{tab:supp_flip_fixed_atlas}
\vspace{-3mm}
\end{table*}

\begin{table*}[t]
	\scriptsize
	\setlength{\tabcolsep}{3pt}
	\centering
	\begin{tabular}{l cccc|cccc|cccc}
		\textbf{Atlas   Size} & \multicolumn{4}{c|}{\textbf{8K}}                                                                    & \multicolumn{4}{c|}{\textbf{4K}}                                                                    & \multicolumn{4}{c}{\textbf{2K}}                                                                    \\
		& \textbf{Static}      & \textbf{SAS}         & \textbf{MSA-P}       & \textbf{Ours}        & \textbf{Static}      & \textbf{SAS}         & \textbf{MSA-P}       & \textbf{Ours}        & \textbf{Static}      & \textbf{SAS}         & \textbf{MSA-P}       & \textbf{Ours}        \\
		 \textbf{Scene} & \textbf{Mean $L^2$/}   & \textbf{Mean $L^2$/}   & \textbf{Mean $L^2$/}   & \textbf{Mean $L^2$/}   & \textbf{Mean $L^2$/}   & \textbf{Mean $L^2$/}   & \textbf{Mean $L^2$/}   & \textbf{Mean $L^2$/}   & \textbf{Mean $L^2$/}   & \textbf{Mean $L^2$/}   & \textbf{Mean $L^2$/}   & \textbf{Mean $L^2$/}   \\
		& \textbf{Mean $L^\infty$} & \textbf{Mean $L^\infty$} & \textbf{Mean $L^\infty$} & \textbf{Mean $L^\infty$} & \textbf{Mean $L^\infty$} & \textbf{Mean $L^\infty$} & \textbf{Mean $L^\infty$} & \textbf{Mean $L^\infty$} & \textbf{Mean $L^\infty$} & \textbf{Mean $L^\infty$} & \textbf{Mean $L^\infty$} & \textbf{Mean $L^\infty$} \\
		\hline
		\textbf{1920x1080}            & \textbf{}            & \textbf{}            & \textbf{}            & \textbf{}            & \textbf{}            & \textbf{}            & \textbf{}            &                      &                      &                      &                      &                      \\
		Breakfast Room                & 1.5 / 2.1            & 12.3 / 26.0          & 1.6 / 2.2            & \textbf{1.0 / 1.0}   & 3.0 / 4.3            & 12.3 / 26.0          & 1.6 / 2.2            & \textbf{1.0 / 1.0}   & 5.8 / 8.4            & 24.6 / 45.9          & 6.0 / 6.4            & \textbf{1.0 / 1.0}   \\
\end{tabular}
	\vspace{1mm}
	\caption{Per-scene comparison of texture stretch measurements for FastAtlas (Ours) against \cite{Neff2022MSA} (MSA-P), \cite{mueller2018shading} (SAS), and static atlases (Static) when shading using fixed-size atlases. A value of * means that atlas packing failed.}
	\label{tab:supp_texture_stretch_fixed_atlas}
	\vspace{-3mm}
\end{table*}

\begin{table*}[t]
\scriptsize
\setlength{\tabcolsep}{3pt}
\centering
\begin{tabular}{l cccc|cccc|cccc}
\textbf{Atlas Size}           & \multicolumn{4}{c|}{\textbf{8K}} & \multicolumn{4}{c|}{\textbf{4K}}  & \multicolumn{4}{c}{\textbf{2K}}  \\
\textbf{Scene}                & \textbf{MSA-P} & \textbf{Our Charts w.} & \textbf{MSA-P Charts}   & \textbf{Ours}           & \textbf{MSA-P} & \textbf{Our Charts w.} & \textbf{MSA-P Charts}   & \textbf{Ours}           & \textbf{MSA-P} & \textbf{Our Charts w.} & \textbf{MSA-P Charts}   & \textbf{Ours}           \\
                              &                & \textbf{MSA Packing}   & \textbf{w. Our Packing} & \textbf{}               &                & \textbf{MSA Packing}   & \textbf{w. Our Packing} &                         &                & \textbf{MSA Packing}   & \textbf{w. Our Packing} &                         \\
\hline
\textbf{1920 $\times$ 1080}   & & & & & & & & & & & & \\
Breakfast Room                & 0.004          & 0.005                  & \textbf{0.002}          & \textbf{0.002}          & 0.004          & 0.005                  & 0.003                   & \textbf{0.002}          & 0.011          & 0.007                  & 0.004                   & \textbf{0.003}          \\
Sibenik                       & 0.039          & 0.041                  & 0.023                   & \textbf{0.018}          & 0.039          & 0.041                  & 0.024                   & \textbf{0.022}          & 0.072          & 0.092                  & 0.037                   & \textbf{0.027}          \\
Bistro Interior               & 0.043          & 0.041                  & 0.027                   & \textbf{0.025}          & 0.051          & 0.043                  & 0.028                   & \textbf{0.026}          & 0.089          & 0.073                  & 0.051                   & \textbf{0.040}          \\
\end{tabular}

\vspace{1mm}
\caption{Per-scene comparison of visual quality for our full method (Ours) against \cite{Neff2022MSA} (MSA-P) and two ablations: supplying our charts to the packing method of \cite{Neff2022MSA}, and applying our packing to the MSA-P charts of \cite{Neff2022MSA}. We report mean \FLIP error vs. forward rendering across different atlas sizes and screen resolutions.}
\label{tab:supp_ablation}
\vspace{-3mm}
\end{table*}

\begin{table*}[t]
\scriptsize
\setlength{\tabcolsep}{3pt}
\centering
\begin{tabular}{l cccc|cccc|cccc}
	\textbf{Atlas   Size} & \multicolumn{4}{c|}{\textbf{8K}}                            & \multicolumn{4}{c|}{\textbf{4K}}                                                  & \multicolumn{4}{c}{\textbf{2K}}                                                  \\
	& \textbf{MSA-P} & \textbf{Our Charts} & \textbf{MSA-P Charts}   & \textbf{Ours} & \textbf{MSA-P} & \textbf{Our Charts} & \textbf{MSA-P Charts}   & \textbf{Ours} & \textbf{MSA-P} & \textbf{Our Charts} & \textbf{MSA-P Charts}   & \textbf{Ours}\\
	& \textbf{}      & \textbf{w. MSA}   & \textbf{w. Our} & \textbf{}     & \textbf{}      & \textbf{w. MSA}   & \textbf{w. Our} & \textbf{}     & \textbf{}      & \textbf{w. MSA}   & \textbf{w. Our} & \textbf{}     \\
	& \textbf{}      & \textbf{Packing}   & \textbf{Packing} & \textbf{}     & \textbf{}      & \textbf{Packing}   & \textbf{Packing} & \textbf{}     & \textbf{}      & \textbf{Packing}   & \textbf{Packing} & \textbf{}     \\

	\textbf{Scene} & \textbf{Mean $L^2$/}   & \textbf{Mean $L^2$/}   & \textbf{Mean $L^2$/}   & \textbf{Mean $L^2$/}   & \textbf{Mean $L^2$/}   & \textbf{Mean $L^2$/}   & \textbf{Mean $L^2$/}   & \textbf{Mean $L^2$/}   & \textbf{Mean $L^2$/}   & \textbf{Mean $L^2$/}   & \textbf{Mean $L^2$/}   & \textbf{Mean $L^2$/}   \\
	& \textbf{Mean $L^\infty$} & \textbf{Mean $L^\infty$} & \textbf{Mean $L^\infty$} & \textbf{Mean $L^\infty$} & \textbf{Mean $L^\infty$} & \textbf{Mean $L^\infty$} & \textbf{Mean $L^\infty$} & \textbf{Mean $L^\infty$} & \textbf{Mean $L^\infty$} & \textbf{Mean $L^\infty$} & \textbf{Mean $L^\infty$} & \textbf{Mean $L^\infty$} \\
	\hline
	\textbf{1920x1080}            & \textbf{}            & \textbf{}            & \textbf{}            & \textbf{}            & \textbf{}            & \textbf{}            & \textbf{}            &                      &                      &                      &                      &                      \\
\end{tabular}
\vspace{1mm}
\caption{Per-scene comparison of texture stretch measurements for our full method (Ours) against \cite{Neff2022MSA} (MSA-P) and two ablations: supplying our charts to the packing method of \cite{Neff2022MSA}, and applying our packing to the MSA-P charts of \cite{Neff2022MSA}.}
\label{tab:supp_texture_stretch_ablation}
\vspace{-3mm}
\end{table*}

\begin{table*}
\scriptsize
\setlength{\tabcolsep}{2pt}
\centering
\begin{tabular}{l cccc|cccc|ccc}
 & \multicolumn{4}{c|}{\FLIP $\downarrow$} & \multicolumn{4}{c|}{Reuse $\uparrow$} & \multicolumn{3}{c}{Shading Time (ms) $\downarrow$} \\
 & \multicolumn{2}{c}{\textbf{RRC}} & \multicolumn{2}{c|}{\textbf{Our}} & \multicolumn{2}{c}{\textbf{RRC}} & \multicolumn{2}{c|}{\textbf{Our}} & \textbf{Ours} & \textbf{Our} & \textbf{Improv.} \\
 & & & \multicolumn{2}{c|}{\textbf{Temporal}} & & & \multicolumn{2}{c|}{\textbf{Temporal}} & \textbf{8K} & \textbf{Temporal} & \textbf{} \\
\textbf{Scene}                & Mean           & Max            & Mean                    & Max                     & Min            & Mean           & Min                     & Mean                    &                 &                          &               \\
\hline
Breakfast Room                & 0.005          & 0.019          & \textbf{0.002}          & \textbf{0.009}          & 0.705          & 0.882          & \textbf{0.714}          & \textbf{0.883}          & 14.258          & \textbf{13.551}          & 5\%           \\
Sibenik                       & 0.034          & 0.050          & \textbf{0.021}          & \textbf{0.035}          & 0.793          & 0.885          & \textbf{0.795}          & \textbf{0.888}          & 20.390          & \textbf{13.615}          & 33\%          \\
Bistro Interior               & 0.036          & 0.051          & \textbf{0.026}          & \textbf{0.033}          & 0.768          & 0.879          & \textbf{0.776}          & \textbf{0.883}          & 53.124          & \textbf{35.93}          & 32\%          \\
Bistro Exterior               & 0.072          & 0.104          & \textbf{0.052}          & \textbf{0.076}          & \textbf{0.795} & \textbf{0.883} & 0.778                   & 0.874                   & 100.225         & \textbf{59.971}         & 40\%          \\
\end{tabular}
\vspace{1mm}
\caption{Per-scene evaluation of our reuse strategy with 8K atlases (Our Temporal) compared to Reverse Reprojection Caching (RRC). Left: Visual quality (mean \FLIP error). Middle: Percentage of fragments reused (number of visible fragments shaded through reuse divided by total visible fragments). Right: Comparison of average per-frame shading time for our method without and with temporal reuse. Results at 1920 $\times$ 1080 screen resolution.}
\label{tab:supp_temporal}
\vspace{-3mm}
\end{table*}

\begin{table*}
\scriptsize
\setlength{\tabcolsep}{3pt}
\centering
\begin{tabular}{l ccc|cccc}
\textbf{Scene}                & \textbf{8K}    & \textbf{4K}    & \textbf{2K}    & \textbf{100\%} & \textbf{50\%}  & \textbf{25\%}  & \textbf{12.50\%} \\
\hline
\textbf{1920 $\times$ 1080}   & & & & & & & \\
Breakfast Room                & 1.006          & 1.002          & 0.960          & 1.006          & 0.255          & 0.065          & 0.017            \\
Sibenik                       & 1.013          & 0.979          & 0.782          & 1.016          & 0.263          & 0.070          & 0.020            \\
Bistro Interior               & 1.037          & 1.004          & 0.503          & 1.053          & 0.291          & 0.088          & 0.031            \\
Bistro Exterior               & 1.119          & 1.066          & 0.421          & 1.139          & 0.357          & 0.129          & 0.058            \\
San Miguel                    & 1.134          & 0.999          & 0.337          & 1.157          & 0.379          & 0.155          & 0.083            \\
Robot Lab                     & 1.025          & 0.977          & 0.429          & 1.047          & 0.286          & 0.084          & 0.027            \\
Sponza                        & 1.018          & 0.946          & 0.419          & 1.034          & 0.274          & 0.076          & 0.022            \\
Greek Villa                   & 1.017          & 0.970          & 0.618          & 1.022          & 0.267          & 0.073          & 0.022            \\
Modern House                  & 1.025          & 0.981          & 0.498          & 1.041          & 0.282          & 0.083          & 0.028            \\
Bistro Interior - Flight Path & 1.020          & 0.963          & 0.455          & 1.038          & 0.279          & 0.081          & 0.026            \\
Bistro Exterior - Flight Path & 1.094          & 1.030          & 0.392          & 1.118          & 0.340          & 0.120          & 0.052            \\
Robot Lab - Flight Path       & 1.029          & 0.981          & 0.381          & 1.055          & 0.291          & 0.088          & 0.030            \\
Modern House - Flight Path    & 1.023          & 0.970          & 0.469          & 1.041          & 0.282          & 0.083          & 0.029            \\
\hline
\textit{avg.}                 & \textit{1.043} & \textit{0.990} & \textit{0.513} & \textit{1.059} & \textit{0.296} & \textit{0.092} & \textit{0.034}   \\
\hline
\textbf{3840 $\times$ 2160}   & & & & & & & \\
Breakfast Room                & 1.000          & 0.957          & 0.251          & 1.003          & 0.252          & 0.064          & 0.016            \\
Sibenik                       & 0.974          & 0.781          & 0.202          & 1.008          & 0.255          & 0.066          & 0.017            \\
Bistro Interior               & 0.980          & 0.496          & 0.128          & 1.026          & 0.266          & 0.075          & 0.022            \\
Bistro Exterior               & 1.008          & 0.399          & 0.109          & 1.084          & 0.307          & 0.098          & 0.036            \\
San Miguel                    & 0.951          & 0.311          & 0.089          & 1.082          & 0.309          & 0.104          & 0.042            \\
Robot Lab                     & 0.950          & 0.420          & 0.108          & 1.022          & 0.261          & 0.072          & 0.021            \\
Sponza                        & 0.940          & 0.412          & 0.106          & 1.019          & 0.259          & 0.070          & 0.019            \\
Greek Villa                   & 0.960          & 0.628          & 0.150          & 1.010          & 0.257          & 0.067          & 0.018            \\
Modern House                  & 0.957          & 0.493          & 0.126          & 1.020          & 0.261          & 0.071          & 0.021            \\
Bistro Interior - Flight Path & 0.943          & 0.451          & 0.115          & 1.019          & 0.260          & 0.071          & 0.020            \\
Bistro Exterior - Flight Path & 0.986          & 0.379          & 0.103          & 1.070          & 0.295          & 0.092          & 0.033            \\
Robot Lab - Flight Path       & 0.952          & 0.367          & 0.096          & 1.027          & 0.264          & 0.074          & 0.022            \\
Modern House - Flight Path    & 0.949          & 0.465          & 0.119          & 1.020          & 0.261          & 0.072          & 0.021            \\
\hline
\textit{avg.}                 & \textit{0.965} & \textit{0.505} & \textit{0.131} & \textit{1.032} & \textit{0.270} & \textit{0.077} & \textit{0.024}  
\end{tabular}

\vspace{1mm}
\caption{Per-scene average shading rates achieved by our method across different fixed atlas sizes (8K, 4K, and 2K) and fixed shading rates (100\%, 50\%, 25\%, and 12.5\%). We measure effective shading rate as the number of shaded fragments on screen divided by the number of texels in the atlas that were read by any fragment. Slight overshading is due to borders for bilinear sampling. Samples (and error) are distributed evenly across the screen due to uniform scaling and perspective projection.}
\label{tab:supp_shading_rate}
\vspace{-3mm}
\end{table*}

\begin{table*}
\scriptsize
\setlength{\tabcolsep}{2pt}
\tiny
\begin{tabular}{lcccccccccc|cccccccccc}
    & \multicolumn{10}{c|}{\textbf{1920 $\times$ 1080}} & \multicolumn{10}{c}{\textbf{3840 $\times$ 2160}} \\
    & Chart          & AABB           & AABB           & Atlas          & Total          & Shading         & Raster         & Other          & Total           & Total           & Chart          & AABB           & AABB           & Atlas          & Total          & Shading          & Raster         & Other          & Total            & Total            \\
    & Extract.       & Comp.          & Pre-Order      & Packing        & Atlasing       &                 &                & Rend.          & Rend.           &                 & Extract.       & Comp.          & Pre-Order      & Packing        & Atlasing       &                  &                & Rend.          & Rend.            &                  \\
\hline
\textbf{8K}        &                &                &                &                &                &                 &                &                &                 &                 &                &                &                &                &                &                  &                &                &                  &                  \\
Breakfast Room     & 0.095          & 0.025          & 0.575          & 0.516          & 1.211          & 14.258          & 0.121          & 0.521          & 14.899          & 16.110          & 0.096          & 0.025          & 0.566          & 0.514          & 1.202          & 46.022           & 0.282          & 2.040          & 48.344           & 49.546           \\
Sibenik            & 0.071          & 0.023          & 0.589          & 0.526          & 1.210          & 20.390          & 0.111          & 0.468          & 20.969          & 22.179          & 0.072          & 0.023          & 0.607          & 0.535          & 1.237          & 60.127           & 0.294          & 1.447          & 61.868           & 63.105           \\
Bistro Interior    & 0.168          & 0.028          & 0.593          & 0.585          & 1.374          & 53.124          & 0.348          & 0.611          & 54.084          & 55.458          & 0.175          & 0.029          & 0.568          & 0.622          & 1.395          & 116.231          & 0.598          & 1.647          & 118.475          & 119.870          \\
Bistro Exterior    & 0.316          & 0.043          & 0.675          & 1.012          & 2.045          & 100.225         & 0.702          & 0.957          & 101.884         & 103.929         & 0.330          & 0.046          & 0.652          & 1.286          & 2.313          & 205.785          & 1.149          & 2.097          & 209.031          & 211.344          \\
San Miguel         & 1.115          & 0.167          & 0.797          & 1.264          & 3.344          & 153.964         & 1.834          & 2.072          & 157.870         & 161.214         & 1.124          & 0.163          & 0.711          & 1.586          & 3.585          & 291.922          & 2.573          & 3.508          & 298.003          & 301.588          \\
Robot Lab          & 0.134          & 0.026          & 0.592          & 0.569          & 1.321          & 54.002          & 0.260          & 0.179          & 54.442          & 55.763          & 0.138          & 0.026          & 0.569          & 0.600          & 1.333          & 111.023          & 0.512          & 0.278          & 111.814          & 113.147          \\
Sponza             & 0.103          & 0.024          & 0.591          & 0.538          & 1.256          & 40.090          & 0.171          & 0.379          & 40.638          & 41.894          & 0.104          & 0.024          & 0.567          & 0.556          & 1.251          & 106.954          & 0.389          & 1.033          & 108.376          & 109.627          \\
Greek Villa        & 0.151          & 0.026          & 0.579          & 0.548          & 1.304          & 30.670          & 0.260          & 0.345          & 31.275          & 32.579          & 0.153          & 0.026          & 0.570          & 0.556          & 1.306          & 76.061           & 0.458          & 0.441          & 76.960           & 78.266           \\
Modern House       & 0.234          & 0.032          & 0.578          & 0.604          & 1.449          & 74.217          & 0.499          & 0.374          & 75.091          & 76.540          & 0.238          & 0.032          & 0.587          & 0.614          & 1.472          & 140.640          & 0.743          & 0.488          & 141.872          & 143.344          \\
Bistro Interior FP & 0.165          & 0.027          & 0.566          & 0.574          & 1.332          & 52.403          & 0.325          & 0.637          & 53.365          & 54.697          & 0.171          & 0.027          & 0.620          & 0.610          & 1.429          & 121.394          & 0.570          & 1.884          & 123.849          & 125.278          \\
Bistro Exterior FP & 0.315          & 0.042          & 0.586          & 1.046          & 1.989          & 95.707          & 0.686          & 0.988          & 97.382          & 99.371          & 0.330          & 0.045          & 0.638          & 1.212          & 2.225          & 183.073          & 1.094          & 2.168          & 186.335          & 188.560          \\
Robot Lab FP       & 0.137          & 0.027          & 0.611          & 0.592          & 1.367          & 62.836          & 0.274          & 0.191          & 63.301          & 64.668          & 0.142          & 0.027          & 0.608          & 0.629          & 1.406          & 123.235          & 0.569          & 0.304          & 124.108          & 125.514          \\
Modern House FP    & 0.234          & 0.031          & 0.578          & 0.608          & 1.452          & 76.257          & 0.500          & 0.377          & 77.134          & 78.586          & 0.237          & 0.033          & 0.600          & 0.632          & 1.501          & 149.517          & 0.754          & 0.487          & 150.757          & 152.258          \\
\hline
\textit{avg.}      & \textit{0.249} & \textit{0.040} & \textit{0.608} & \textit{0.691} & \textit{1.589} & \textit{63.703} & \textit{0.469} & \textit{0.623} & \textit{64.795} & \textit{66.384} & \textit{0.255} & \textit{0.040} & \textit{0.605} & \textit{0.766} & \textit{1.666} & \textit{133.230} & \textit{0.768} & \textit{1.371} & \textit{135.369} & \textit{137.034} \\
\hline
\textbf{4K}        &                &                &                &                &                &                 &                &                &                 &                 &                &                &                &                &                &                  &                &                &                  &                  \\
Breakfast Room     & 0.094          & 0.025          & 0.405          & 0.516          & 1.040          & 14.107          & 0.120          & 0.520          & 14.747          & 15.787          & 0.096          & 0.025          & 0.396          & 0.517          & 1.034          & 44.100           & 0.271          & 2.042          & 46.412           & 47.446           \\
Sibenik            & 0.072          & 0.023          & 0.403          & 0.521          & 1.019          & 19.753          & 0.110          & 0.468          & 20.331          & 21.350          & 0.071          & 0.023          & 0.406          & 0.533          & 1.032          & 49.176           & 0.274          & 1.447          & 50.897           & 51.929           \\
Bistro Interior    & 0.167          & 0.028          & 0.397          & 0.583          & 1.175          & 52.183          & 0.343          & 0.611          & 53.137          & 54.312          & 0.176          & 0.029          & 0.397          & 0.601          & 1.202          & 72.610           & 0.524          & 1.646          & 74.780           & 75.982           \\
Bistro Exterior    & 0.315          & 0.043          & 0.481          & 1.021          & 1.860          & 97.643          & 0.701          & 0.955          & 99.299          & 101.159         & 0.329          & 0.046          & 0.464          & 1.119          & 1.958          & 113.343          & 0.949          & 2.098          & 116.389          & 118.347          \\
San Miguel         & 1.103          & 0.158          & 0.610          & 1.249          & 3.120          & 150.549         & 1.825          & 2.066          & 154.440         & 157.560         & 1.121          & 0.163          & 0.531          & 1.340          & 3.156          & 168.594          & 2.215          & 3.507          & 174.317          & 177.473          \\
Robot Lab          & 0.134          & 0.026          & 0.399          & 0.566          & 1.125          & 52.861          & 0.253          & 0.179          & 53.293          & 54.418          & 0.138          & 0.026          & 0.401          & 0.581          & 1.146          & 67.725           & 0.431          & 0.279          & 68.436           & 69.582           \\
Sponza             & 0.102          & 0.024          & 0.395          & 0.537          & 1.059          & 38.178          & 0.170          & 0.382          & 38.730          & 39.789          & 0.104          & 0.024          & 0.400          & 0.546          & 1.075          & 55.244           & 0.326          & 1.033          & 56.603           & 57.678           \\
Greek Villa        & 0.151          & 0.026          & 0.397          & 0.545          & 1.120          & 29.673          & 0.257          & 0.349          & 30.279          & 31.399          & 0.153          & 0.026          & 0.400          & 0.553          & 1.133          & 54.195           & 0.413          & 0.456          & 55.063           & 56.196           \\
Modern House       & 0.235          & 0.031          & 0.421          & 0.581          & 1.268          & 74.401          & 0.496          & 0.386          & 75.283          & 76.551          & 0.239          & 0.032          & 0.398          & 0.603          & 1.272          & 99.260           & 0.684          & 0.488          & 100.432          & 101.704          \\
Bistro Interior FP & 0.165          & 0.027          & 0.396          & 0.571          & 1.159          & 50.975          & 0.323          & 0.636          & 51.934          & 53.093          & 0.171          & 0.027          & 0.421          & 0.589          & 1.208          & 71.077           & 0.492          & 1.884          & 73.453           & 74.661           \\
Bistro Exterior FP & 0.315          & 0.042          & 0.415          & 1.027          & 1.799          & 93.465          & 0.682          & 0.991          & 95.138          & 96.937          & 0.331          & 0.046          & 0.437          & 1.088          & 1.902          & 107.277          & 0.928          & 2.171          & 110.376          & 112.278          \\
Robot Lab FP       & 0.138          & 0.027          & 0.407          & 0.584          & 1.156          & 61.820          & 0.272          & 0.191          & 62.282          & 63.438          & 0.142          & 0.027          & 0.407          & 0.599          & 1.175          & 72.741           & 0.470          & 0.305          & 73.516           & 74.691           \\
Modern House FP    & 0.233          & 0.031          & 0.398          & 0.592          & 1.254          & 75.205          & 0.495          & 0.386          & 76.086          & 77.340          & 0.236          & 0.033          & 0.419          & 0.613          & 1.301          & 99.660           & 0.688          & 0.487          & 100.835          & 102.136          \\
\hline
\textit{avg.}      & \textit{0.248} & \textit{0.039} & \textit{0.425} & \textit{0.684} & \textit{1.396} & \textit{62.370} & \textit{0.465} & \textit{0.625} & \textit{63.460} & \textit{64.856} & \textit{0.254} & \textit{0.041} & \textit{0.421} & \textit{0.714} & \textit{1.430} & \textit{82.692}  & \textit{0.667} & \textit{1.373} & \textit{84.731}  & \textit{86.162}  \\
\hline
\textbf{2K}        &                &                &                &                &                &                 &                &                &                 &                 &                &                &                &                &                &                  &                &                &                  &                  \\
Breakfast Room     & 0.094          & 0.025          & 0.340          & 0.513          & 0.972          & 13.628          & 0.118          & 0.519          & 14.266          & 15.238          & 0.096          & 0.025          & 0.340          & 0.514          & 0.975          & 13.981           & 0.201          & 2.040          & 16.222           & 17.197           \\
Sibenik            & 0.072          & 0.023          & 0.335          & 0.521          & 0.952          & 16.884          & 0.105          & 0.468          & 17.457          & 18.409          & 0.072          & 0.023          & 0.333          & 0.520          & 0.947          & 17.151           & 0.238          & 1.447          & 18.835           & 19.782           \\
Bistro Interior    & 0.168          & 0.028          & 0.339          & 0.572          & 1.107          & 39.755          & 0.327          & 0.611          & 40.694          & 41.801          & 0.175          & 0.029          & 0.342          & 0.571          & 1.116          & 40.641           & 0.490          & 1.646          & 42.777           & 43.893           \\
Bistro Exterior    & 0.315          & 0.043          & 0.418          & 0.955          & 1.732          & 70.196          & 0.668          & 0.955          & 71.819          & 73.551          & 0.329          & 0.046          & 0.415          & 0.957          & 1.747          & 72.406           & 0.904          & 2.095          & 75.405           & 77.152           \\
San Miguel         & 1.117          & 0.162          & 0.527          & 1.141          & 2.947          & 105.550         & 1.783          & 2.069          & 109.402         & 112.349         & 1.122          & 0.169          & 0.454          & 1.128          & 2.873          & 122.248          & 2.142          & 3.505          & 127.895          & 130.768          \\
Robot Lab          & 0.135          & 0.026          & 0.342          & 0.556          & 1.059          & 40.608          & 0.229          & 0.178          & 41.015          & 42.074          & 0.138          & 0.026          & 0.341          & 0.550          & 1.055          & 40.837           & 0.400          & 0.277          & 41.514           & 42.569           \\
Sponza             & 0.103          & 0.024          & 0.337          & 0.531          & 0.994          & 24.629          & 0.152          & 0.374          & 25.154          & 26.148          & 0.104          & 0.024          & 0.340          & 0.527          & 0.996          & 24.611           & 0.305          & 1.035          & 25.951           & 26.947           \\
Greek Villa        & 0.151          & 0.026          & 0.343          & 0.541          & 1.061          & 23.453          & 0.246          & 0.367          & 24.065          & 25.126          & 0.153          & 0.026          & 0.341          & 0.536          & 1.056          & 23.548           & 0.380          & 0.445          & 24.374           & 25.430           \\
Modern House       & 0.233          & 0.032          & 0.342          & 0.577          & 1.183          & 62.504          & 0.479          & 0.385          & 63.368          & 64.551          & 0.237          & 0.033          & 0.342          & 0.571          & 1.182          & 64.288           & 0.648          & 0.488          & 65.424           & 66.606           \\
Bistro Interior FP & 0.165          & 0.027          & 0.340          & 0.561          & 1.093          & 37.013          & 0.310          & 0.635          & 37.958          & 39.051          & 0.171          & 0.027          & 0.341          & 0.559          & 1.098          & 37.233           & 0.469          & 1.883          & 39.585           & 40.683           \\
Bistro Exterior FP & 0.315          & 0.042          & 0.362          & 0.984          & 1.703          & 70.537          & 0.651          & 0.991          & 72.179          & 73.882          & 0.330          & 0.045          & 0.357          & 0.967          & 1.699          & 72.001           & 0.884          & 2.168          & 75.053           & 76.752           \\
Robot Lab FP       & 0.137          & 0.027          & 0.340          & 0.570          & 1.075          & 46.658          & 0.251          & 0.190          & 47.099          & 48.174          & 0.142          & 0.027          & 0.341          & 0.563          & 1.074          & 47.073           & 0.438          & 0.304          & 47.815           & 48.889           \\
Modern House FP    & 0.234          & 0.031          & 0.339          & 0.581          & 1.185          & 60.945          & 0.479          & 0.386          & 61.810          & 62.995          & 0.236          & 0.032          & 0.341          & 0.578          & 1.188          & 63.653           & 0.655          & 0.487          & 64.795           & 65.983           \\
\hline
\textit{avg.}      & \textit{0.249} & \textit{0.040} & \textit{0.362} & \textit{0.662} & \textit{1.313} & \textit{47.105} & \textit{0.446} & \textit{0.625} & \textit{48.176} & \textit{49.488} & \textit{0.254} & \textit{0.041} & \textit{0.356} & \textit{0.657} & \textit{1.308} & \textit{49.205}  & \textit{0.627} & \textit{1.371} & \textit{51.203}  & \textit{52.512} 
\end{tabular}
\vspace{1mm}
\caption{Per-scene average runtimes (milliseconds) for FastAtlas when shading using fixed-size atlases, across both 1920 $\times$ 1080 and 3840 $\times$ 2160 screen resolutions. Column labels from left to right: chart extraction, AABB computation, AABB pre-ordering, atlas packing, total atlasing (sum of previous four columns), shading, rasterization, other rendering/presentation, total rendering (sum of previous three columns), total time (sum of total rendering and total atlasing).}
\label{tab:supp_runtime_fixed_atlas}
\vspace{-3mm}
\end{table*}

\begin{table*}
\scriptsize
\setlength{\tabcolsep}{2pt}
\tiny
\begin{tabular}{lcccccccccc|cccccccccc}
    & \multicolumn{10}{c|}{\textbf{1920 $\times$ 1080}} & \multicolumn{10}{c}{\textbf{3840 $\times$ 2160}} \\
    & Chart          & AABB           & AABB           & Atlas          & Total          & Shading         & Raster         & Other          & Total           & Total           & Chart          & AABB           & AABB           & Atlas          & Total          & Shading         & Raster         & Other          & Total           & Total           \\
    & Extract.       & Comp.          & Pre-Order      & Packing        & Atlasing       &                 &                & Rend.          & Rend.           &                 & Extract.       & Comp.          & Pre-Order      & Packing        & Atlasing       &                 &                & Rend.          & Rend.           &                 \\
\hline
\textbf{8K 50\%}   &                &                &                &                &                &                 &                &                &                 &                 &                &                &                &                &                &                 &                &                &                 &                 \\
Breakfast Room     & 0.095          & 0.025          & 0.566          & 0.514          & 1.199          & 6.044           & 0.101          & 0.521          & 6.666           & 7.865           & 0.096          & 0.025          & 0.590          & 0.517          & 1.227          & 14.309          & 0.205          & 2.039          & 16.553          & 17.780          \\
Sibenik            & 0.071          & 0.023          & 0.575          & 0.525          & 1.194          & 9.448           & 0.101          & 0.468          & 10.017          & 11.211          & 0.072          & 0.023          & 0.578          & 0.533          & 1.206          & 20.312          & 0.245          & 1.444          & 22.001          & 23.207          \\
Bistro Interior    & 0.168          & 0.028          & 0.610          & 0.583          & 1.389          & 35.149          & 0.321          & 0.612          & 36.083          & 37.472          & 0.175          & 0.029          & 0.609          & 0.619          & 1.432          & 54.098          & 0.501          & 1.646          & 56.245          & 57.677          \\
Bistro Exterior    & 0.317          & 0.043          & 0.653          & 1.028          & 2.041          & 69.524          & 0.664          & 0.955          & 71.143          & 73.184          & 0.330          & 0.046          & 0.678          & 1.293          & 2.346          & 105.788         & 0.941          & 2.094          & 108.823         & 111.169         \\
San Miguel         & 1.104          & 0.167          & 0.772          & 1.258          & 3.301          & 109.991         & 1.787          & 2.071          & 113.849         & 117.150         & 1.139          & 0.165          & 0.695          & 1.564          & 3.563          & 173.065         & 2.217          & 3.505          & 178.787         & 182.350         \\
Robot Lab          & 0.134          & 0.026          & 0.572          & 0.569          & 1.301          & 37.575          & 0.226          & 0.179          & 37.980          & 39.281          & 0.138          & 0.026          & 0.591          & 0.598          & 1.354          & 54.817          & 0.413          & 0.279          & 55.509          & 56.863          \\
Sponza             & 0.102          & 0.024          & 0.567          & 0.539          & 1.232          & 21.065          & 0.152          & 0.376          & 21.593          & 22.825          & 0.104          & 0.024          & 0.589          & 0.560          & 1.277          & 40.334          & 0.313          & 1.033          & 41.680          & 42.957          \\
Greek Villa        & 0.151          & 0.026          & 0.568          & 0.547          & 1.292          & 17.730          & 0.236          & 0.354          & 18.320          & 19.612          & 0.153          & 0.026          & 0.589          & 0.556          & 1.324          & 31.163          & 0.387          & 0.444          & 31.994          & 33.318          \\
Modern House       & 0.234          & 0.031          & 0.588          & 0.581          & 1.435          & 57.913          & 0.472          & 0.385          & 58.770          & 60.205          & 0.239          & 0.032          & 0.608          & 0.616          & 1.495          & 77.593          & 0.666          & 0.489          & 78.747          & 80.242          \\
Bistro Interior FP & 0.165          & 0.027          & 0.569          & 0.573          & 1.334          & 32.935          & 0.305          & 0.636          & 33.876          & 35.210          & 0.171          & 0.027          & 0.607          & 0.606          & 1.410          & 53.014          & 0.479          & 1.883          & 55.375          & 56.785          \\
Bistro Exterior FP & 0.315          & 0.042          & 0.588          & 1.033          & 1.979          & 69.294          & 0.649          & 0.987          & 70.929          & 72.908          & 0.330          & 0.046          & 0.625          & 1.215          & 2.216          & 99.239          & 0.916          & 2.169          & 102.324         & 104.540         \\
Robot Lab FP       & 0.137          & 0.027          & 0.571          & 0.587          & 1.322          & 44.335          & 0.248          & 0.191          & 44.775          & 46.097          & 0.142          & 0.027          & 0.589          & 0.629          & 1.387          & 63.734          & 0.459          & 0.304          & 64.497          & 65.884          \\
Modern House FP    & 0.234          & 0.031          & 0.588          & 0.593          & 1.446          & 56.184          & 0.472          & 0.386          & 57.042          & 58.488          & 0.238          & 0.032          & 0.589          & 0.633          & 1.492          & 79.377          & 0.673          & 0.488          & 80.537          & 82.029          \\
\hline
\textit{avg.}      & \textit{0.248} & \textit{0.040} & \textit{0.599} & \textit{0.687} & \textit{1.574} & \textit{43.630} & \textit{0.441} & \textit{0.625} & \textit{44.696} & \textit{46.270} & \textit{0.256} & \textit{0.041} & \textit{0.611} & \textit{0.765} & \textit{1.671} & \textit{66.680} & \textit{0.647} & \textit{1.371} & \textit{68.698} & \textit{70.369} \\
\hline
\textbf{8K 25\%}   &                &                &                &                &                &                 &                &                &                 &                 &                &                &                &                &                &                 &                &                &                 &                 \\
Breakfast Room     & 0.094          & 0.025          & 0.566          & 0.514          & 1.200          & 4.117           & 0.097          & 0.521          & 4.734           & 5.934           & 0.096          & 0.025          & 0.587          & 0.517          & 1.225          & 6.117           & 0.185          & 2.033          & 8.336           & 9.561           \\
Sibenik            & 0.071          & 0.023          & 0.611          & 0.526          & 1.232          & 6.534           & 0.099          & 0.468          & 7.101           & 8.333           & 0.072          & 0.023          & 0.579          & 0.536          & 1.210          & 9.431           & 0.230          & 1.446          & 11.107          & 12.317          \\
Bistro Interior    & 0.168          & 0.028          & 0.567          & 0.584          & 1.347          & 28.738          & 0.303          & 0.610          & 29.652          & 30.999          & 0.175          & 0.029          & 0.608          & 0.620          & 1.432          & 35.938          & 0.482          & 1.646          & 38.066          & 39.498          \\
Bistro Exterior    & 0.315          & 0.043          & 0.651          & 1.021          & 2.030          & 59.166          & 0.641          & 0.954          & 60.762          & 62.792          & 0.330          & 0.046          & 0.679          & 1.287          & 2.341          & 72.362          & 0.903          & 2.094          & 75.358          & 77.699          \\
San Miguel         & 1.118          & 0.164          & 0.765          & 1.249          & 3.297          & 94.084          & 1.754          & 2.066          & 97.904          & 101.201         & 1.117          & 0.160          & 0.712          & 1.574          & 3.562          & 127.817         & 2.154          & 3.503          & 133.474         & 137.036         \\
Robot Lab          & 0.134          & 0.026          & 0.622          & 0.567          & 1.349          & 32.187          & 0.205          & 0.178          & 32.570          & 33.919          & 0.138          & 0.026          & 0.589          & 0.599          & 1.352          & 37.693          & 0.394          & 0.278          & 38.365          & 39.717          \\
Sponza             & 0.103          & 0.024          & 0.566          & 0.538          & 1.231          & 16.110          & 0.145          & 0.382          & 16.637          & 17.868          & 0.104          & 0.024          & 0.588          & 0.558          & 1.275          & 21.034          & 0.305          & 1.032          & 22.371          & 23.646          \\
Greek Villa        & 0.151          & 0.026          & 0.620          & 0.545          & 1.342          & 14.965          & 0.223          & 0.356          & 15.544          & 16.886          & 0.153          & 0.026          & 0.589          & 0.555          & 1.324          & 18.213          & 0.370          & 0.441          & 19.024          & 20.348          \\
Modern House       & 0.235          & 0.031          & 0.588          & 0.583          & 1.437          & 52.303          & 0.451          & 0.385          & 53.140          & 54.577          & 0.237          & 0.032          & 0.569          & 0.619          & 1.458          & 59.241          & 0.638          & 0.488          & 60.366          & 61.824          \\
Bistro Interior FP & 0.165          & 0.027          & 0.569          & 0.573          & 1.334          & 27.472          & 0.288          & 0.636          & 28.395          & 29.729          & 0.171          & 0.027          & 0.606          & 0.605          & 1.410          & 32.896          & 0.463          & 1.883          & 35.242          & 36.652          \\
Bistro Exterior FP & 0.316          & 0.042          & 0.588          & 1.040          & 1.985          & 62.138          & 0.625          & 0.987          & 63.749          & 65.734          & 0.330          & 0.045          & 0.624          & 1.209          & 2.208          & 71.423          & 0.884          & 2.168          & 74.476          & 76.684          \\
Robot Lab FP       & 0.138          & 0.027          & 0.570          & 0.586          & 1.320          & 38.046          & 0.228          & 0.191          & 38.465          & 39.785          & 0.143          & 0.027          & 0.589          & 0.629          & 1.389          & 44.823          & 0.433          & 0.303          & 45.560          & 46.949          \\
Modern House FP    & 0.234          & 0.031          & 0.588          & 0.593          & 1.446          & 50.166          & 0.452          & 0.385          & 51.003          & 52.449          & 0.237          & 0.033          & 0.601          & 0.635          & 1.505          & 58.666          & 0.646          & 0.486          & 59.798          & 61.303          \\
\hline
\textit{avg.}      & \textit{0.249} & \textit{0.040} & \textit{0.605} & \textit{0.686} & \textit{1.581} & \textit{37.387} & \textit{0.424} & \textit{0.625} & \textit{38.435} & \textit{40.016} & \textit{0.254} & \textit{0.040} & \textit{0.609} & \textit{0.765} & \textit{1.669} & \textit{45.820} & \textit{0.622} & \textit{1.369} & \textit{47.811} & \textit{49.480} \\
\hline
\textbf{8K 12.5\%} &                &                &                &                &                &                 &                &                &                 &                 &                &                &                &                &                &                 &                &                &                 &                 \\
Breakfast Room     & 0.094          & 0.025          & 0.566          & 0.514          & 1.199          & 3.463           & 0.094          & 0.520          & 4.077           & 5.276           & 0.096          & 0.025          & 0.590          & 0.517          & 1.228          & 4.220           & 0.181          & 2.039          & 6.440           & 7.668           \\
Sibenik            & 0.071          & 0.023          & 0.607          & 0.524          & 1.224          & 5.639           & 0.098          & 0.467          & 6.204           & 7.428           & 0.072          & 0.023          & 0.580          & 0.535          & 1.210          & 6.519           & 0.234          & 1.445          & 8.198           & 9.408           \\
Bistro Interior    & 0.167          & 0.028          & 0.570          & 0.584          & 1.350          & 26.800          & 0.292          & 0.612          & 27.703          & 29.053          & 0.175          & 0.029          & 0.607          & 0.618          & 1.429          & 29.474          & 0.463          & 1.645          & 31.582          & 33.011          \\
Bistro Exterior    & 0.315          & 0.043          & 0.651          & 1.041          & 2.050          & 54.147          & 0.624          & 0.954          & 55.725          & 57.775          & 0.329          & 0.046          & 0.677          & 1.286          & 2.338          & 61.380          & 0.876          & 2.095          & 64.350          & 66.688          \\
San Miguel         & 1.115          & 0.167          & 0.775          & 1.255          & 3.313          & 85.547          & 1.731          & 2.066          & 89.343          & 92.656          & 1.120          & 0.170          & 0.712          & 1.574          & 3.575          & 111.764         & 2.108          & 3.580          & 117.452         & 121.027         \\
Robot Lab          & 0.135          & 0.026          & 0.571          & 0.569          & 1.301          & 29.735          & 0.194          & 0.178          & 30.107          & 31.408          & 0.139          & 0.026          & 0.589          & 0.598          & 1.352          & 32.693          & 0.376          & 0.278          & 33.348          & 34.700          \\
Sponza             & 0.102          & 0.024          & 0.617          & 0.540          & 1.284          & 15.372          & 0.143          & 0.381          & 15.896          & 17.180          & 0.104          & 0.024          & 0.589          & 0.556          & 1.273          & 16.069          & 0.305          & 1.035          & 17.408          & 18.681          \\
Greek Villa        & 0.151          & 0.026          & 0.568          & 0.547          & 1.292          & 14.334          & 0.217          & 0.348          & 14.899          & 16.191          & 0.153          & 0.026          & 0.589          & 0.555          & 1.324          & 15.502          & 0.360          & 0.443          & 16.305          & 17.629          \\
Modern House       & 0.234          & 0.031          & 0.588          & 0.584          & 1.437          & 51.546          & 0.440          & 0.385          & 52.372          & 53.809          & 0.239          & 0.032          & 0.600          & 0.616          & 1.488          & 54.438          & 0.614          & 0.487          & 55.539          & 57.027          \\
Bistro Interior FP & 0.165          & 0.027          & 0.566          & 0.572          & 1.331          & 25.869          & 0.282          & 0.635          & 26.786          & 28.117          & 0.171          & 0.027          & 0.606          & 0.605          & 1.410          & 27.364          & 0.448          & 1.883          & 29.695          & 31.105          \\
Bistro Exterior FP & 0.315          & 0.042          & 0.588          & 1.043          & 1.989          & 58.394          & 0.611          & 0.988          & 59.993          & 61.982          & 0.331          & 0.046          & 0.624          & 1.215          & 2.215          & 63.309          & 0.858          & 2.172          & 66.339          & 68.554          \\
Robot Lab FP       & 0.137          & 0.027          & 0.571          & 0.591          & 1.326          & 35.193          & 0.214          & 0.190          & 35.597          & 36.923          & 0.143          & 0.027          & 0.590          & 0.629          & 1.388          & 38.707          & 0.408          & 0.304          & 39.419          & 40.807          \\
Modern House FP    & 0.234          & 0.031          & 0.588          & 0.594          & 1.446          & 49.004          & 0.442          & 0.385          & 49.831          & 51.277          & 0.236          & 0.033          & 0.601          & 0.631          & 1.501          & 52.957          & 0.625          & 0.486          & 54.068          & 55.569          \\
\hline
\textit{avg.}      & \textit{0.249} & \textit{0.040} & \textit{0.602} & \textit{0.689} & \textit{1.580} & \textit{35.003} & \textit{0.414} & \textit{0.624} & \textit{36.041} & \textit{37.621} & \textit{0.254} & \textit{0.041} & \textit{0.612} & \textit{0.764} & \textit{1.672} & \textit{39.569} & \textit{0.604} & \textit{1.376} & \textit{41.549} & \textit{43.221}
\end{tabular}
\vspace{1mm}
\caption{Per-scene average runtimes (milliseconds) for FastAtlas when targeting fixed shading rates, across both 1920 $\times$ 1080 and 3840 $\times$ 2160 screen resolutions. Column labels from left to right: chart extraction, AABB computation, AABB pre-ordering, atlas packing, total atlasing (sum of previous four columns), shading, rasterization, other rendering/presentation, total rendering (sum of previous three columns), total time (sum of total rendering and total atlasing).}
\label{tab:supp_runtime_fixed_sr}
\vspace{-3mm}
\end{table*}

\end{document}